\pdfoutput=1

\documentclass[11pt,twoside,a4paper,cmspaper,final,collab]{cms-tdr}
\def\svnVersion{257340}\def\cmsCernNo{2013-037}\def\cmsCernDate{\today}\def\cmsMessage{Published in Physical Review D as \href{http://dx.doi.org/10.1103/PhysRevD.90.032004}{\doi{10.1103/PhysRevD.90.032004.}}}

\begin{document}\cmsNoteHeader{SMP-12-021}

\hyphenation{had-ron-i-za-tion}
\hyphenation{cal-or-i-me-ter}
\hyphenation{de-vices}

\RCS$Revision: 257277 $
\RCS$HeadURL: svn+ssh://svn.cern.ch/reps/tdr2/papers/SMP-12-021/trunk/SMP-12-021.tex $
\RCS$Id: SMP-12-021.tex 257277 2014-08-22 15:05:41Z alverson $
\newlength\cmsFigWidth
\ifthenelse{\boolean{cms@external}}{\setlength\cmsFigWidth{0.85\columnwidth}}{\setlength\cmsFigWidth{0.45\textwidth}}
\ifthenelse{\boolean{cms@external}}{\providecommand{\cmsLeft}{top}}{\providecommand{\cmsLeft}{left}}
\ifthenelse{\boolean{cms@external}}{\providecommand{\cmsRight}{bottom}}{\providecommand{\cmsRight}{right}}
\ifthenelse{\boolean{cms@external}}{}{\newcolumntype{d}{D{.}{.}{1}}}

\ifthenelse{\boolean{cms@external}}{\providecommand{\CL}{C.L.\xspace}}{\providecommand{\CL}{CL\xspace}}
\ifthenelse{\boolean{cms@external}}{\providecommand{\CLend}{C.L.\xspace}}{\providecommand{\CLend}{CL.\xspace}}
\ifthenelse{\boolean{cms@external}}{\providecommand{\suppMaterial}{as supplemental material [URL will be inserted by publisher]}}
{\providecommand{\suppMaterial}{in the Appendix}}

\providecommand{\rX}{\ensuremath{\cmsSymbolFace{X}}\xspace}
\providecommand{\EmT}{\ensuremath{\Em_\mathrm{T}}\xspace}
\providecommand{\vecEmT}{\ensuremath{\vec{\Em}_\mathrm{T}}\xspace}
\providecommand{\qT}{\ensuremath{q_\mathrm{T}}\xspace}
\providecommand{\vecqT}{\ensuremath{\vec{q}_\mathrm{T}}\xspace}

\newcommand{\wlnu}{\PW\to \ell\Pgn\xspace}
\newcommand{\wmunu}{\PW\to \Pgm\Pgn\xspace}
\newcommand{\wtaunu}{\PW\to  \Pgt\Pgn\xspace}
\newcommand{\wenu}{\PW\to\Pe\Pgn}
\newcommand{\zll}{ \cPZ/\gamma^*\to \ell^+ \ell^-\xspace}
\newcommand{\zmumu}{\cPZ/\gamma^*\to \Pgmp\Pgmm\xspace}
\newcommand{\zee}{\cPZ/\gamma^*\to \Pep\Pem\xspace}
\newcommand{\ztautau}{\cPZ/\gamma^*\to\TT\xspace}
\newcommand{\upar}{\ensuremath{u_{||}}}
\newcommand{\uper}{\ensuremath{u_{\perp}}}
\newcommand\difsig{\ensuremath{ \frac{ \cPqd\sigma}{\cPqd\eta}}}

\newcommand{\myttbar}{ \cPqt\cPaqt}

\newcommand{\ADATA}{{\mathcal{A}}^{\text{data}}}
\newcommand{\ATHEORY}{{\mathcal{A}}^{\text{theory}}}
\newcommand{\SIGDATATOT}{{\sigma^{\text{data}}_{\text{tot}}}}

\newcommand{\NWp}   {\ensuremath{N^{\PWp}}\xspace}
\newcommand{\NWm}   {\ensuremath{N^{\PWm}}\xspace}
\newcommand{\ARAW}  {\ensuremath{\mathcal{A}^{\text{raw}}}\xspace}
\newcommand{\ATRUE} {{\mathcal{A}}^{\text{true}}}
\newcommand{\etaabs} {\ensuremath{\abs{ \eta }}\xspace}
\newcommand{\uparmean}  {\ensuremath{\langle\upar\rangle}\xspace}
\newcommand{\upermean} {\ensuremath{\langle\uper\rangle}\xspace}

\cmsNoteHeader{SMP-12-021} 
\title{Measurement of the muon charge asymmetry in inclusive \texorpdfstring{$\Pp\Pp\to\PW+\rX$}{pp to WX} production at \texorpdfstring{$\sqrt{s} = 7$\TeV}{sqrt(s)=7 TeV} and an improved determination of light parton distribution functions}

\date{\today}

\abstract{Measurements of the muon charge asymmetry in inclusive $\Pp\Pp\to\PW+\rX$ production at $\sqrt{s}=7\TeV$ are presented. The data sample corresponds to  an integrated luminosity of 4.7\fbinv recorded with the CMS detector at the LHC.   With a sample of more than 20 million  $\PW\to\Pgm\Pgn$  events, the  statistical precision is greatly improved  in comparison to previous measurements.  These new results provide additional constraints on the
parton distribution functions of the proton in the range of the Bjorken scaling variable $x$ from  $10^{-3}$ to $10^{-1}$. These measurements and the recent CMS measurement of associated  $\PW + \text{charm}$ production are used together with the cross sections for inclusive deep inelastic $\Pe^\pm \Pp$ scattering at HERA in a next-to-leading-order QCD analysis. The determination of the valence quark distributions is improved, and the strange-quark distribution is probed directly through the leading-order process $\Pg + \cPqs\to\PW + \cPqc$ in proton-proton collisions at the LHC.}

\hypersetup{%
pdfauthor={CMS Collaboration},%
pdftitle={Measurement of the muon charge asymmetry in inclusive pp to WX production at sqrt(s)=7 TeV and an improved determination of light parton distribution functions},%
pdfsubject={CMS},%
pdfkeywords={CMS, physics, charge asymmetry, W boson, QCD analysis}}
\maketitle 
\section{Introduction}
\label{sec:introduction}

In the standard model~(SM), the dominant processes for inclusive
$\PW$-boson production in $\Pp\Pp$ collisions are annihilation
 processes:
$\cPqu\, \cPaqd \to \PWp$ and $\cPqd\, \cPaqu \to \PWm$ involving a valence quark from one proton and a sea antiquark
from the other.
 Since there are two valence $\cPqu$ quarks and one valence $\cPqd$ quark in the proton, $\PWp$ bosons are produced
more often than  $\PWm$ bosons. The Compact Muon Solenoid~(CMS) experiment at  the Large Hadron Collider~(LHC)
has investigated
this production asymmetry in inclusive $\PW$-boson production and
 measured the
 inclusive ratio of total cross sections for $\PWp$ and $\PWm$ boson
production at $\sqrt{s}=7\TeV$ to be
$1.421 \pm 0.006\stat \pm 0.032\syst$~\cite{CMS:WZ}.
 This result is in agreement with SM  predictions based on various
parton distribution functions~(PDFs) such as the MSTW2008
 and {CT10} PDF sets~\cite{Martin:2009ad, CTEQ:1007}.
 Measurements of the  production asymmetry
between $\PWp$ and $\PWm$ bosons as a function of boson rapidity can
provide additional constraints on  the $\cPqd/\cPqu$ ratio and on the sea antiquark densities
in the proton.  For $\Pp\Pp$ collisions at $\sqrt{s}=7\TeV$, these measurements explore the
PDFs for the proton for  Bjorken $x$ from  $10^{-3}$ to $10^{-1}$~\cite{BJORKEN}.
However,  it is difficult
to measure the boson rapidity production asymmetry because
of the energy carried away by neutrinos in leptonic $\PW$-boson decays.
A quantity more directly accessible experimentally
is  the lepton charge asymmetry, defined as
\begin{equation}
  \mathcal{A}(\eta) = \frac{\difsig(\PWp\to\ell^+\Pgn)-\difsig(\PWm\to\ell^-\Pagn)}
   {\difsig(\PWp\to\ell^+\Pgn)+\difsig(\PWm\to\ell^-\Pagn)},
\end{equation}
where  $\cPqd\sigma/\cPqd\eta$ is the differential
cross section for $\PW$-boson production and subsequent leptonic decay and  $\eta= -\ln{[\tan{(\theta/2)}]}$ is the charged lepton
pseudorapidity in the laboratory frame, with
 $\theta$ being the polar angle measured with respect to the beam axis.

High precision measurements of the $\PW$-boson lepton charge asymmetry
can improve the determination of the PDFs.
Both the  $\PW$-boson lepton charge asymmetry and the $\PW$-boson
production charge
asymmetry were studied in  $\Pp \Pap$ collisions by
the CDF and D0 experiments at the Tevatron collider~\cite{CDF:wasym, D0:asym, Abazov:2013rja}.
The ATLAS,  CMS, and LHCb experiments also reported measurements of the
lepton charge asymmetry using  data
collected at the LHC in 2010~\cite{ATLAS:asym:2010, PhysRevD.85.072004, CMS:asym:2010, LHCb:wz:2010}.
An earlier
measurement of the $\PW$-boson electron charge asymmetry is based on
 2011 CMS data corresponding to  an integrated luminosity of
0.84\fbinv~\cite{CMS-PAS-SMP-12-001}.

The impact of CMS measurements of the lepton charge
asymmetry on the global PDF fits has been studied by several
groups~\cite{Ball:2010gb, Ball:2011gg,  MSTW:12, arxiv_1202.4642, Alekhin:2013nda},
who concluded that
 improvements  in the PDF uncertainties
for several quark flavors could be
achieved
 with more precise data.
In this paper, we report  a measurement of the muon charge asymmetry
using a data sample corresponding to
  an integrated luminosity of 4.7\fbinv collected with the
CMS detector at the LHC in  2011.  The number of $\wmunu$  events (more than 20 million)   in this data sample is
 2 orders of magnitude larger than for our previous measurement~\cite{CMS:asym:2010}.

This precise measurement of the muon charge asymmetry and the recent CMS measurement of associated $\PW + \text{charm}$
production~\cite{wpluscharm_cms}  are combined with the cross
sections for inclusive deep inelastic $\Pe^\pm \Pp$
scattering at HERA~\cite{Aaron:2009aa} in a quantum chromodynamics~(QCD) analysis at next-to-leading
order~(NLO).
The impact of these measurements of $\PW$-boson production at CMS on the determination of light-quark distributions
in the proton is studied and the strange-quark density is determined.

This paper is organized as follows. A brief description of the CMS detector
is given in Section~\ref{sec:cms}. The selection of  $\wmunu$ candidates
is described in Section~\ref{sec:selection}. The  corrections for residual
charge-specific bias in the measurement of the  muon
transverse momentum~($\pt$)
 and in the  muon trigger, reconstruction,  and selection
efficiencies are discussed in Section~\ref{sec:eff_scale}. The extraction of the
   $\wmunu$ signal  is described in detail in
Section~\ref{sec:signal_estimation}. Systematic uncertainties and
the full correlation matrix are given in
Section~\ref{sec:systematics}.
The final measurements are  presented in
Section~\ref{sec:results} and the QCD analysis is discussed in detail in Section~\ref{sec:QCD}. The summary and conclusion
follow in  Section~\ref{sec:summary}.

\section{The CMS experiment}
\label{sec:cms}

The central feature of the CMS apparatus is
a superconducting solenoid  6\unit{m} in diameter and 13\unit{m} long,
which
provides an axial field of
3.8\unit{T}. Within the field volume are a silicon pixel and strip tracker,
a crystal electromagnetic calorimeter (ECAL), and a brass/scintillator
hadron calorimeter. Muons are measured in gas-ionization detectors
embedded in the steel flux return yoke.
 The ECAL consists
of nearly 76\,000 lead tungstate crystals that provide coverage in pseudorapidity $\abs{\eta} <1.479$  in the
 barrel region and $1.479 < \abs{\eta} < 3.0$ in the two endcap regions.
A preshower detector consisting of two planes of silicon sensors interleaved with a total of three radiation lengths
 of lead is located in front of the ECAL endcaps.
Muons are selected in the pseudorapidity range $\abs{\eta} < 2.4$,
with detection planes constructed of  drift tubes~(DT), cathode strip chambers~(CSC), and resistive plate chambers, and matched to the
tracks measured in the silicon tracker resulting in an $\eta$-dependent \pt resolution of about 1--5\%  for muon \pt up to 1\TeV. The inner tracker,
consisting of 1440 silicon pixel and 15\,148 silicon strip detector modules,
measures charged particles within the pseudorapidity range $\abs{\eta}< 2.5$.
It provides an impact parameter resolution of ${\sim}15\mum$ and a \pt resolution of about 1.5\% for 100\GeV particles.

The CMS experiment
uses a right-handed coordinate system, with the origin at the
 nominal interaction point, the $x$ axis pointing toward the center of the LHC,
the $y$ axis pointing up (perpendicular to the LHC plane), and the $z$ axis
along the counterclockwise-beam direction. The polar angle, $\theta$, is
 measured from the positive $z$ axis and the azimuthal angle, $\phi$, is
measured in the $x$-$y$ plane.

A detailed description of the CMS experiment can be found in Ref.~\cite{Chatrchyan:2008zzk}.

\section{Event reconstruction}
\label{sec:selection}

The  signature of a  $\wmunu$ event is a
high-$\pt$
muon accompanied by missing transverse momentum~$\vecEmT$  due
to the escaping neutrino. The CMS experiment has utilized a
 particle-flow algorithm in event reconstruction, and the
$\vecEmT$ used by this analysis
is determined as
 the negative vector sum of the
 transverse momenta of
all  particles reconstructed by this algorithm~\cite{pfmet}.
The  $\wmunu$ candidates were
collected with a set of isolated
single-muon triggers with different \pt thresholds,
which is the major difference with respect to  the previous
CMS measurement where nonisolated single-muon triggers were used~\cite{ CMS:asym:2010}. The isolated muon
trigger requires that in the neighboring region of the muon
trigger
 candidate
both the transverse
energy deposits in calorimeters and the scalar sum of the \pt of the
reconstructed
tracks are small, and it reduces
 the trigger rate while maintaining a relatively low muon \pt threshold. We use all the data-taking periods during which the isolated muon
triggers were not prescaled~(\ie they were exposed to the full integrated luminosity).

Other physics processes can produce
 high-$\pt$ muons and mimic
$\wmunu$  signal candidates.
We consider the SM background contributions from
  multijet production~(QCD background),
Drell--Yan~($\zll$) production, $\wtaunu$ production~[electroweak~(EW) background],
and top-quark pair ($\myttbar$) production. In addition,
cosmic-ray muons can penetrate through the center of the CMS detector and
also mimic $\wmunu$ candidates.

Monte Carlo~(MC) simulations are used to help evaluate the background
contributions in the data sample and to study systematic uncertainties.
Primarily,  we use NLO MC
 simulations based on the \POWHEG event generator~\cite{POWHEG}
 where the NLO {CT10} PDF model~\cite{CTEQ:1007} is used. The generated events are interfaced with the \PYTHIA~(v.6.422) event generator~\cite{PYTHIA6} for simulating the electromagnetic finite-state radiation~(FSR) and the parton showering.
The $\Pgt$ lepton decay in the  $\wtaunu$ process is simulated by the \TAUOLA MC package~\cite{TAUOLA}.
We simulate the QCD background with the \PYTHIA event generator where the CTEQ6L PDF model~\cite{CTEQ6L} is used.
The CMS detector simulation is interfaced with \GEANTfour~\cite{GEANT4}.
All generated events are first passed
through the detector simulation and then reconstructed in the same way
as the collision data.
Pileup is the presence of multiple interactions recorded in the same
 event.
 For the data used in this analysis, there are an
average of about 7 reconstructed primary interaction vertices
 for each beam crossing.
The MC simulation is generated with a different pileup distribution than we observe in the data.
  Therefore, the
MC simulation is  weighted such that the mean number of interactions per crossing
matches that in data, using the inelastic $\Pp\Pp$ cross section measured by the CMS experiment~\cite{Chatrchyan:2012nj}.

The selection criteria for muon reconstruction and identification
are described in detail in a previous report~\cite{CMS-PAPERS-MUO-10-004}.
Therefore, only a  brief summary is given here.
A muon candidate is reconstructed using two different algorithms: one starts
with a track measured by the silicon tracker
 and then requires a minimum number of matching hits in the muon detectors,
and the other starts by finding a track in the muon system and  then matching it to a track measured by the silicon tracker. Muons used in this measurement
are required to be reconstructed by both algorithms.
A global track fit, including both the silicon tracker hits
and muon chamber hits,
is performed to improve the quality of the
reconstructed muon candidate. The track \pt measured by the silicon tracker
  is used as the muon \pt and
the muon charge is identified from the signed curvature.
Cosmic-ray contamination is reduced by requiring that the distance of the closest approach
 to the leading primary vertex is small: $|d_{xy}| < 0.2$\unit{cm}.
 The remaining cosmic-ray background yield is estimated
  to be about $10^{-5}$ of the expected $\wmunu$ signal,
and  is therefore  neglected~\cite{CMS:asym:2010}.
The track-based muon isolation, $\text{Iso}_{\text{track}}$,  is defined to be the
scalar sum of the \pt of additional tracks in a cone
with a radius of 0.3 around the
muon candidate~($R = \sqrt{\smash[b]{(\Delta \eta)^2 + (\Delta \phi)^2}} < 0.3$, with
$\Delta \phi$ and $\Delta \eta$ being the differences between the muon
candidate
and the track in the $\eta$-$\phi$ plane). Muons are required to have
$\text{Iso}_{\text{track}}/\pt <0.1$. Only muons within   $\abs{\eta} <2.4$ are included in the
data sample.

In each event, muons passing the above selection criteria are ordered according to
$\pt$, and the leading muon is selected as the
$\wmunu$ candidate.  The  leading muon is
required to be the particle  that triggered the event. In addition, the muon
is required to have $\pt > 25\GeV$, which is safely above
the trigger turn-on thresholds.
Events that have
a second muon
with $\pt > 15\GeV$ are rejected to reduce the background
from Drell--Yan dimuon events (``Drell--Yan veto'').
The rejected events, predominantly  $\zmumu$ events,
 are used as a Drell--Yan control sample to study the
modeling of the $\vecEmT$ and also
 to provide constraints on the
modeling of the \pt  spectrum of  $\PW$ and $\cPZ$ bosons.
In addition, this sample is used to
estimate the level of background from  Drell--Yan  events
where the second muon is not identified.
The muon 
is  corrected for a bias
in the measurement of the momentum
(discussed below) prior to the application of  the \pt selection.

The $\wmunu$ candidates that pass the above selection criteria are divided
into 11 bins
in  absolute value of  muon pseudorapidity $\abs{\eta}$. The bin width is 0.2,
except that the last three $\abs{\eta}$ bins are [1.6--1.85],\ [1.85--2.1],  and  [2.1--2.4], respectively.
The muon charge asymmetry is measured in each of the $\abs{\eta}$ bins,
along with the determination of the correlation matrix of the systematic
uncertainties between   different $\abs{\eta}$ bins.

\section{Muon momentum correction and efficiency studies}
\label{sec:eff_scale}

The measured momentum of the muon depends critically on the correct alignment of the
tracker system and the details of the magnetic field. Even
after the alignment of the tracker detector a residual
misalignment remains, which is not
perfectly reproduced in  the MC simulation.
  This misalignment
leads to a charge dependent bias in the reconstruction of  muon momenta, which
 is removed by using a muon momentum correction.
The detailed description of the  method for the extraction
of the correction factors using  $\zmumu$ events is given
in Ref.~\cite{bib:momcor}.  Here we provide only a short summary
of the method.
First, corrections to muon
momentum  in bins of $\eta$ and $\phi$
are extracted separately for
positively and negatively charged muons using the average of the $1/\pt$ spectra
of muons
in $\zmumu$ events.
The mean values of the $1/\pt$ spectra
at the MC generator level, varied by the reconstruction  resolution,
are used as a ``reference''. The mean values of the
 reconstructed $1/\pt$ spectra in data or simulation are tuned to match the reference.
Second, the correction factors derived in the previous step are  tuned further
by  comparing the  dimuon invariant mass in each bin of muon charge $Q$,
$\eta$, and $\phi$ to the  ones at the MC generator level varied by the reconstruction
resolution.
The same procedure is performed for both data and reconstructed
MC events, and
correction factors are determined separately.
The correction factors are extracted using the same $\eta$ binning
defined above in order to avoid correlations
 between  different $\eta$ bins.

The dataset used to derive the corrections was collected with a double-muon trigger with asymmetric
 \pt thresholds of 17 and 8\GeV. Both muons are required to have $\pt >25\GeV$, which exceeds significantly the trigger \pt thresholds. The simulation has been
corrected for the muon efficiency difference between data and MC simulation as
discussed below. We illustrate the relative size of the
derived corrections using a 40\GeV muon as an example. For muons within
$\abs{\eta} <0.2$, the corrections derived using the $1/\pt$ spectra are less than 1.5\% and 0.4\% for data and MC simulation, respectively. A $\phi$ modulation
of these corrections is observed. The maximum corrections are larger in
high-$\eta$ region, and for muons with $\abs{\eta}>2.1$ these corrections can be as large as
 3.5\% and 1.4\% for data and MC simulation, respectively. The additional
corrections derived using the dimuon invariant mass are smaller. For muons
within the complete detector acceptance, the additional corrections are less than
0.5\% and 0.2\% for data and simulation, respectively. These additional
corrections show no evidence of $\eta$-$\phi$ dependence and fluctuate around
zero within the statistical uncertainties of the final
corrections. The statistical uncertainties of the corrections for
 various $\eta$-$\phi$ bins are uncorrelated.
By comparing the correction factors
 for positively and negatively charged  muons
 in each bin, we can determine  relative corrections from
misalignment and from mismodeling of
the magnetic field in the tracker system.
The mismodeling  corrections for muons with $\abs{\eta} >2.1$,
where maximum deviations from zero are evident, are less than 0.3\% and 0.4\%
for data and MC simulation, respectively. In contrast, in the same detector
region, the corrections due to misalignment are about 4.4\% and 1.7\% for
data and MC simulation, respectively. Hence, the bias comes predominantly
from misalignment.

\begin{figure*}[tbhp]
  \begin{center}
     \includegraphics[width=\cmsFigWidth]{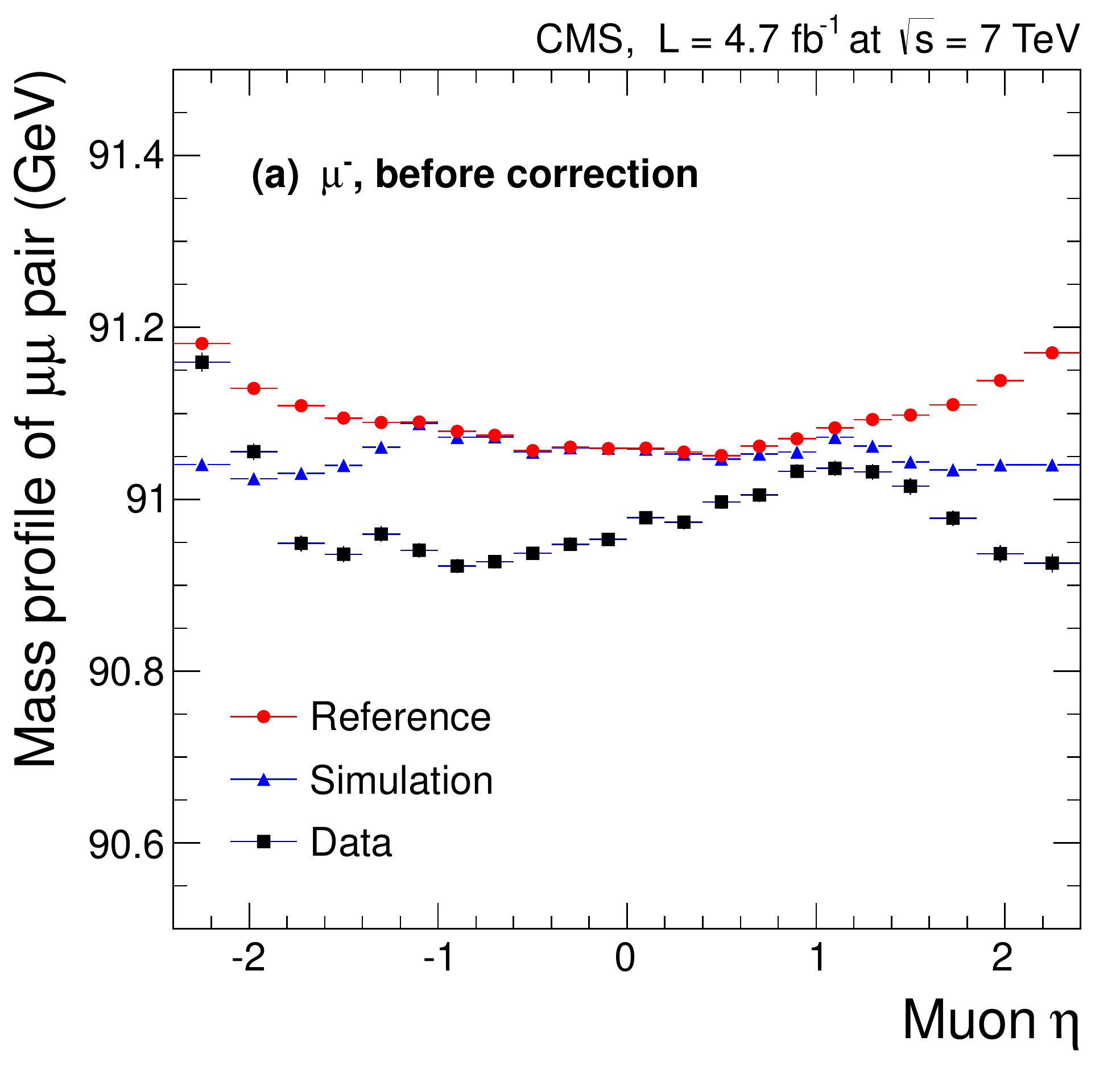}
     \includegraphics[width=\cmsFigWidth]{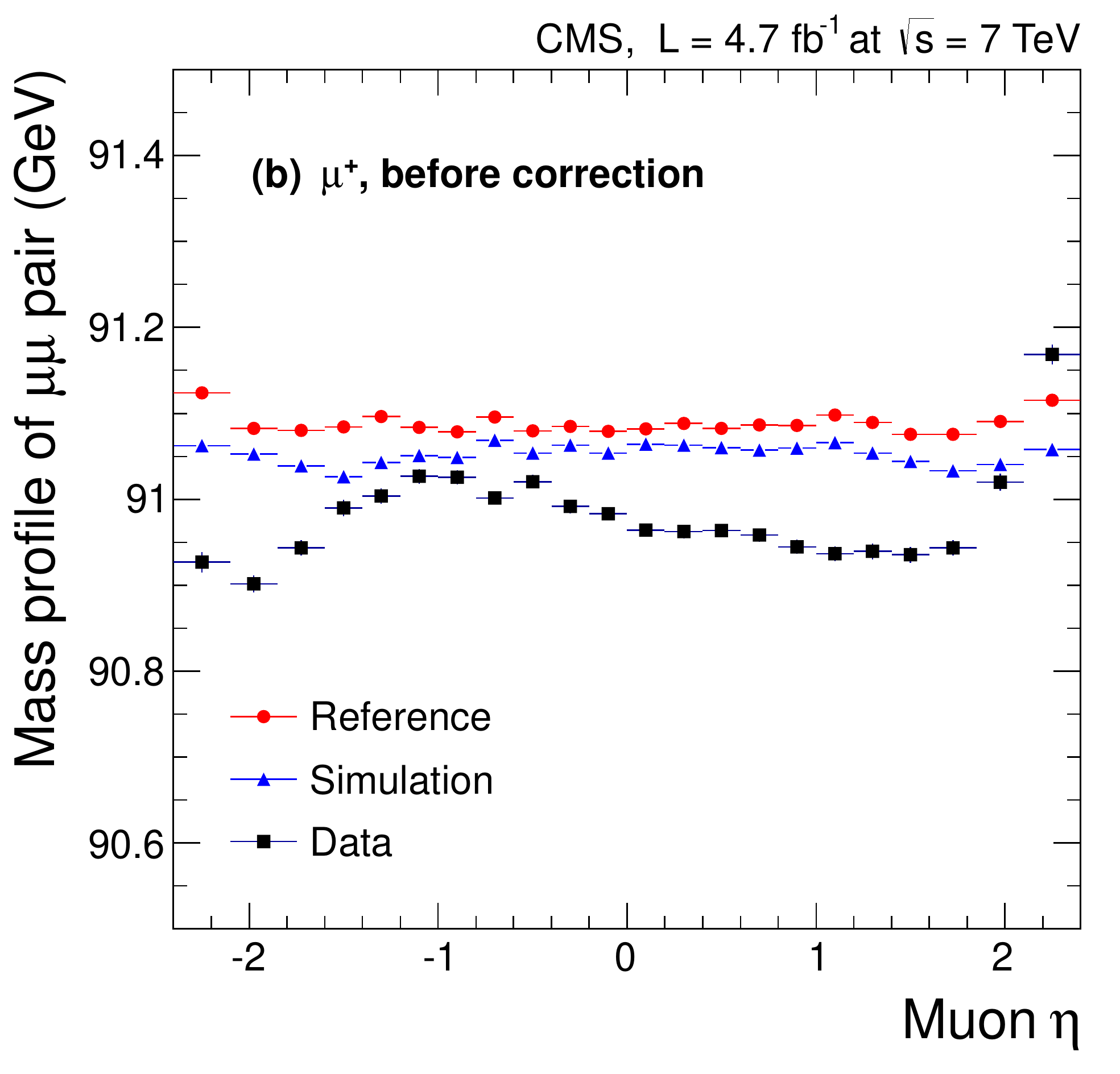}
     \includegraphics[width=\cmsFigWidth]{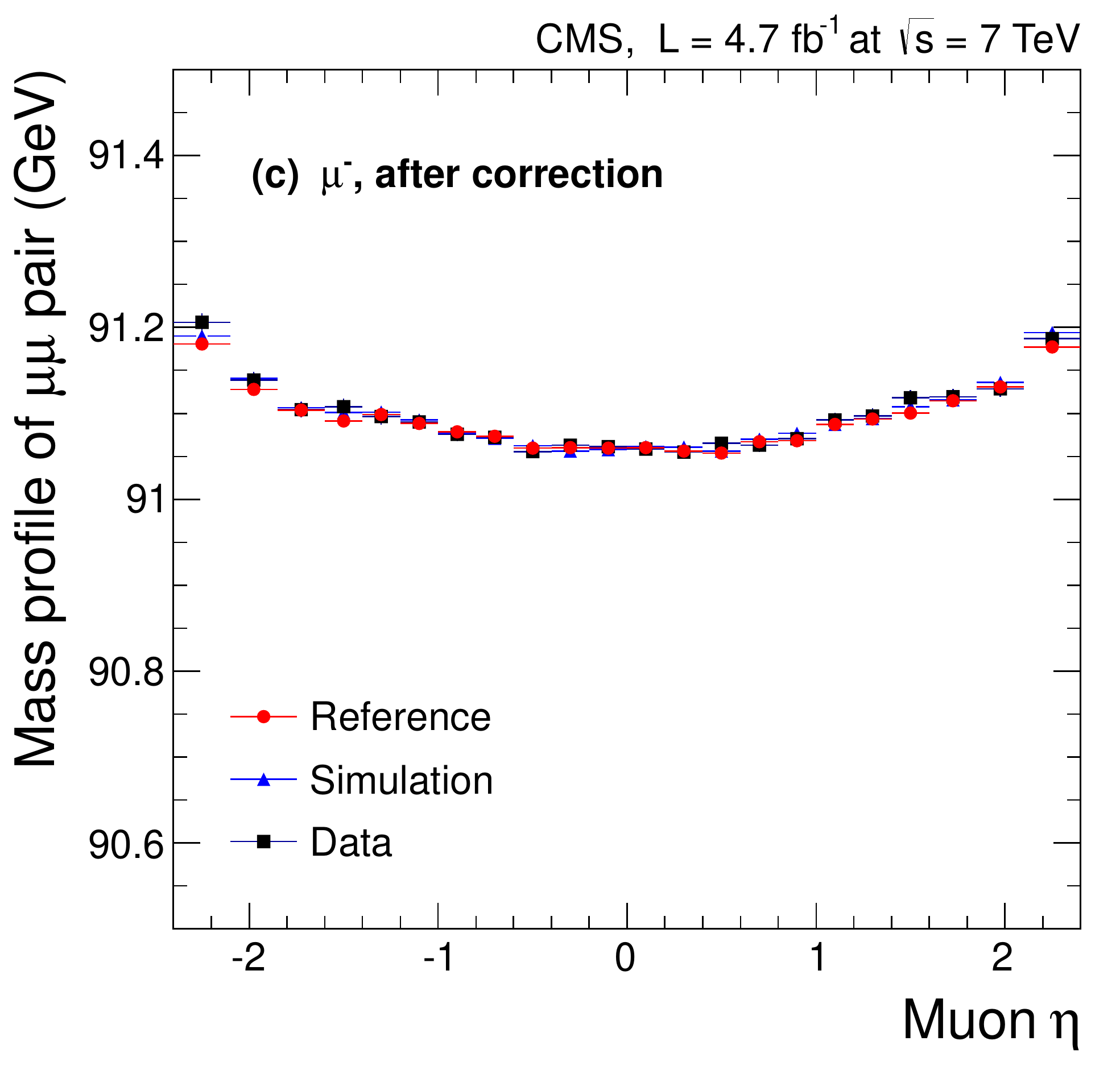}
     \includegraphics[width=\cmsFigWidth]{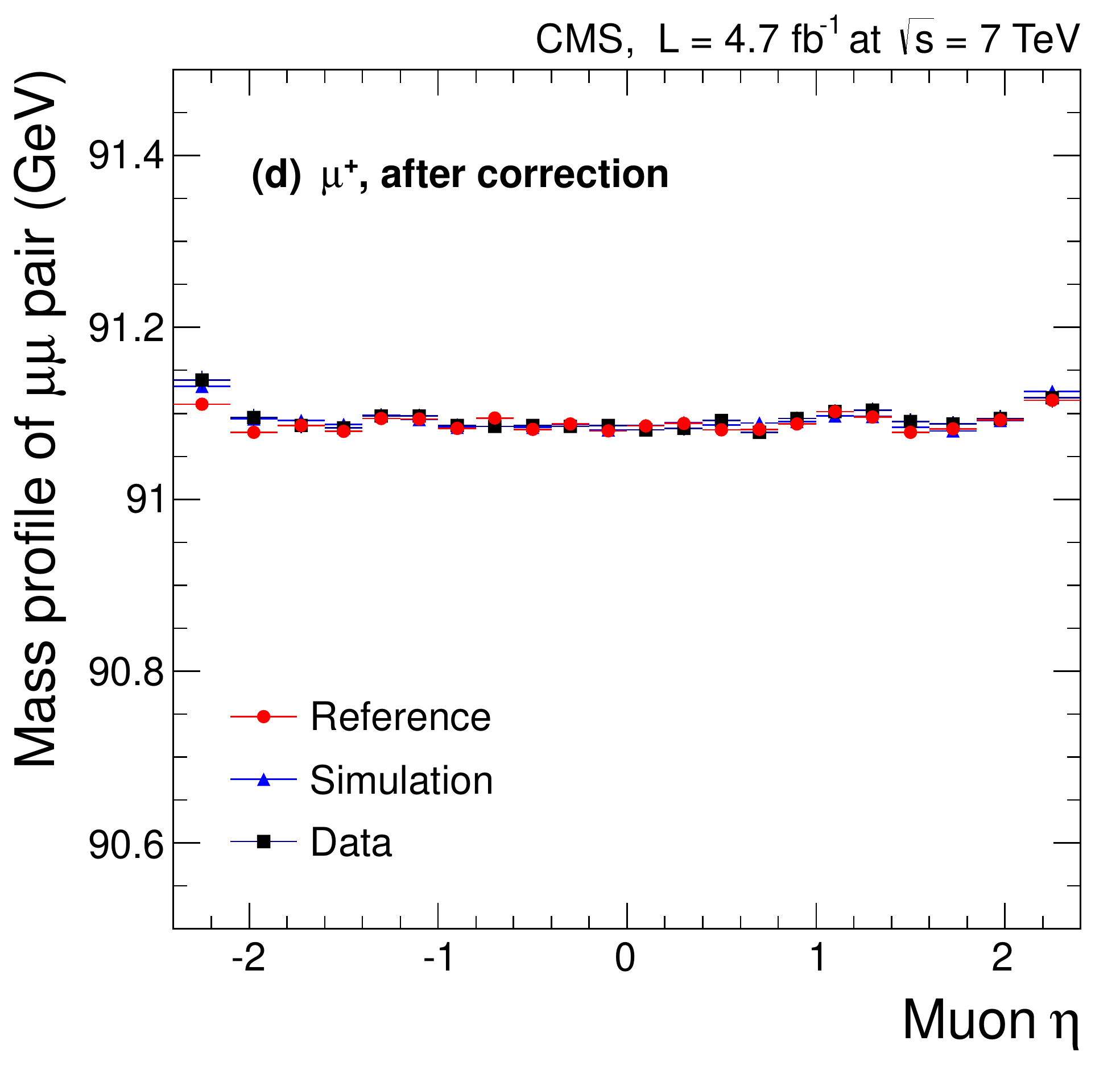}
    \caption{The dimuon
mass profile as a function of muon
  $\eta$ for   $\Pgmm$~(a,~c) and  $\Pgmp$~(b,~d),
where (a) and (b) are before the correction and (c) and (d) are
after the correction. The generated muon \pt varied by reconstruction
resolution in data is used to obtain the dimuon invariant mass of the
reference.
}
    \label{fig:momcor}
  \end{center}
\end{figure*}

Figure~\ref{fig:momcor} shows the average dimuon invariant
mass~(mass profile)  as a function of
muon $Q$ and $\eta$  before and after the correction, which includes both the contributions from
 tracker misalignment and mismodeling of the magnetic field.
The dimuon mass profiles
 after the correction
are compared to the reference mass profile for data and MC simulation.
They agree well with the reference,  so
the muon momentum bias is largely removed.
The reference mass profile
is expected
 to be a function of $\eta$ because of the  \pt  requirements for the two daughter muons in $\zmumu$ decays.
Values of the dimuon mass profile
as a function of muon $\eta$ are averaged over $\phi$,
while the muon scale corrections correct for muon momentum bias in
both $\eta$ and $\phi$.

The  overall efficiency in the selection of muon candidates includes
contributions from  reconstruction, identification (including isolation), and
trigger efficiencies.
The muon reconstruction efficiency includes contributions from the reconstruction efficiency in
the tracker system~(``tracking'') and in the muon system.
 The muon ``offline'' efficiency is the product of reconstruction and
identification efficiencies.
The contribution of each component
 to the overall efficiency~(tracking, muon standalone reconstruction, identification, and trigger)
is measured directly from the   $\zmumu$ events using the  tag-and-probe method
~\cite{CMS:WZ, CMS-PAPERS-MUO-10-004}.
In  this method one of the daughter muons
 is used to tag
the $\zmumu$ event and the other  muon
candidate is used as a probe to
study the  muon efficiencies as a function of $Q$, $\eta$,  and $\pt$.
For every event a positively charged muon can be selected as the tag and
a negatively charged
probe candidate is used to study the efficiencies for negatively charged muons.
The same procedure is repeated by selecting a negatively charged muon as the
tag to study efficiencies for positively charged muons. Each individual efficiency is
determined in 22 bins of muon $\eta$,
as defined above, and  7 bins of
 \pt~(15--20, 20--25, 25--30, 30--35, 35--40, 40--45, and $>$45\GeV) for
both $\Pgmp$ and $\Pgmm$.
The same  procedure is applied to
both data and MC simulation and scale factors are determined to match the
simulation efficiencies  to the data.

\begin{figure}[hbtp]
  \begin{center}
    \includegraphics[width=\cmsFigWidth]{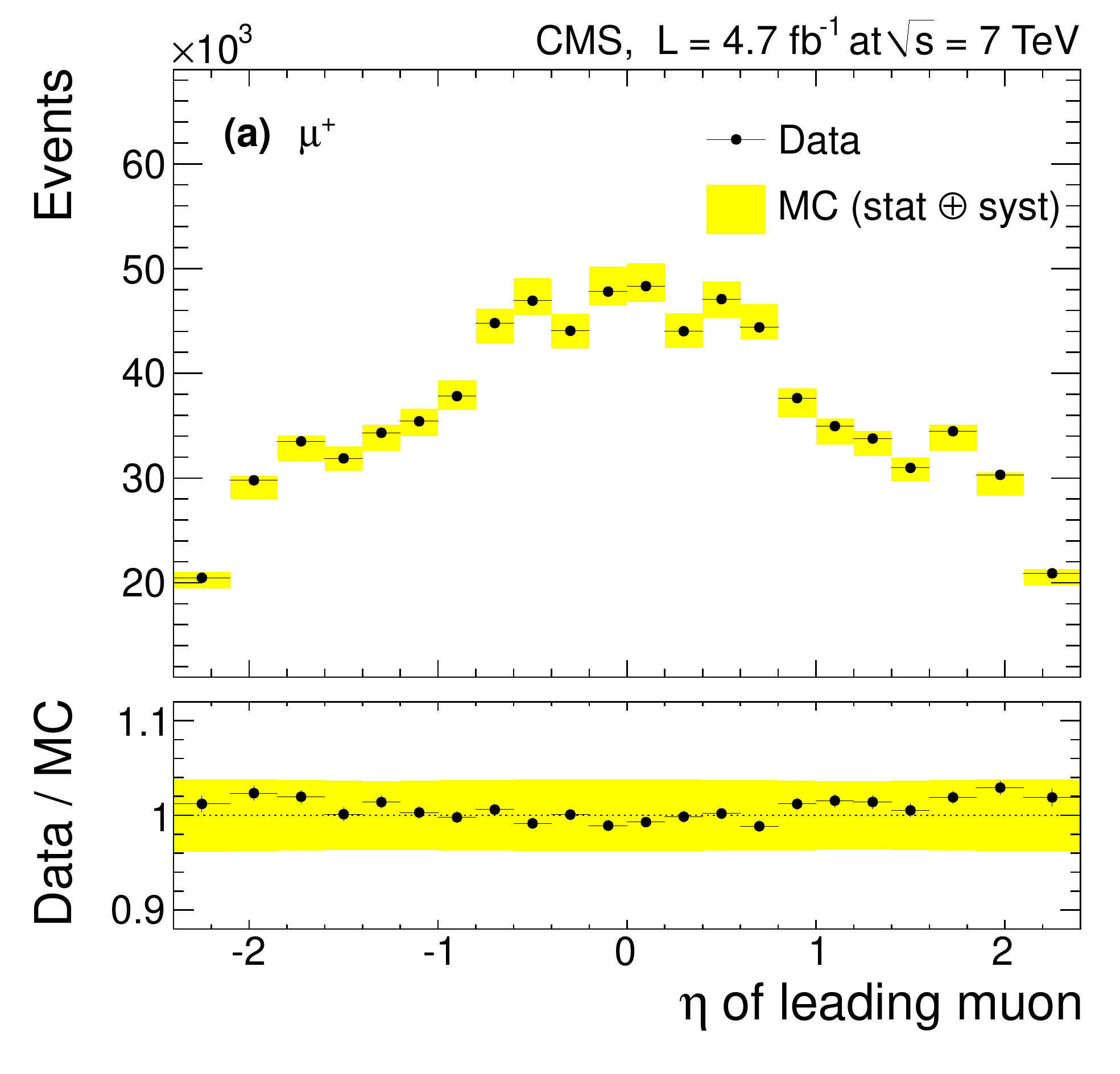}
    \includegraphics[width=\cmsFigWidth]{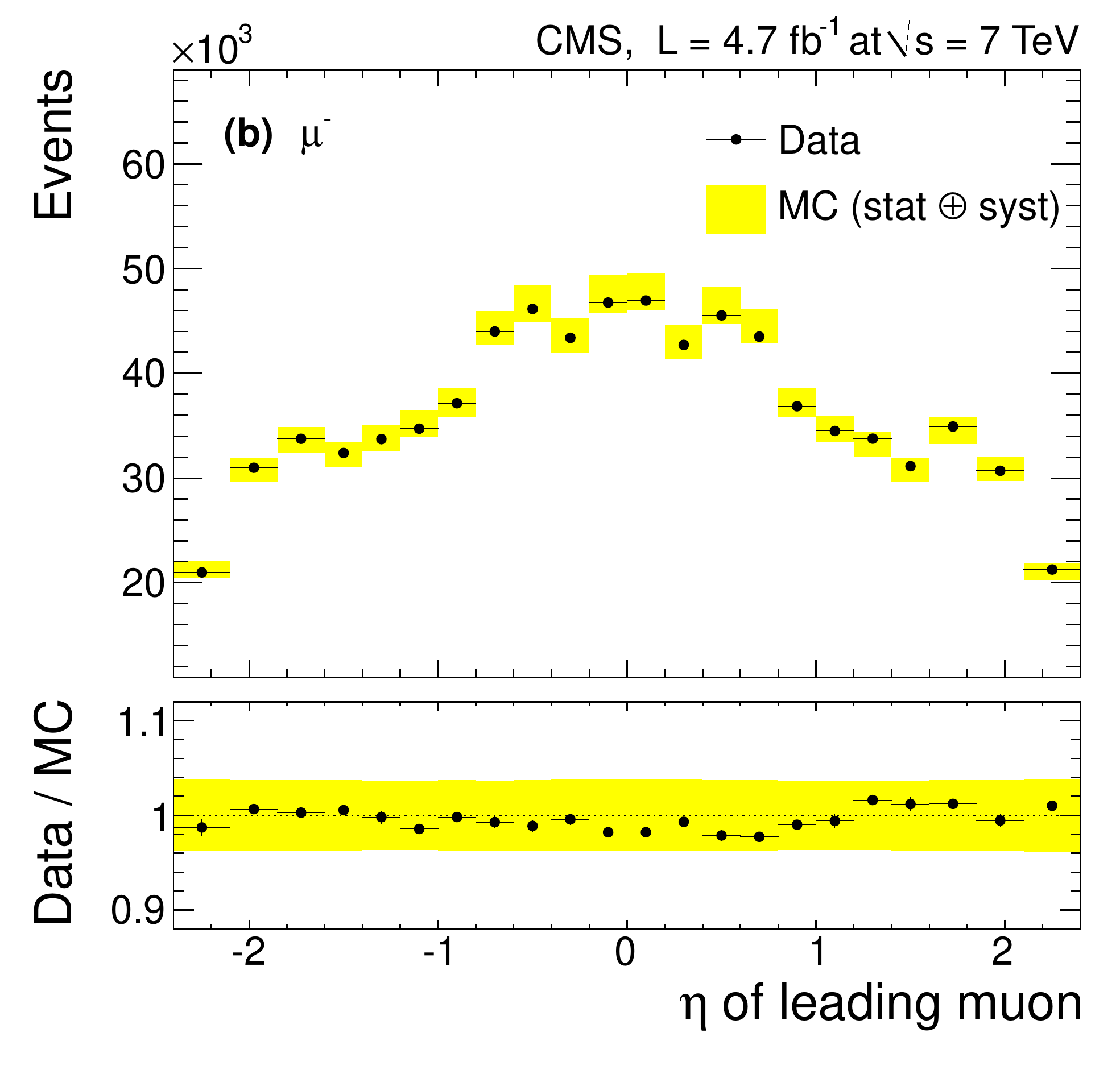}
    \caption{The $\eta$ distribution of the leading
$\Pgmp$~(a)  and
 $\Pgmm$~(b) in the $\zmumu$ sample. The dimuon invariant mass is within
$60 < m_{\Pgm\Pgm} < 120\GeV$.  The MC simulation is normalized to the data.
The light shaded band is
the total uncertainty in predicting  the $\zmumu$ event yields using MC
simulation, as described in Section~\ref{sec:systematics}. }
    \label{fig:zmumu:norm}
  \end{center}
\end{figure}

The measured average tracking efficiency in each $\eta$ bin
 varies from  99.6 to 99.9\% with a slight  inefficiency in
the transition regions from the barrel to the endcap segments
 and  at the edge of the
 tracker system.
The ratio of  tracking efficiencies for  $\Pgmp$ and $\Pgmm$ is consistent with unity within
statistical uncertainty. In the transition regions from the DT to the CSC,
there is evidence that the muon
offline efficiency has a slight asymmetry between  $\Pgmp$ and $\Pgmm$. The
 ratio of efficiencies for
positively and negatively charged muons  differs from unity
by up to $1.0\pm0.3$\%.
 The trigger efficiency ratio is
also  found to
differ from unity in some $\eta$ regions.  The  maximum deviation is at
$\eta>2.1$ where the
efficiency for  $\Pgmp$ is about $2.0\pm0.5$\% higher than that
for $\Pgmm$.
Figure~\ref{fig:zmumu:norm} shows the $\eta$ distribution for the  leading
 $\Pgmp$ and $\Pgmm$
in the $\zmumu$ sample.  The dimuon invariant mass is within
$60 < m_{\Pgm\Pgm} < 120\GeV$. Here, the MC simulation is corrected for
muon momentum bias, efficiency, and modeling
of the $\cPZ$-boson transverse momentum~($\vecqT$) before normalizing to the
measured data. The modeling of $\cPZ$-boson $\vecqT$ spectrum is discussed in detail in Section~\ref{sec:sys:Wqt}. The $\eta$ dependence effect in data and MC simulation are
in good agreement.

\section{Extraction of the asymmetry}
\label{sec:signal_estimation}

The asymmetry is calculated in bins of~$\etaabs$ from the yields of~$\PWp$ and~$\PWm$.
In this section, we explain how the yields are obtained from the $\EmT$ distributions,
and we discuss corrections to the~$\EmT$ needed in the accurate estimation of the yields.
Finally, we explain how backgrounds are taken into account.

The raw charged asymmetry ($\ARAW$) is defined in terms of the numbers
$\NWp$ and $\NWm$ of $\PWp$ and $\PWm$ signal events:
\begin{equation}
\label{eq:asym}
  \ARAW = \frac{ \NWp - \NWm }{ \NWp + \NWm } .
\end{equation}
The yields $\NWp$ and~$\NWm$ are obtained from simultaneous binned maximum-likelihood
fits of the $\EmT$~distributions; the signal yields and the normalization of
the QCD background are free parameters.
The likelihood is constructed following the Barlow--Beeston
method~\cite{Barlow:Beeston} to take into account the limited size of the MC
signal event sample.
The shapes of the $\EmT$ distributions for the $\wmunu$ signal and the
background contributions are taken from MC simulations after correcting for
mismodeling of the detector response and the~$\vecqT$ distribution of the
$\PW$~bosons, as discussed further in Section~\ref{sec:sys:Wqt} below.
The pileup of each MC sample is matched to the data using an ``accept--reject''
technique based on the observed and simulated pileup distributions.
This technique avoids a large spread of weights that would come from
simply reweighting the MC events; the $\EmT$ templates are constructed
using the accepted MC events.

A total of 12.9 million $\PWp \to \Pgmp \Pgn$ and 9.1 million $\PWm \to \Pgmm\Pagn$
candidate events are selected.  The expected backgrounds from QCD, EW, and
$\ttbar$ events are about  8\%, 8\%, and 0.5\%, respectively.
The single top-quark and diboson~($\PW\PW/\PW\cPZ/\cPZ\cPZ$)
production is  less than 0.1\% and is neglected.
The variation of the background composition as a function of~$\etaabs$
is taken into account.

The estimate for the Drell-Yan background is based on the
observed yields in a Drell-Yan control sample. The $\wtaunu$ background
scales with the $\wmunu$ signal using a factor determined from
a MC simulation.  The $\ttbar$ background is normalized to the NLO
cross section obtained from
\textsc{mcfm}~\cite{Campbell:1999ah, Campbell:2010ff, Campbell:2012uf}.
Efficiency correction factors are applied to the simulation
before determining the background normalization.

The level of the QCD background is determined by the fit.  A
constraint on the relative amount of QCD background in the
$\PWp$ and $\PWm$ samples is obtained from a QCD-enriched control
sample collected using a muon trigger with no isolation requirement.
This constraint induces a correlation of $\NWp$ and $\NWm$, and
the resulting covariance is taken into account when evaluating the
statistical uncertainty on~$\ARAW$.

In the following sections, we discuss the corrections to the~$\EmT$
and then report the results of the fit to the~$\EmT$ distributions.

\subsection{Corrections of the missing energy measurement}
\label{sec:signal:procedure}

The analysis depends critically on the control of the $\EmT$ distributions.
Several corrections are needed to bring the simulation into agreement with
the observed distributions.  The $\EmT$ depends on both the measured
muon kinematics and the kinematics of the hadrons recoiling against
the $\PW$~boson.  The corrections for the calibration of the muon momentum,
discussed in Section~\ref{sec:eff_scale} above, are applied by adding the
$\ptvec$~correction to $\vecEmT$ vectorially.
The kinematic corrections for the so-called ``hadronic recoil,'' which are based on
the control sample of $\zmumu$ events, are explained in detail here.

By definition, the hadronic recoil, $\vec{u}$, is the vector sum of
transverse momenta of all reconstructed particles except for the muon(s).
For $\zmumu$ events,
\begin{equation}
\label{eq:recoil}
\vec{u} = -\vecEmT - \vecqT,
\end{equation}
where $\vecqT$ is the transverse momentum of the dimuon system
and $\vecEmT \approx 0$.
The components of $\vec{u}$  parallel and perpendicular to $\vecqT$ are
$\upar$ and $\uper$, respectively.
The mean of $\uper$, $\upermean$, is approximately zero, while
the mean of $\upar$, $\uparmean$, is close to the mean of the boson~$\qT$.
Differences in the distributions from data and MC are ascribed
to detector effects, the simulation of jets, pileup and the underlying
event, all of which should be nearly the same for $\zmumu$ and $\wmunu$ events.
The distributions of $\upar$ and $\uper$ in $\zmumu$ events are used
to derive corrections for the simulation that improve the modeling
of~$\EmT$ for $\wmunu$ signal events as well as for backgrounds;
this technique was employed previously by the Tevatron experiments
and by CMS~\cite{PhysRevD.77.112001, Abazov:2009tra,  RECOIL}.
We correct both the scale and resolution of~$\EmT$.

A comparison of the $\vecEmT$ distributions for $\zmumu$ events
in data and MC shows that the agreement is not perfect.
Both show a small $\phi$ modulation, but the phase and amplitude
of the modulation are not the same.  This modulation follows from
the fact that collisions, including hard interactions that produce
$\PW$ events as well as pileup events, do not occur exactly at the
origin of the coordinate system.
This modulation can be  characterized by a cosine function,
$C \cos{(\phi - \phi_{0})}$.
The dependence of the amplitude $C$ and phase term $\phi_0$ on
the number of primary vertices
is  extracted from the  $\zmumu$ event sample by
fitting a $\phi$-dependent profile of $\upar - \uparmean$.
The amplitude~$C$ is observed to depend linearly
on the number of primary vertices, while the phase~$\phi_0$
is almost independent of pileup.  The $\phi$ modulation of $\vecEmT$
can be removed by  adding a vector in the transverse plane,
$\Delta\vecEmT = C\cos\phi_0 \hat{x} +  C\sin\phi_0 \hat{y}$ to $\vecEmT$
for each event.

The dependence of $\uparmean$ with $\cPZ$-boson~$\qT$ should
be approximately linear, and this behavior is indeed
observed in both data and MC.  This dependence is further studied according to
the direction of the leading jet, namely, in four bins of jet~$\etaabs$:
[0.0--1.2], [1.2--2.4], [2.4--3.0], and [3.0--5.0].
The jets are formed by clustering particle-flow candidates using the
anti-$k_{\mathrm{T}}$ jet clustering algorithm~\cite{Cacciari:2008gp}
with a distance  parameter of~0.5, and the muons are not included
in the reconstruction of jets.
The $\uparmean$ behavior with~$\qT$ for MC and data agrees very well
when the leading jet is in the central region of the detector.
When the leading jet is in the forward direction, a modest difference
is observed, amounting to less than 10\% in the highest $\etaabs$ bin.

The distributions of $\upar - \uparmean$ and $\uper$ are fit to Gaussian functions
whose widths are parametrized as a functions of~$\qT$.   They depend
strongly on the pileup, so they are also fit as functions of the
number of vertices in the event.  The weak dependence of $\uparmean$
on the leading jet~$\etaabs$ is neglected.  The widths of
the $\upar - \uparmean$ and $\uper$ distributions are slightly
larger in data than in MC.   For example, when there are seven
reconstructed vertices in the event (which corresponds to the mean
number for this data set), the widths are $4$--$10\%$ larger.

\begin{figure}[hbtp]
  \begin{center}
    \includegraphics[width=\cmsFigWidth]{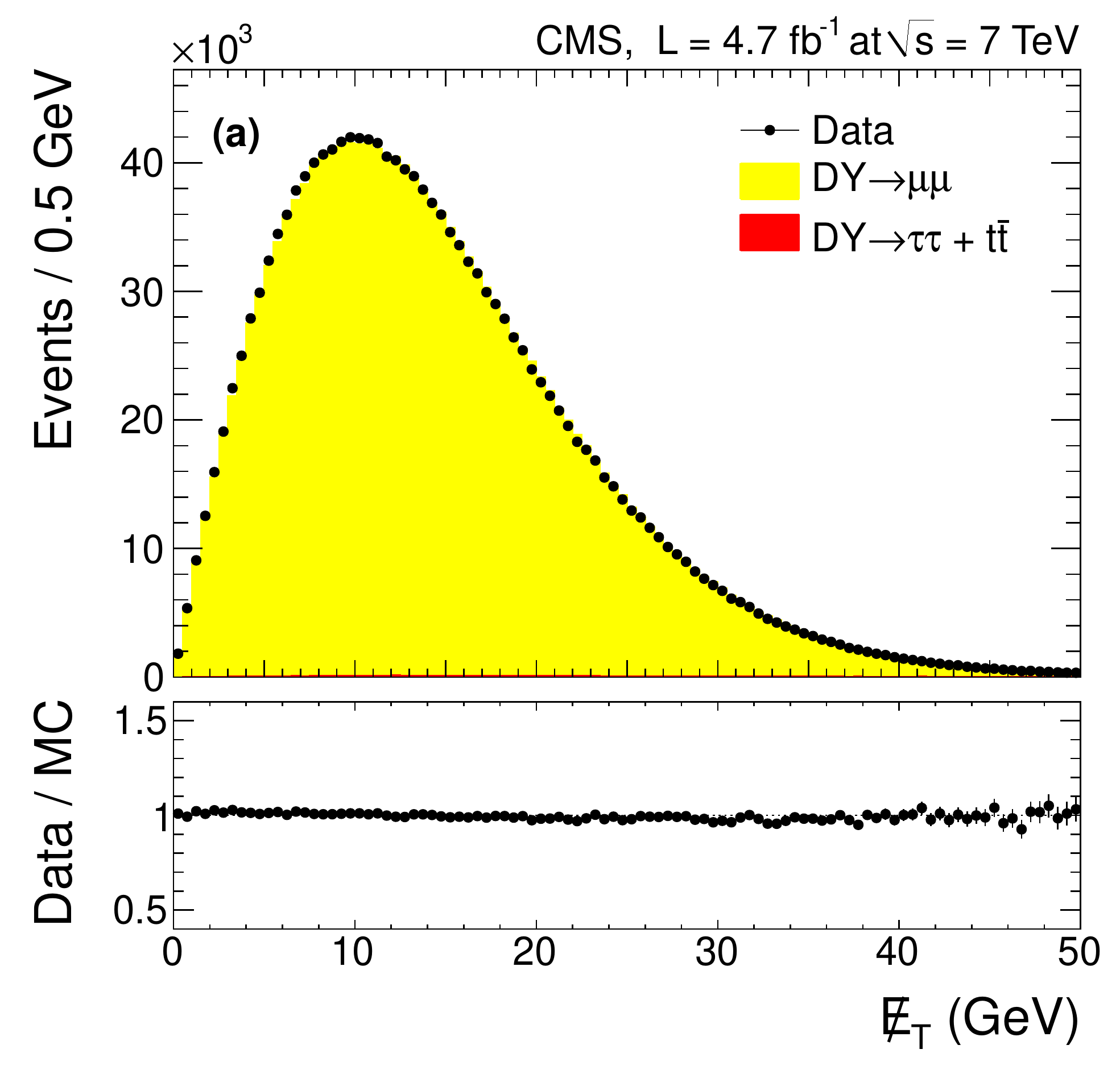}
    \includegraphics[width=\cmsFigWidth]{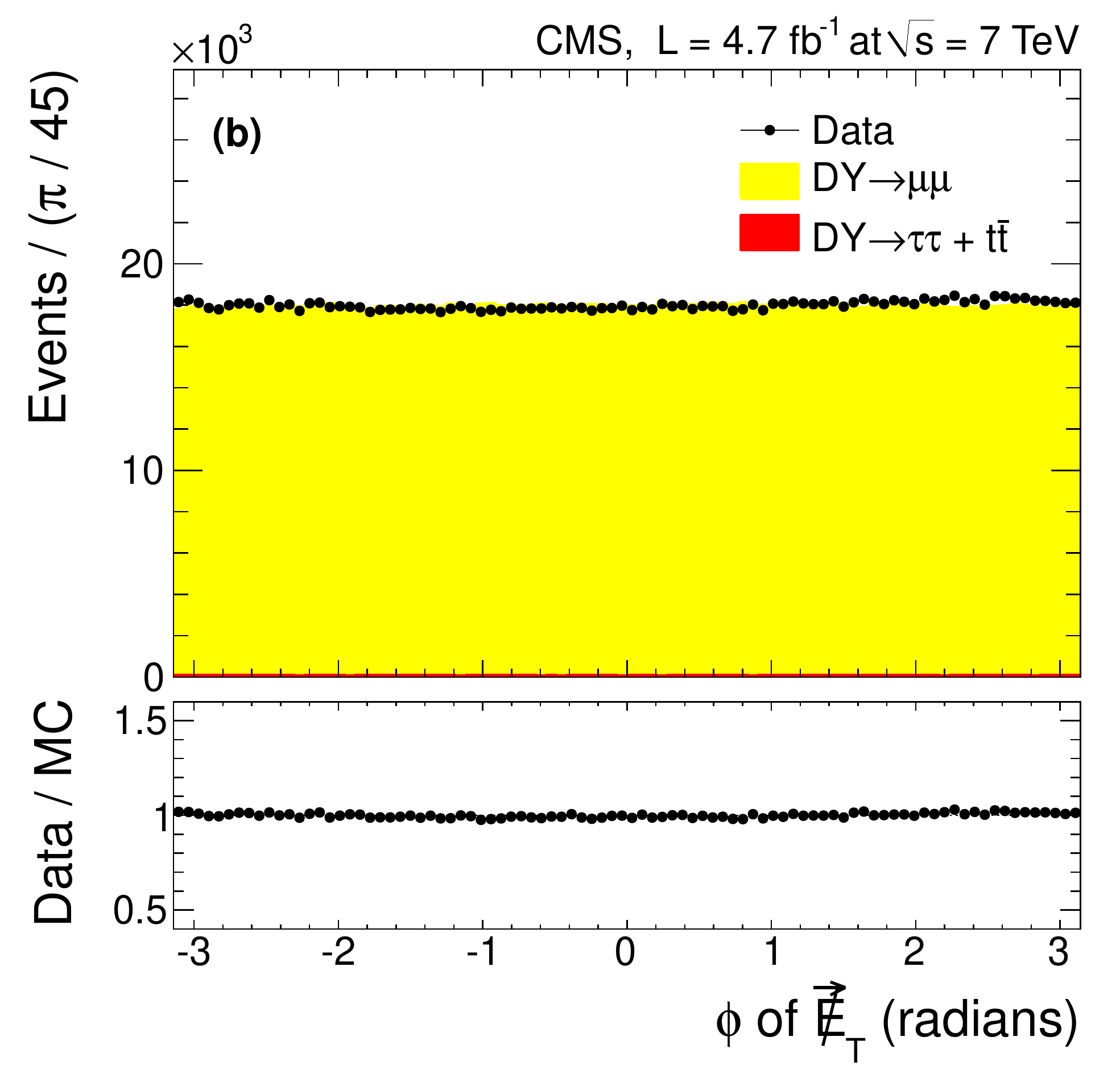}
    \caption{Data to simulation comparison for  $\EmT$~(a) and
  $\phi$ distribution of $\vec{\EmT}$~(b)  in  the Drell--Yan control sample.  Here, the hadronic recoil  derived from the  data was  used to correct
 the MC simulation. The $\ztautau + \ttbar$ contribution~(dark shaded region)
in data is
normalized to the integrated luminosity of the data sample using a MC simulation, and the
normalization of the $\zmumu$ MC
simulation~(light shaded region) is  taken as the difference between the data and the
estimated $\ztautau + \ttbar$ contribution. In this data sample, the
$\ztautau + \ttbar$ contribution is negligible.
}
    \label{fig:met:zmumu}
  \end{center}
\end{figure}

A test of the hadronic recoil corrections is carried out with
$\zmumu$ events.  The hadronic recoil $\vec{u}$ is calculated in
each MC event, and the parallel component $\upar$ is rescaled
by the ratio of $\uparmean$ in data and in MC.  Furthermore, the
smearing of $\upar$ and $\uper$ is adjusted to match the resolutions
measured with the data.  The $\vecEmT$ is recalculated according
to Eq.~(\ref{eq:recoil}).  Figure~\ref{fig:met:zmumu} shows
$\EmT$ and the $\phi$ of the $\vecEmT$ after applying
the hadronic recoil corrections.
The data and MC simulation are in  excellent agreement,
demonstrating  that this empirical correction to $\EmT$
works very well for $\zmumu$ events.

To apply the hadronic recoil correction determined
in $\zmumu$ events to other
MC simulations, such as $\wmunu$ events,  requires defining
a variable equivalent to  the boson $\vecqT$ in
$\zmumu$ events.
In $\wmunu$ events, the hadronic recoil is defined to be
\begin{equation}
\label{eq:recoil:w}
\vec{u} = -\vecEmT -\ptvec,
\end{equation}
where $\ptvec$ is the muon transverse momentum.
The hadronic recoil is decomposed into $\upar$ and $\uper$ components
relative to~$\vecqT$.   The hadronic recoil correction is applied in the
manner above, and $\vecEmT$ is recalculated.
For the $\wmunu$ signal events, $\vecqT$ is
the vector sum of the transverse momentum of
the reconstructed muon  and the generated neutrino.
For $\PW \to\tau\Pgn$ events, the generated $\PW$-boson $\vecqT$ is used.
For selected Drell--Yan background events, one muon is not
reconstructed or not identified, so $\vecqT$ is calculated
using the $\ptvec$ of the lost muon at the generator level.
For the QCD background events, $\vecqT$ is identified with
the $\ptvec$ of the reconstructed muon.

\begin{figure}[hbtp]
  \begin{center}
    \includegraphics[width=\cmsFigWidth]{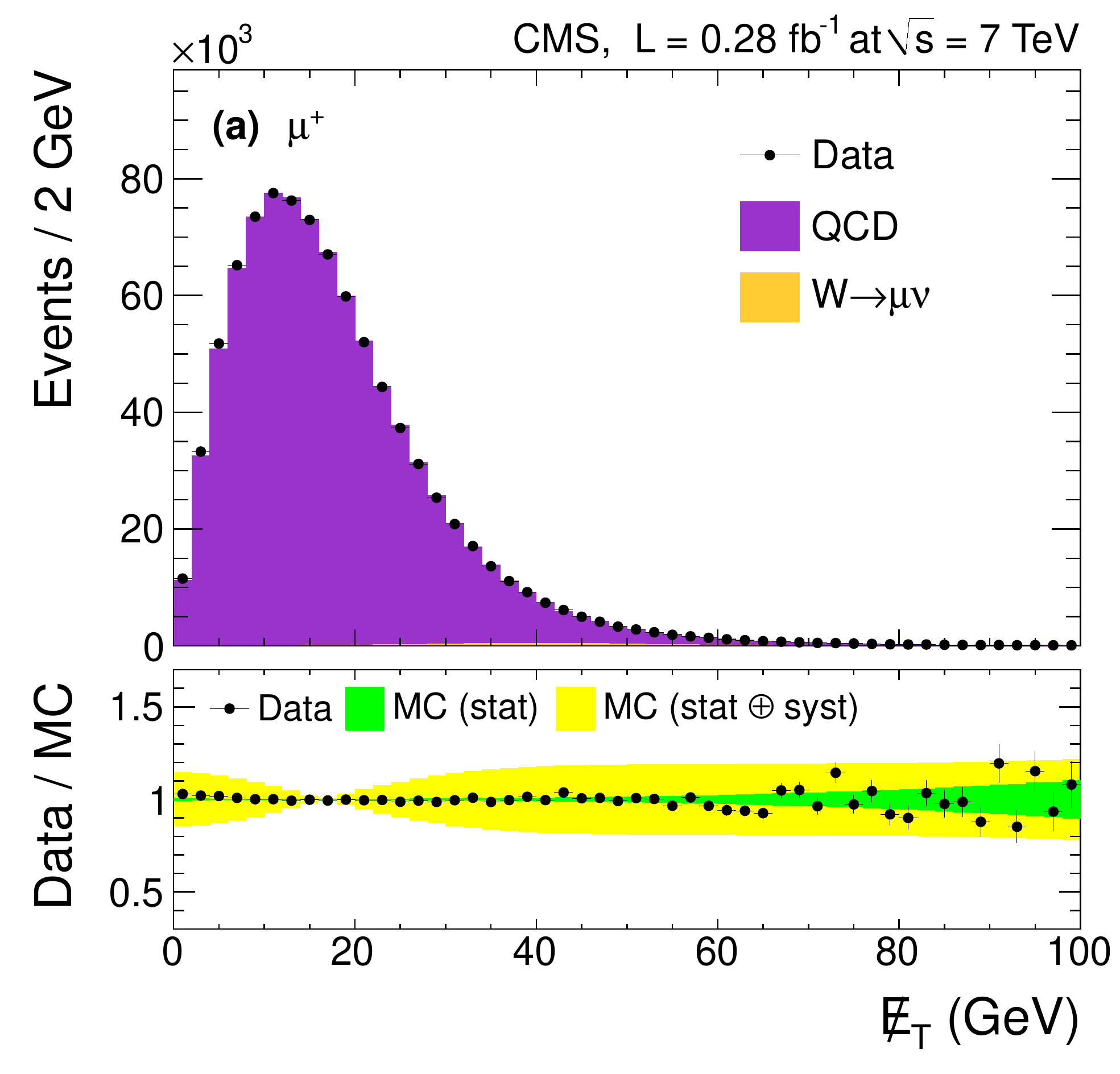}
    \includegraphics[width=\cmsFigWidth]{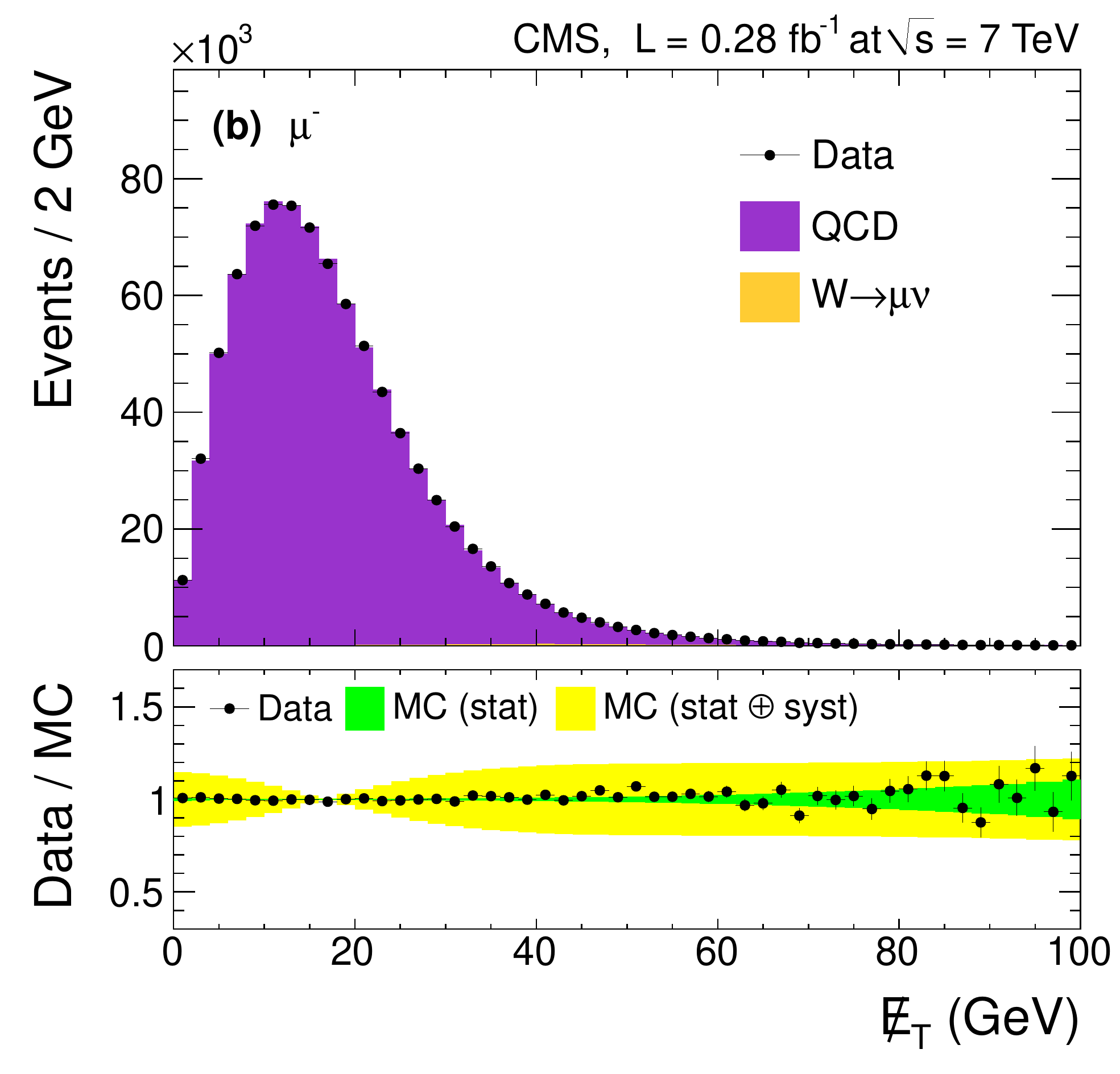}
    \caption{The  $\EmT$ distribution for
$\Pgmp$~(a) and
$\Pgmm$~(b)  in the data sample dominated by the QCD
background.
The hadronic recoil  derived from data
has been  used to correct the MC simulation. The $\wmunu$ contribution~(light shaded region)
is normalized to the integrated luminosity of the data sample using a MC simulation,
and the normalization of the QCD simulation~(dark shaded region) is
taken as the difference between the data and the estimated
$\wmunu$ contribution. The $\wmunu$ contribution in this data sample is
negligible.
 The dark shaded band in each ratio plot shows
the statistical uncertainty in the QCD MC $\EmT$ shape, and
the light shaded band shows the total uncertainty,  including  the  systematic uncertainties due to QCD  $\EmT$  modeling
as discussed in Section~\ref{sec:systematics}.
}
    \label{fig:met:qcd}
  \end{center}
\end{figure}

Figure~\ref{fig:met:qcd} shows the $\EmT$  distribution for the
QCD control sample.  We have selected only those events that
pass a non-isolated muon trigger but that fail the isolated
muon trigger. We also impose an anti-isolation selection cut:
$\textrm{Iso}_{\text{track}}/\pt > 0.1$.   With the application
of the hadronic recoil corrections, the data and simulation
are in very good agreement.

\subsection{Extraction of the asymmetry from fits to the missing transverse energy}

\begin{figure*}[hbtp]
  \begin{center}
    \includegraphics[width=\cmsFigWidth]{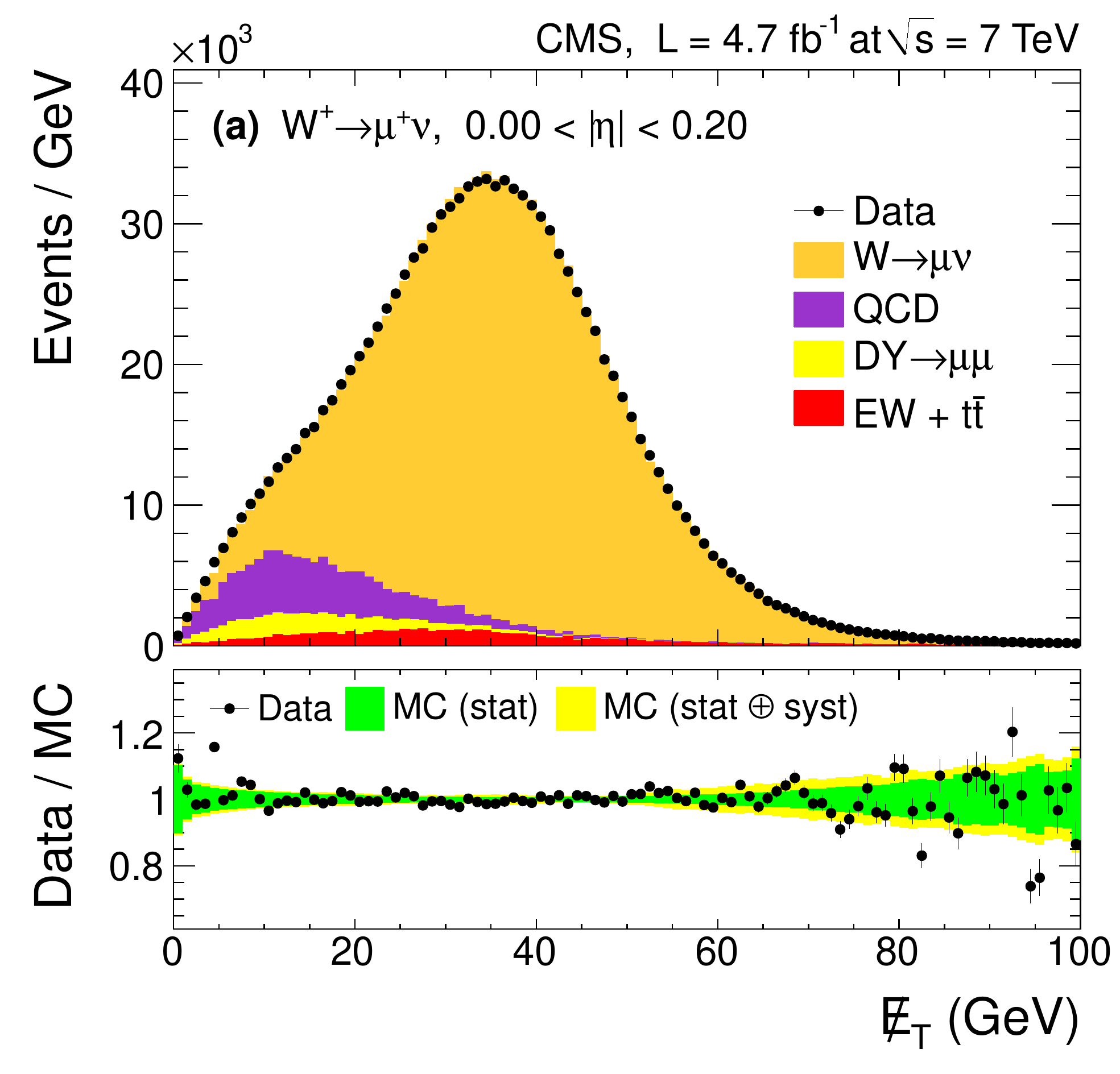}
    \includegraphics[width=\cmsFigWidth]{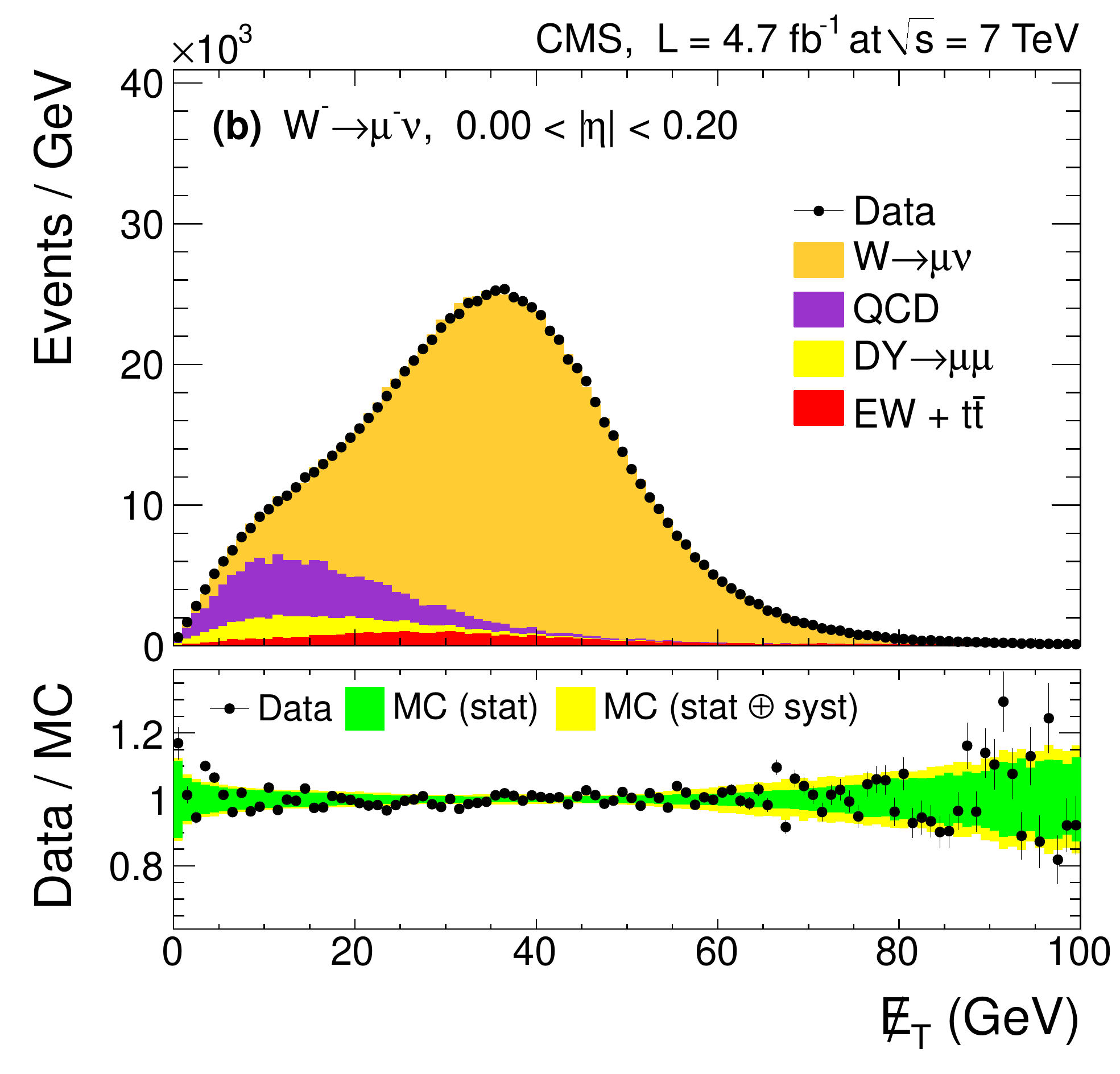}
    \includegraphics[width=\cmsFigWidth]{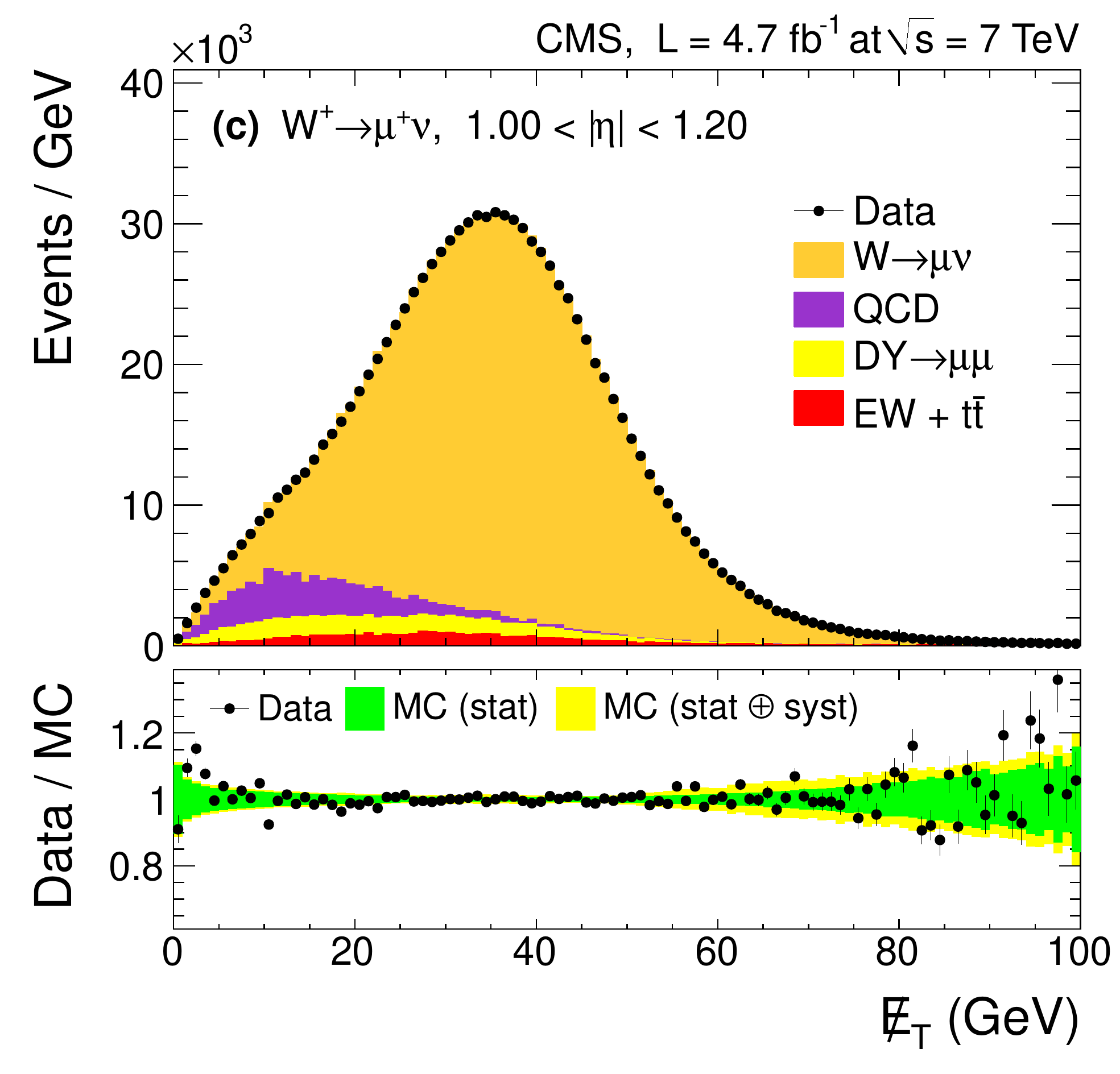}
    \includegraphics[width=\cmsFigWidth]{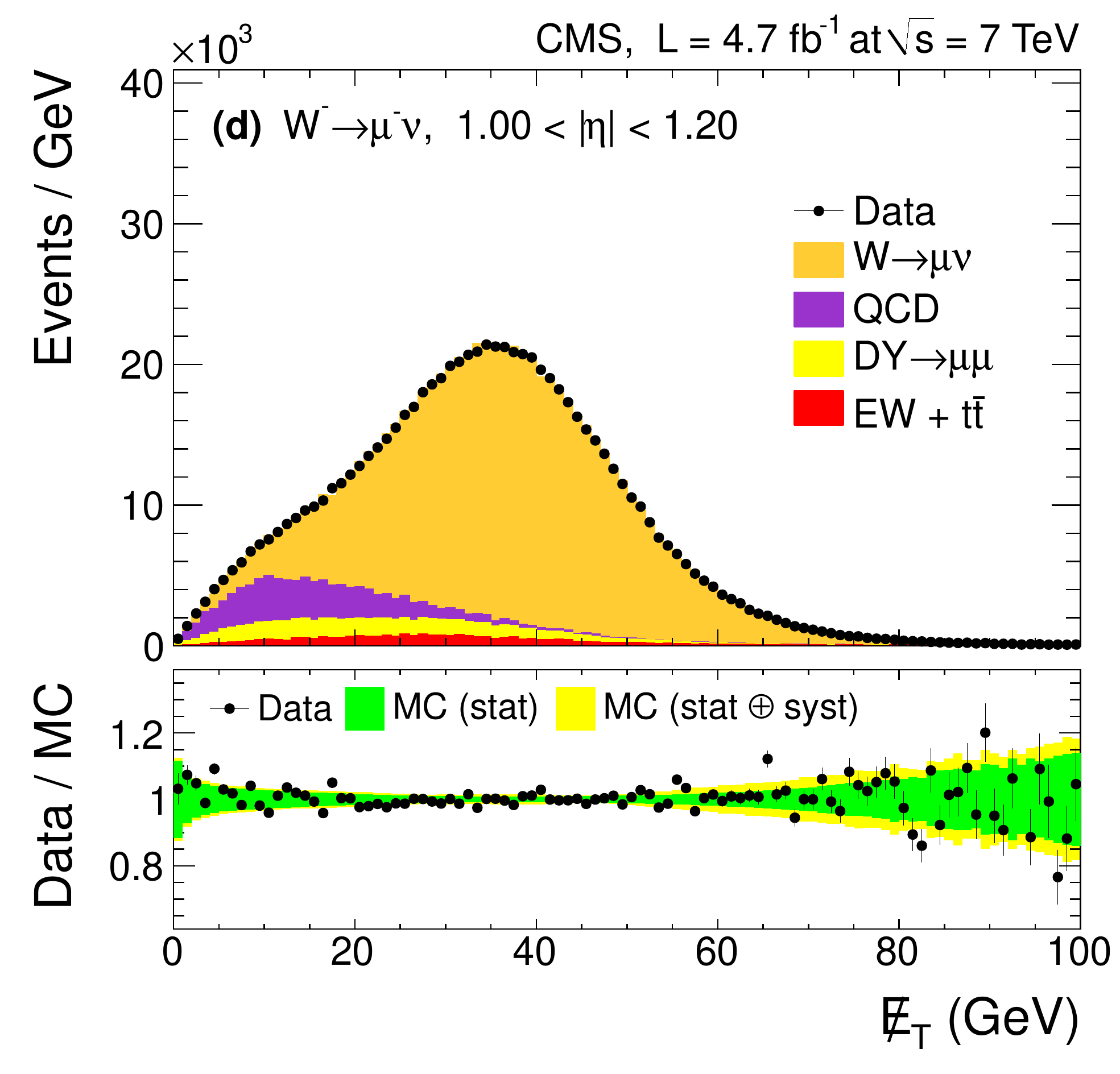}
    \includegraphics[width=\cmsFigWidth]{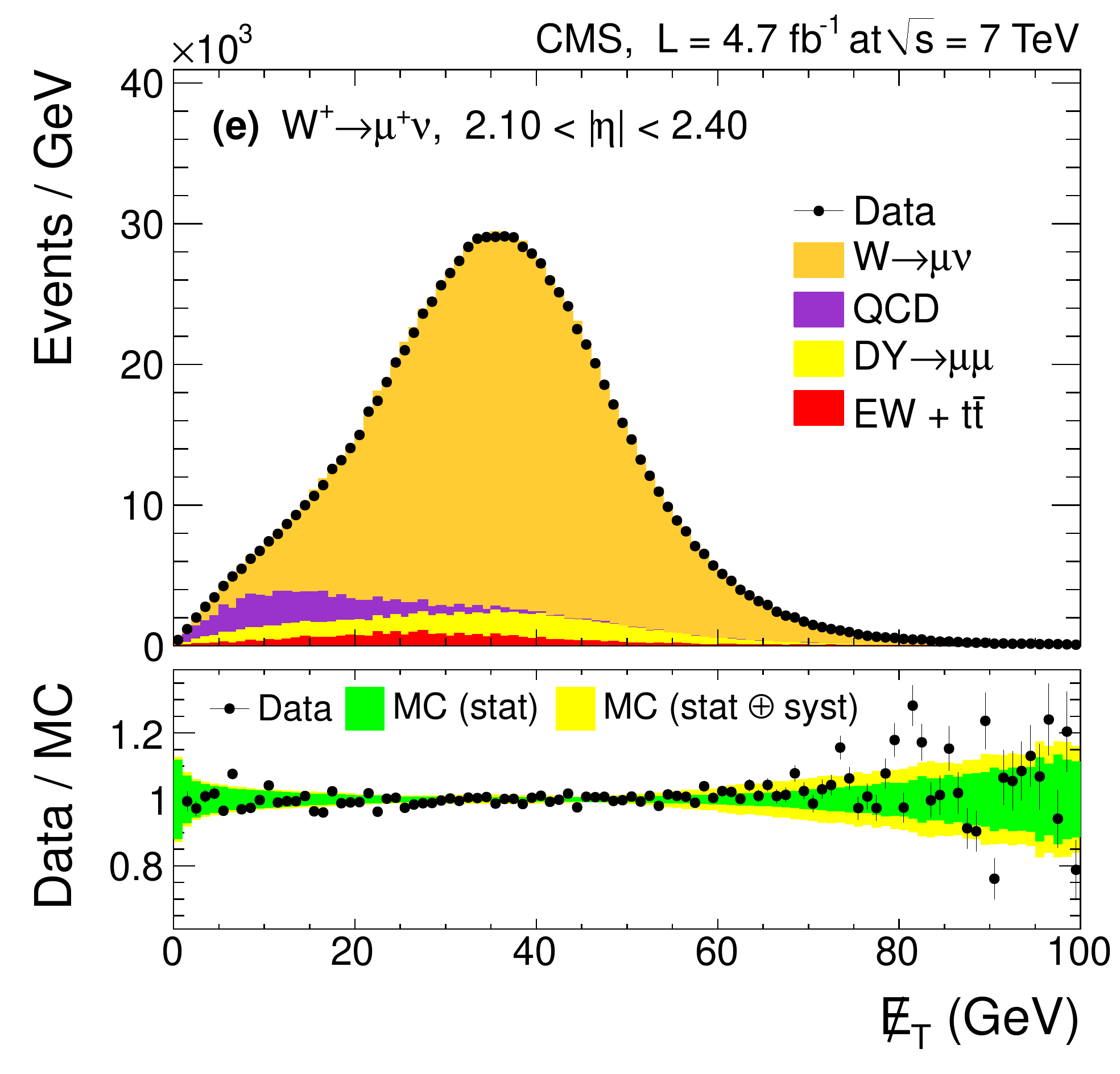}
    \includegraphics[width=\cmsFigWidth]{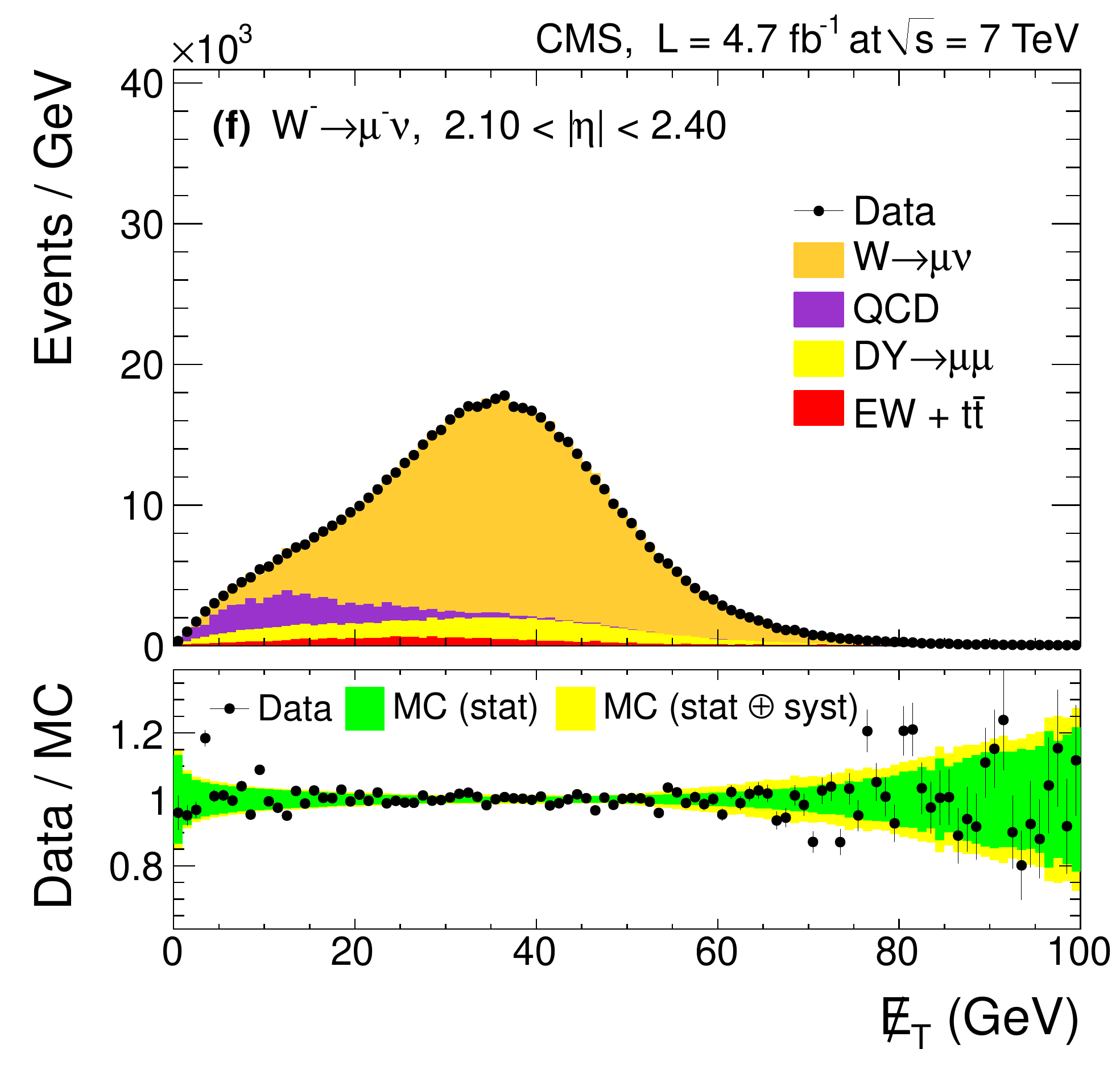}
    \caption{ Muon  $\pt > 25\GeV$ data  sample.  Examples of the extraction
    of the $\wmunu$ signal from fits to $\EmT$  distributions of $\wmunu$
candidates in data:
 $0.0\leq \abs{\eta} <0.2$~(a,~b),   $1.0\leq \abs{\eta} <1.2$~(c,~d), and  $2.1\leq \abs{\eta} <2.4$~(e,~f). The fits to
 $\PWp \to \Pgmp\Pgn$  and
$\PWm \to \Pgmm\Pagn$ candidates are in panels~(a,~c,~e) and (b,~d,~f), respectively.
The ratios between the data points  and the
final fits are
shown at the bottom of each panel.
 The dark shaded band in each ratio plot shows
the statistical uncertainty in the shape of the  MC  $\EmT$ distribution, and
the light shaded band shows the total uncertainty, including all systematic uncertainties as discussed in Section~\ref{sec:systematics}.
}
    \label{fig:results:fits}
  \end{center}
\end{figure*}

The $\wmunu$ signal yields are obtained by fitting the $\EmT$
distributions with all corrections applied.  The events are selected
with the default muon \pt threshold of~$25\GeV$.
The fits for $\PWp$ and $\PWm$ are shown in Fig.~\ref{fig:results:fits}
for three ranges of $\etaabs$, namely,
$0.0 \leq \etaabs < 0.2$,  $1.0 \le \etaabs  < 1.2$, and $2.1 \leq \etaabs < 2.4$.
The ratio of the data to the fit result is shown below each distribution,
demonstrating good agreement of the fits with the data.

Table~\ref{table:fitresults} summarizes the fitted yields
$\NWp$ and $\NWm$, the correlation coefficient, the $\chi^2$ value
for each fit, and the raw asymmetry~$\ARAW$.   The $\chi^2$ values
indicate that the fits are good.  The uncertainty in~$\ARAW$ takes
the covariance of $\NWp$ and~$\NWm$ into account.
Corrections to $\ARAW$ for potential bias are discussed in
the next section.

As an important cross-check, we repeat the analysis with  a higher
muon \pt threshold of~$35\GeV$.  The background compositions
change significantly; the QCD background is reduced to about~1\%.
Furthermore, the predicted asymmetry differs from that predicted for
the default analysis with the $25\GeV$ threshold.  The results are
summarized in Table~\ref{table:fitresults}; they can be compared directly
to the earlier measurement done with electrons~\cite{CMS-PAS-SMP-12-001}.

\begin{table*}[htb]
\centering
\topcaption{Summary of the fitted $N^{\PWp}$,  $N^{\PWm}$,  the
correlation between  the uncertainties in $N^{\PWp}$ and
 $N^{\PWm}$~($\rho_{(N^{\PWp},\, N^{\PWm} )} $), the $\chi^{2}$ of the fit,
and the extracted
 $ \mathcal{A}^{\text{raw}}$ for each $\abs{\eta}$ bin. The number of degrees of
freedom~($n_{\text{doff}}$) in each fit is 197.  Here,
$\rho_{(N^{\PWp},\, N^{\PWm} )} $ and $ \mathcal{A}^{\text{raw}}$ are
 expressed as percentages.  }
\label{table:fitresults}
\resizebox{\textwidth}{!}{
\begin{scotch}{  l | r  r  c c c  }
$\abs{\eta}$ bin 	&   $N^{\PWp}$~($10^3$)	&   $N^{\PWm}$~($10^3$)   &  $\rho_{(N^{\PWp}, N^{\PWm} )} $~(\%) &   $\chi^{2}$ ($n_{\text{doff}}$ = 197)&  $ \mathcal{A}^{\text{raw}}$~(\%)  \\
\hline
\multicolumn{6}{c}{$\pt > 25\GeV$}\\\hline
0.00--0.20 & $ 1033.0\pm1.4 $  & $ 764.9\pm1.2 $ & 14.5 & 255  & $14.912\pm0.096 $  \\
0.20--0.40 & $  970.2\pm1.3 $  & $ 713.9\pm1.2 $ & 14.9 & 190  & $15.216\pm0.098 $ \\
0.40--0.60 & $ 1060.3\pm1.4 $  & $ 771.5\pm1.2 $ & 14.7 & 220  & $15.766\pm0.094 $ \\
0.60--0.80 & $ 1055.1\pm1.4 $  & $ 752.4\pm1.2 $ & 14.6 & 213  & $16.745\pm0.093 $ \\
0.80--1.00 & $  935.8\pm1.3 $  & $ 652.1\pm1.1 $ & 14.5 & 245  & $17.866\pm0.098 $ \\
1.00--1.20 & $  931.0\pm1.3 $  & $ 625.4\pm1.1 $ & 13.9 & 231  & $19.636\pm0.099 $ \\
1.20--1.40 & $  949.0\pm1.3 $  & $ 621.6\pm1.1 $ & 14.2 & 209  & $20.848\pm0.099 $ \\
1.40--1.60 & $  957.1\pm1.3 $  & $ 607.3\pm1.1 $ & 13.7 & 202  & $22.365\pm0.099 $ \\
1.60--1.85 & $ 1131.8\pm1.4 $  & $ 687.6\pm1.2 $ & 14.7 & 225  & $24.417\pm0.093 $ \\
1.85--2.10 & $ 1113.4\pm1.4 $  & $ 656.8\pm1.1 $ & 12.9 & 237  & $25.797\pm0.094 $ \\
2.10--2.40 & $  843.6\pm1.2 $  & $ 481.3\pm1.0 $ & 11.8 & 244  & $27.341\pm0.106 $ \\

\hline
\multicolumn{6}{c}{$\pt > 35\GeV$}\\\hline

0.00--0.20 & $  574.3\pm1.0 $  & $ 459.7\pm0.9 $ & 18.9 & 203  & $11.083\pm0.116 $ \\
0.20--0.40 & $  538.9\pm0.9 $  & $ 428.9\pm0.9 $ & 17.4 & 202  & $11.371\pm0.119 $ \\
0.40--0.60 & $  588.3\pm1.0 $  & $ 462.8\pm0.9 $ & 18.5 & 187  & $11.935\pm0.114 $ \\
0.60--0.80 & $  582.9\pm1.0 $  & $ 453.7\pm0.9 $ & 18.7 & 205  & $12.472\pm0.114 $ \\
0.80--1.00 & $  513.7\pm0.9 $  & $ 392.3\pm0.8 $ & 18.7 & 218  & $13.406\pm0.124 $ \\
1.00--1.20 & $  509.1\pm0.9 $  & $ 379.2\pm0.8 $ & 15.7 & 226  & $14.620\pm0.121 $ \\
1.20--1.40 & $  520.2\pm0.9 $  & $ 376.9\pm0.8 $ & 16.2 & 191  & $15.970\pm0.123 $ \\
1.40--1.60 & $  522.7\pm0.9 $  & $ 370.2\pm0.8 $ & 14.7 & 195  & $17.074\pm0.123 $ \\
1.60--1.85 & $  614.6\pm1.0 $  & $ 418.8\pm0.9 $ & 17.5 & 239  & $18.945\pm0.118 $ \\
1.85--2.10 & $  604.7\pm1.0 $  & $ 395.8\pm0.9 $ & 15.0 & 192  & $20.885\pm0.123 $ \\
2.10--2.40 & $  464.3\pm0.9 $  & $ 288.5\pm0.8 $ & 14.7 & 234  & $23.357\pm0.141 $ \\
\end{scotch}
}
\end{table*}

\section{Systematic uncertainties and corrections}
\label{sec:systematics}

The systematic uncertainties arise from many sources, including the
measurement of the muon kinematics (efficiency, scale and resolution),
the modeling of the $\EmT$ distributions, backgrounds, the boson
$\vecqT$ distribution and final-state radiation.
In general, the total systematic uncertainty is 2--2.5 times larger
than the statistical uncertainty (see Table~\ref{table:systematics})
and the main contributions come from the muon efficiency and
from the QCD background.    In the sections below, we discuss
each source of systematic uncertainty, starting with muon-related
quantities, followed by the $\EmT$ measurement, backgrounds, and
boson-related modeling issues.

We evaluate many of these uncertainties using a MC method,
in which 400 sets of pseudo-data are fitted to obtain the
distribution of~$\ARAW$ values.  This method allows us to propagate the
uncertainties of the corrections to the measurement in a rigorous manner.

Several sources of potential bias are considered.  To evaluate
the bias, we defined a ``true'' muon charge asymmetry, $\ATRUE$,
calculated by taking the muon four-vectors and charge directly
from the MC generator.

\subsection{Muon kinematics}

One source of potential bias for $\ARAW$ is the charge of the muon.
The rate of charge mis-measurement, $w$, is very small but not zero.
The measured asymmetry will differ from the true asymmetry by
a factor $(1 - 2w)$ assuming that the rate of mismeasurement
is the same for positive and negative muons.
The muon charge misidentification rate has been studied in detail
and  shown to have a negligible effect on the measured  asymmetry~\cite{CMS:asym:2010}.

The muon \pt resolution can induce a spread of the measured
asymmetry from $\ATRUE$, which varies from 1.5 to 5.0\%~\cite{CMS-PAPERS-MUO-10-004}
as a function of~$\etaabs$.
The resolution of $\etaabs$ is several orders of magnitude
smaller than the bin widths used in this measurement; consequently,
event migration around $\pt$-$\eta$ thresholds has a
negligible effect on the measured asymmetry.

The muon momentum correction
affects both the yields and the shapes of the $\EmT$ distributions.
To estimate the systematic uncertainty from this source,
the muon $1/\pt$ correction parameters in each $\eta$--$\phi$ bin and
the muon scale global correction
parameters are varied 400 times within their uncertainties.
Each time the event yields can  be slightly different in both data and MC
simulation,
and the extraction of the asymmetry is done for each of the 400 cases.
The root mean square~(RMS) of the measured $\ARAW$ variations
in  each muon $\etaabs$ bin is taken as the systematic uncertainty and the
bin-to-bin correlations are assumed to be zero.

The systematic uncertainties resulting from the muon momentum corrections
 are typically less than  40\% of those from the  uncertainties in the
muon efficiencies (discussed below) for the    $\pt > 25\GeV$ sample.
However, the two uncertainties are  comparable for the $\pt > 35\GeV$
 sample for two reasons:
 first, the charge-dependent bias from the alignment
increases with  $\pt$;
second, the Jacobian peak of the $\wmunu$ events is close to 35\GeV.

\subsection{Muon efficiency ratio}
\label{sec:sys:muon_eff}

A difference in the muon efficiencies for positively and negatively charged
muons will  cause the ratio of the selection efficiencies for $\PWp$ and $\PWm$
to differ from unity.
This would bias the measured charge asymmetry and we correct the
$\ARAW$ for this bias.

As discussed previously, the muon offline and trigger efficiencies
are measured  in 7 bins  in \pt  and 22 bins in $\etaabs$ for
both $\Pgmp$ and $\Pgmm$.
The offline efficiency ratio between  $\Pgmp$ and $\Pgmm$ is
very close to unity in most of the detector regions.  However, there is
evidence that this ratio deviates from unity in the  transition regions between
the DT and CSC detectors.

We correct for this  bias using efficiencies for $\mu^+$ and $\mu^-$
extracted from the $\zmumu$ data and MC samples.
For each $\etaabs$ bin, an average $\PW$ selection
efficiency $\epsilon (\PW^{\pm})$ is  obtained from the expression
\begin{equation}
\epsilon (\PW^{\pm}) = \frac{\Sigma (k \epsilon^{\pm}_{\text{data}}(\pt, \eta)/\epsilon^{\pm}_{\textrm{MC}}(\pt, \eta) ) }{\Sigma ( k/\epsilon^{\pm}_{\textrm{MC}}(\pt, \eta) )  },
\end{equation}
where $\epsilon^{\pm}_{\text{data}}(\pt, \eta),\ \epsilon^{\pm}_{\mathrm{MC}}(\pt, \eta)$
are total muon efficiencies, $k$ are additional event-by-event weights
introduced by $\PW$-boson $\qT$ weighting described
below, and the sum is over the selected $\wmunu$ events.
The efficiency ratio~($r^{\PWp/\PWm} = \epsilon^+ / \epsilon^-$) is
used to correct the $\ARAW$ for the efficiency bias using
\begin{equation}
\label{eq:eff_corr}
 \ATRUE = \ARAW - \frac{1-(\ARAW)^{2}}{2}\left(r^{\PWp/\PWm} -1\right),
\end{equation}
which is an expansion to leading order in $\left( r^{\PWp/\PWm} -1 \right)$.
In addition, all MC samples are corrected for
any data/MC efficiency difference.

To estimate the systematic uncertainty due to  the  muon efficiencies,
the muon efficiency values in data and MC simulation are modified
according to their errors in each $\pt$--$\eta$ bin independently and
400 pseudo-efficiency tables are generated.
In each pseudo-experiment the efficiency values are used to correct the MC
simulation and $\ARAW$.
The raw asymmetry is further corrected for
the efficiency ratio $r^{\PWp/\PWm}$ described above.
The RMS of the resulting asymmetries in each $\abs{\eta}$ bin is taken
to be the systematic uncertainty originating from the determination of
the ratio of the muon efficiencies.
In this study, the variations for different $\etaabs$
bins are completely independent from each other, so
the systematic uncertainties due to the efficiency ratio
have zero correlation between different $\etaabs$ bins.

\begin{figure}[hbtp]
  \begin{center}
    \includegraphics[width=\cmsFigWidth]{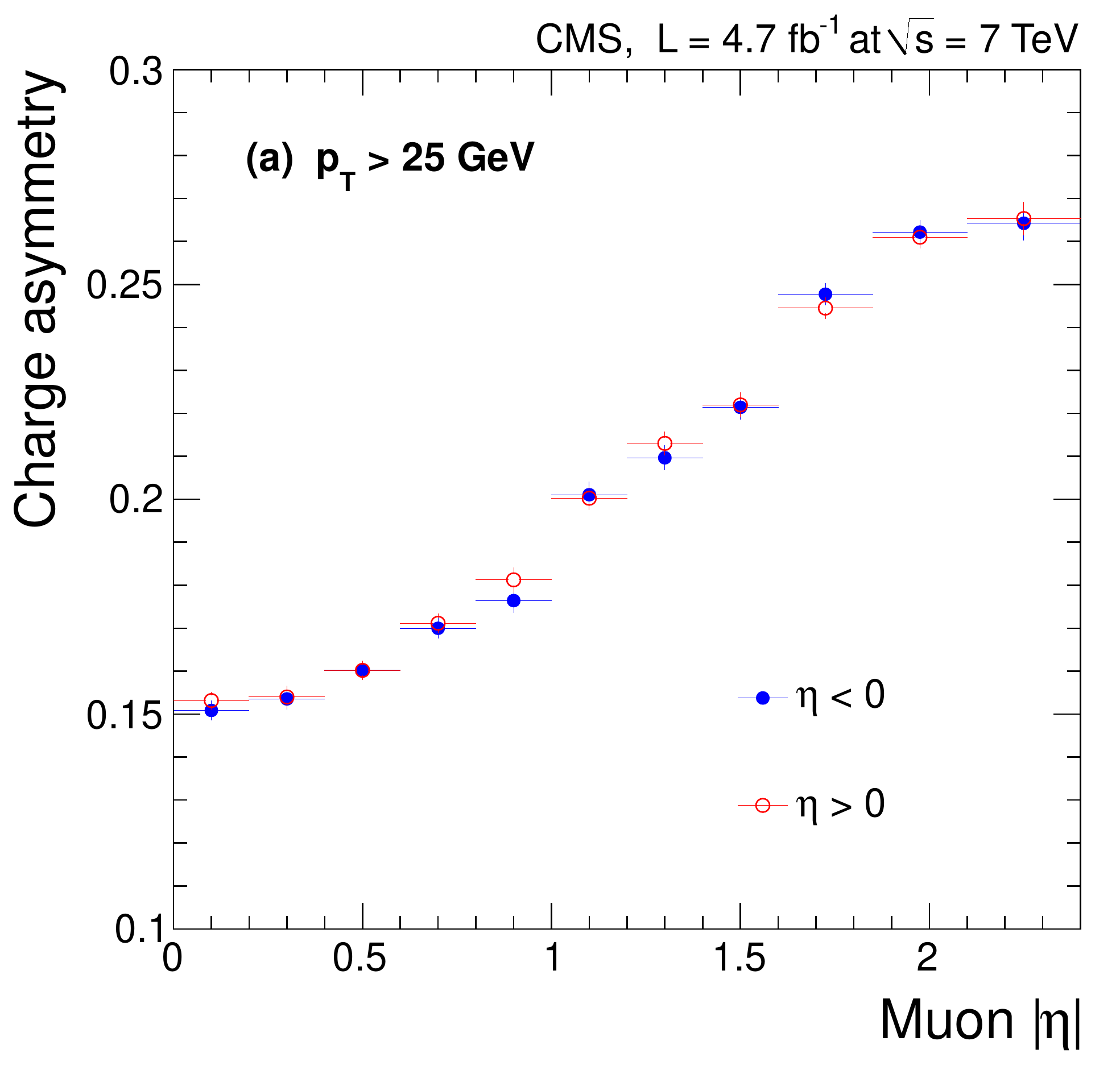}
    \includegraphics[width=\cmsFigWidth]{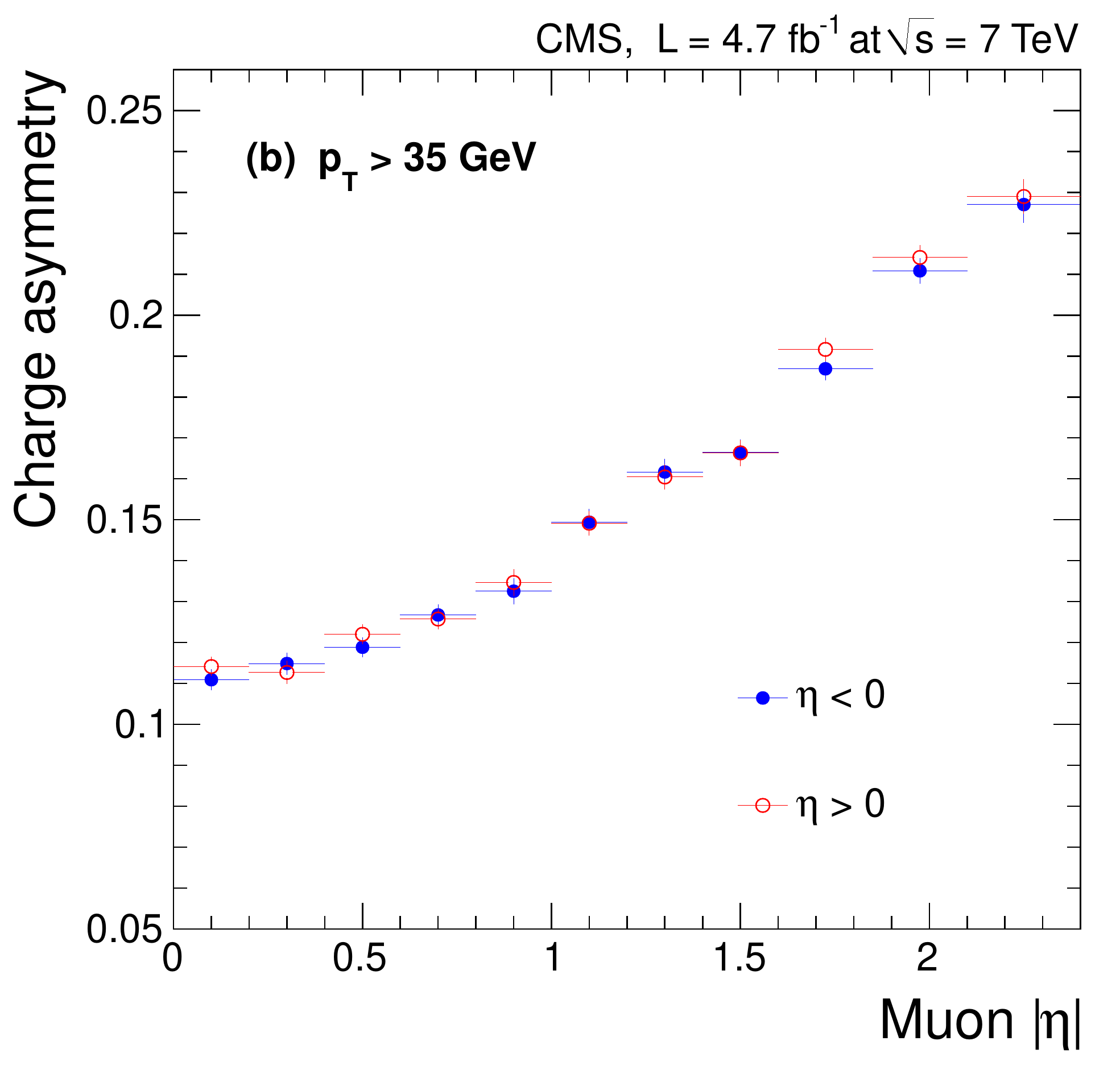}
    \caption{Comparison of the final muon charge
asymmetry~($ \mathcal{A}$) extracted for
the positive pseudorapidity~($\eta>0$) and negative pseudorapidity~($\eta<0$) regions
with muon $\pt > 25\GeV$~(a)  and  muon $\pt > 35\GeV$~(b) samples.
The uncertainties include only the statistical uncertainty from the signal extraction and uncertainty
in the determination of the efficiencies for positive and negative muons.
}
    \label{fig:results:etap_etam}
  \end{center}
\end{figure}

As a cross-check, Fig.~\ref{fig:results:etap_etam} shows a comparison of
the measured muon charge asymmetry between positive and negative $\eta$ regions,
taken separately and then overlaid.  They are in very good agreement
with each other, for both muon~$\pt$ thresholds.

\subsection{Backgrounds}

The QCD background is estimated in part from the data.
Nonetheless, a nonnegligible systematic uncertainty remains.
We also discuss the uncertainty from the Drell-Yan background,
and from the $\ttbar$ and $\wtaunu$ backgrounds.   The luminosity
uncertainty enters in the estimation of these backgrounds,
as discussed below.

\subsubsection{QCD background}

The total QCD background normalization is a parameter
 in the signal fit.  The ratio of the QCD backgrounds in the
$\PWp$ and $\PWm$ samples is fixed to the ratio observed in
the QCD control region for each muon $\eta$ bin.
The ratios are about 1.02 for the first ten $\eta$ bins and approximately
 1.05 for the last $\eta$ bin, similar for both muon $\pt >$~25 and 35\GeV.
There are two sources of the systematic uncertainties
 in the QCD background.  The first  is related to
 the ratio of the backgrounds in the
$\PWp$ and $\PWm$ samples~(``QCD $+/-$''), and the second is related to the modeling of
the shape of the
$\EmT$ distribution in QCD events~(``QCD shape'').

To evaluate the systematic
uncertainties  ``QCD  $+/-$'', the QCD ratio is varied by $\pm$5 and
$\pm$15\% for muon \pt thresholds of 25 and 35\GeV, respectively.
The resulting shifts in  the $\mathcal{A}^{\text{raw}}$ are taken as the uncertainties.
For the last $\abs{\eta}$ bin, the variations are 10\%~(25\GeV) and
20\%~(35\GeV).
These variations of the QCD ratio span the maximum range indicated by the
QCD MC simulation.   As an  additional cross-check,
we fix the QCD shape
to be the same for  $\Pgmp$ and
$\Pgmm$ and allow the two  QCD normalizations  to float
 in the extraction of the signal. We find that the fitted values for the ratio of the  QCD
backgrounds for
$\PWp$ and $\PWm$ are within the uncertainties quoted above.
The
bin-to-bin correlation of these uncertainties
in the asymmetries is assumed to be zero.

The second source of systematic uncertainties is a difference in the shape of
the QCD background  for $\PWp$ and $\PWm$.
 The QCD  $\EmT$ shape is taken from the MC simulation and
the recoil correction is applied as discussed in Section~\ref{sec:signal:procedure}. Two types of variations
in the shape of the  QCD
  $\EmT$ distribution are  considered. First,
the shape of the QCD  $\EmT$ distribution
without the hadronic recoil correction is used in the extraction of the
signal. This is done in a correlated way for the $\PWp$ and $\PWm$ samples.
Second, the shape of the  $\EmT$ distribution for the QCD background is varied separately
for the $\PWp$ and $\PWm$ samples~(within the statistical uncertainties)
 and the resulting shapes are used in
the signal extraction. These  two contributions to the uncertainties
from the  ``QCD shape'' are then
added in quadrature. The bin-to-bin correlation
of the systematic uncertainties due to each shape variation is
assumed to be 100\%.

\subsubsection{Drell--Yan background}
\label{sec:sys:Drell-Yan}

The $\zmumu$ events in the Drell--Yan control region  are used to check the
Drell--Yan normalization. This is done in bins of dimuon
invariant mass: 15--30, 30--40, 40--60, 60--120, 120--150, and $>$150\GeV. The
$\zmumu$ MC simulation in each bin is compared to the data yields after
correcting the simulation
for the data/simulation difference in pileup, $\cPZ$-boson $\qT$,  $\EmT$
modeling, and efficiencies. After correcting for the detector bias and
physics mismodeling, the MC simulation describes the data well,
as shown in Fig.~\ref{fig:zmumu:norm} for the
dimuon invariant mass between 60 and 120\GeV. The data yield in
this mass bin is about
3\% higher than the predictions from the next-to-next-to-leading-order~(NNLO) cross section
as calculated with \textsc{fewz} 3.1~\cite{FEWZ}.

The ratios of data to MC simulation of the $\zmumu$ event  yields as a function
of the dimuon mass
are used to   rescale the MC prediction of the Drell--Yan background. We take the shift in the
$\mathcal{A}^{\text{raw}}$ with and without this rescaling as the
systematic uncertainty. This and the PDF uncertainties in the
$\zmumu$ yields are considered as systematic uncertainties due to
``Drell--Yan background normalization''. This uncertainty
is almost negligible at  central $\abs{\eta}$ bins and increases in the
forward $\abs{\eta}$ bins.
The Drell--Yan background is larger
in the forward region
because of the lower efficiency of
the ``Drell--Yan veto'' due to less
 detector coverage. The
 systematic uncertainties in the Drell--Yan background are assumed to have
  100\% correlation from bin to bin.

\subsubsection{The \texorpdfstring{$\ttbar$}{t t-bar} and \texorpdfstring{$\wtaunu$}{W tau nu} backgrounds}

The $\ttbar$ and $\ztautau$  backgrounds are normalized to the integrated luminosity
of the data sample after correcting for the muon efficiency difference between data and MC simulation.
The uncertainty of the integrated luminosity is 2.2\%~\cite{CMS-PAS-SMP-12-008}. The
normalization of all the MC backgrounds is varied by $\pm 2.2\%$,
and the resulting maximum shift in $\ARAW$ is taken
as the systematic uncertainty in the determination of the  luminosity.
The  bin-to-bin correlations are 100$\%$.

The $\ttbar$ background estimate also depends on the
theoretical prediction~\cite{Campbell:1999ah, Campbell:2010ff, Campbell:2012uf},
to which we assign an additional 15\%.   The bin-to-bin
correlation is~100\%.

The $\wtaunu$ background is normalized to the $\wmunu$ yields in
data with a ratio obtained from a MC simulation. This ratio
is largely determined by the branching fraction of $\Pgt$ decaying to $\Pgm$.
A 2\% uncertainty is assigned to the  $\wtaunu$ to $\wmunu$ ratio~\cite{PhysRevD.86.010001}.
The correlation of this uncertainty is $100\%$ bin-to-bin.

\subsection{Modeling Uncertainties}

The remaining systematic uncertainties pertain to the modeling
of the detector and the signal process $\wmunu$.
We discuss first the issues concerning the $\EmT$ distribution,
then FSR and finally the $\qT$ distribution.

\subsubsection{Modeling of missing transverse momentum}

To evaluate the systematic uncertainty due to the $\phi$ modulation of
 $\vecEmT$, the correction for  the $\phi$ modulation
 is removed and the shift in the $\mathcal{A}^{\text{raw}}$ is taken
as the systematic uncertainty.

The hadronic recoil  correction changes
the shape of the  $\EmT$ distribution of all MC
samples. To calculate the uncertainties resulting from this source,
the average recoil  and resolution parameters are varied
 within their uncertainties, taking into account the correlations between
them. This is done 400 times, the RMS of the resulting
$ \mathcal{A}^{\text{raw}}$ variations is  taken as
 systematic uncertainty and bin-to-bin correlations are calculated.

Pileup can affect the $\EmT$ shapes. To estimate the effect of
 mismodeling the pileup in the simulation, the minimum-bias cross section
is  varied  by $\pm$5\% and the pileup distributions expected in data are
regenerated. The MC simulation is then weighted to match to data and the
resulting shift in $ \mathcal{A}^{\text{raw}}$ is treated as a systematic uncertainty due to
 the  pileup.
Pileup affects the $\EmT$ shapes for all muon $\eta$ bins in the same
direction with a  correlation of   100\%.

\subsubsection{Final-State Radiation}
\label{sec:sys:FSR}

The emission of FSR photons in $\PW$ decays reduces the muon~$\pt$
and can cause a difference in acceptance between $\PWp$ and $\PWm$.
We studied the impact of the FSR on the muon charge asymmetry using the
\POWHEG $\wmunu$ MC sample. In this sample,
 FSR is implemented using a similar approach to parton showering and
is approximate at the leading order~(LO).
 We compare the muon charge asymmetry before and after FSR, and the
difference is found to be within 0.07--0.12\% and 0.03--0.11\% for muon \pt selections
 of  25 and 35\GeV, respectively.
The raw asymmetry values are not corrected for FSR. Instead, the full
shift in the muon charge asymmetry predicted by the  \POWHEG MC
is taken as an additional systematic uncertainty
and the bin-to-bin correlation is assumed to be 100\%.

\subsubsection{PDF uncertainty}

The evaluation of PDF uncertainties follows the PDF4LHC recommendation~\cite{PDF4LHC}.
The NLO  MSTW2008~\cite{Martin:2009ad}, {CT10}~\cite{CTEQ:1007},  and
NNPDF2.1~\cite{Ball:2011mu} PDF sets are used.
All simulated events are weighted to a given PDF set
and the overall normalization is allowed to vary.
In this way  both the uncertainties in the total cross sections, as well as
in the shape of the $\EmT$  distribution  are considered.
To estimate the systematic uncertainty resulting from the uncertainties in the
CT10 and MSTW2008  PDFs,
asymmetric master equations  are used~\cite{Martin:2009ad, CTEQ:1007}. For the
CT10, the
90\% confidence level (\CL) uncertainty is
rescaled to 68\% \CL  by dividing by a factor of 1.64485. For the
NNPDF2.3
PDF set, the RMS of the  $ \mathcal{A}^{\text{raw}}$ distributions is taken.
The half-width of the maximum deviation from combining all three
PDF uncertainty bands is taken as the PDF uncertainty.
The  CT10 error set is
used to estimate  the bin-to-bin correlations.
The PDF uncertainties are about~10\% of the total experimental
uncertainty.

\subsubsection{\texorpdfstring{$\PW$}{W}-boson \texorpdfstring{$\qT$}{qT} modeling}
\label{sec:sys:Wqt}

To improve the agreement between data and simulation,
the $\PW$-boson $\qT$ spectrum is  weighted using
 weight factors determined by the ratios
 of the distribution  of boson $\qT$ for  $\zmumu$  events  in
 data and MC simulation.
We assume that the corrections are the same for $\PW$ and $\cPZ$ events.
This assumption is tested using two different sets of MC simulations: one from the
\POWHEG event generator and the other
from \MADGRAPH~\cite{MADGRAPH}.  Here,
the  \MADGRAPH simulation
is  treated as the ``data'', and the ratio of $\cPZ$-boson $\qT$
of the  \MADGRAPH and  \POWHEG simulations is
compared to the same ratio in simulated $\PW$-boson events.
This double ratio is parametrized using an
empirical function to smooth the statistical fluctuations, and
additional weights are obtained using the fitted function.
We weight the  \POWHEG simulation to be close to
the \MADGRAPH simulation and measure the asymmetry again.
The deviation of  $\ARAW$
is taken as the systematic uncertainty due to mismodeling of $\PW$-boson $\qT$.
The default boson $\qT$ weighting is based on the \POWHEG simulation.

\subsection{Total systematic uncertainty}

Table~\ref{table:systematics} summarizes the systematic
uncertainties in all $\etaabs$ bins.  For comparison,
the statistical uncertainty in each $\etaabs$ bin
is also shown. The dominant systematic uncertainties
come from muon efficiencies, QCD background, and the
muon momentum correction. The correlation matrix of
systematic uncertainty among  $\etaabs$ bins is
reported in Table~\ref{table:correlation}.
The correlations among $\etaabs$ bins are small
and do not exceed 37 and 14\% for muon \pt thresholds of~25 and~35\GeV, respectively.  Much of the correlation
is due to the systematic uncertainties in FSR and QCD background.
The total covariance matrix, including both statistical and systematic
uncertainties, is provided \suppMaterial.
\begin{table*}[htm]
\centering
\topcaption{ Systematic
uncertainties in  $\mathcal{A}$ for each $\abs{\eta}$ bin.
The statistical uncertainty in each $\abs{\eta} $ bin
is also shown for comparison. A detailed description of each systematic uncertainty is
given in the text. The values  are
 expressed as percentages, the same as for the
asymmetries.  }
\label{table:systematics}
\resizebox{\textwidth}{!}{
\begin{scotch}{  l | c  c  c  c  c  c  c  c  c  c  c  }
$\abs{\eta}$ bin   &   0.0--0.2 & 0.2--0.4 & 0.4--0.6 & 0.6--0.8 & 0.8--1.0 & 1.0--1.2 & 1.2--1.4 & 1.4--1.6 & 1.6--1.85 & 1.85--2.1 & 2.1--2.4 \\
\hline

\multicolumn{12}{c}{$\pt > 25\GeV$}\\\hline

Stat. unc.			    &  0.096 &   0.098 &   0.094 &   0.093 &   0.098 &   0.099 &   0.099 &   0.099 &   0.093 &   0.094 &   0.106 \\
Efficiency			    &  0.111 &   0.133 &   0.121 &   0.122 &   0.170 &   0.175 &   0.170 &   0.168 &   0.165 &   0.175 &   0.268 \\
QCD  $+/-$				    &  0.120 &   0.113 &   0.110 &   0.105 &   0.102 &   0.103 &   0.097 &   0.104 &   0.108 &   0.094 &   0.183 \\
QCD shape			    &  0.070 &   0.065 &   0.065 &   0.067 &   0.068 &   0.069 &   0.078 &   0.082 &   0.092 &   0.083 &   0.087 \\
Muon scale			    &  0.045 &   0.050 &   0.050 &   0.049 &   0.051 &   0.054 &   0.054 &   0.058 &   0.054 &   0.054 &   0.055 \\
FSR				    &  0.074 &   0.077 &   0.104 &   0.109 &   0.089 &   0.113 &   0.107 &   0.091 &   0.118 &   0.087 &   0.077 \\
PDF				    &  0.028 &   0.026 &   0.023 &   0.025 &   0.018 &   0.020 &   0.027 &   0.031 &   0.042 &   0.050 &   0.069 \\
Drell--Yan bkg.			    &  0.002 &   0.001 &   0.002 &   0.003 &   0.000 &   0.007 &   0.001 &   0.013 &   0.019 &   0.038 &   0.046 \\
$\EmT \ \phi$ modul.    &  0.011 &   0.009 &   0.033 &   0.012 &   0.029 &   0.034 &   0.044 &   0.045 &   0.055 &   0.049 &   0.038 \\
Recoil				    &  0.003 &   0.003 &   0.003 &   0.003 &   0.003 &   0.003 &   0.003 &   0.003 &   0.003 &   0.004 &   0.003 \\
Pileup				    &  0.017 &   0.013 &   0.011 &   0.005 &   0.014 &   0.025 &   0.022 &   0.031 &   0.019 &   0.028 &   0.000 \\
Luminosity			    &  0.002 &   0.003 &   0.004 &   0.004 &   0.006 &   0.009 &   0.012 &   0.017 &   0.024 &   0.033 &   0.040 \\
$\ttbar$ bkg.			    &  0.012 &   0.013 &   0.012 &   0.012 &   0.011 &   0.011 &   0.010 &   0.009 &   0.008 &   0.007 &   0.005 \\
$\wtaunu$ bkg.			    &  0.026 &   0.026 &   0.026 &   0.026 &   0.026 &   0.025 &   0.025 &   0.025 &   0.025 &   0.025 &   0.024 \\
$\PW \ \qT$			    &  0.003 &   0.004 &   0.004 &   0.005 &   0.008 &   0.011 &   0.008 &   0.009 &   0.006 &   0.003 &   0.000 \\ \hline
Total syst. unc.		    &  0.203 &   0.212 &   0.217 &   0.216 &   0.238 &   0.255 &   0.251 &   0.250 &   0.266 &   0.256 &   0.364 \\ \hline
Total unc.			    &  0.225 &   0.233 &   0.236 &   0.235 &   0.258 &   0.274 &   0.270 &   0.269 &   0.282 &   0.273 &   0.379 \\
\hline
\multicolumn{12}{c}{$\pt > 35\GeV$}\\
\hline
Stat. unc.			    &  0.116 &   0.119 &   0.114 &   0.114 &   0.124 &   0.121 &   0.123 &   0.123 &   0.118 &   0.123 &   0.141 \\
Efficiency			    &  0.120 &   0.138 &   0.116 &   0.107 &   0.159 &   0.164 &   0.171 &   0.176 &   0.186 &   0.194 &   0.325 \\
QCD  $+/-$				    &  0.151 &   0.138 &   0.135 &   0.128 &   0.133 &   0.118 &   0.116 &   0.122 &   0.137 &   0.120 &   0.168 \\
QCD shape			    &  0.030 &   0.025 &   0.017 &   0.023 &   0.024 &   0.022 &   0.018 &   0.017 &   0.031 &   0.031 &   0.037 \\
Muon scale			    &  0.122 &   0.135 &   0.134 &   0.141 &   0.146 &   0.154 &   0.162 &   0.170 &   0.161 &   0.172 &   0.189 \\
FSR				    &  0.028 &   0.050 &   0.057 &   0.078 &   0.022 &   0.041 &   0.076 &   0.055 &   0.090 &   0.109 &   0.105 \\
PDF				    &  0.008 &   0.008 &   0.007 &   0.011 &   0.012 &   0.010 &   0.017 &   0.022 &   0.031 &   0.040 &   0.058 \\
Drell--Yan bkg.			    &  0.010 &   0.009 &   0.009 &   0.003 &   0.006 &   0.010 &   0.008 &   0.009 &   0.009 &   0.020 &   0.040 \\
$\EmT \ \phi$ modul.    &  0.002 &   0.009 &   0.010 &   0.003 &   0.008 &   0.028 &   0.037 &   0.035 &   0.022 &   0.022 &   0.001 \\
Recoil				    &  0.005 &   0.006 &   0.005 &   0.004 &   0.005 &   0.004 &   0.005 &   0.004 &   0.004 &   0.006 &   0.008 \\
Pileup				    &  0.015 &   0.003 &   0.005 &   0.018 &   0.019 &   0.002 &   0.007 &   0.003 &   0.013 &   0.014 &   0.032 \\
Luminosity			    &  0.001 &   0.002 &   0.000 &   0.000 &   0.000 &   0.001 &   0.004 &   0.010 &   0.016 &   0.025 &   0.039 \\
$\ttbar$ bkg.			    &  0.011 &   0.013 &   0.012 &   0.011 &   0.011 &   0.010 &   0.010 &   0.009 &   0.007 &   0.006 &   0.005 \\
$\wtaunu$ bkg.			    &  0.013 &   0.012 &   0.013 &   0.012 &   0.012 &   0.012 &   0.011 &   0.012 &   0.011 &   0.011 &   0.011 \\
$\PW \ \qT$			    &  0.004 &   0.002 &   0.004 &   0.004 &   0.007 &   0.005 &   0.006 &   0.009 &   0.009 &   0.001 &   0.014 \\ \hline
Total syst. unc.		    &  0.234 &   0.245 &   0.232 &   0.234 &   0.258 &   0.261 &   0.278 &   0.283 &   0.301 &   0.313 &   0.436 \\ \hline
Total unc.			    &  0.261 &   0.272 &   0.259 &   0.260 &   0.286 &   0.288 &   0.304 &   0.308 &   0.323 &   0.336 &   0.458 \\
\end{scotch}
}
\end{table*}
\begin{table*}[htm]
\centering
\topcaption{Correlation matrix of  systematic uncertainties
 between different $\abs{\eta}$ bins. All
systematic uncertainties are treated as additive.
 The values  are
 expressed as percentages.}
\label{table:correlation}
\resizebox{\textwidth}{!}{
\begin{scotch}{  l | d  d  d  d  d  d  d  d  d  d  d  }
\multicolumn{1}{l}{$\abs{\eta}$ bin}   &   \multicolumn{1}{c}{0.0--0.2} & \multicolumn{1}{c}{0.2--0.4} & \multicolumn{1}{c}{0.4--0.6} & \multicolumn{1}{c}{0.6--0.8} & \multicolumn{1}{c}{0.8--1.0} & \multicolumn{1}{c}{1.0--1.2} & \multicolumn{1}{c}{1.2--1.4} & \multicolumn{1}{c}{1.4--1.6} & \multicolumn{1}{c}{1.6--1.85} & \multicolumn{1}{c}{1.85--2.1} & \multicolumn{1}{c}{2.1--2.4} \\\hline

\multicolumn{12}{c}{$\pt > 25\GeV$}\\\hline
0.00--0.20 & 100.0 &  28.1 &  32.4 &  32.9 &  27.1 &  29.0 &  29.5 &  28.0 &  30.5 &  26.1 &  16.7 \\
0.20--0.40 &       & 100.0 &  30.7 &  31.4 &  25.6 &  27.5 &  27.9 &  26.3 &  28.9 &  24.5 &  15.8 \\
0.40--0.60 &       &       & 100.0 &  37.4 &  30.9 &  33.8 &  34.5 &  32.1 &  36.1 &  30.3 &  19.3 \\
0.60--0.80 &       &       &       & 100.0 &  31.1 &  34.0 &  34.4 &  32.0 &  36.3 &  30.4 &  20.0 \\
0.80--1.00 &       &       &       &       & 100.0 &  28.5 &  29.5 &  28.0 &  31.2 &  26.9 &  17.3 \\
1.00--1.20 &       &       &       &       &       & 100.0 &  32.6 &  31.1 &  34.8 &  30.2 &  19.3 \\
1.20--1.40 &       &       &       &       &       &       & 100.0 &  32.8 &  36.9 &  32.2 &  20.8 \\
1.40--1.60 &       &       &       &       &       &       &       & 100.0 &  36.0 &  32.7 &  21.3 \\
1.60--1.85 &       &       &       &       &       &       &       &       & 100.0 &  37.1 &  24.9 \\
1.85--2.10 &       &       &       &       &       &       &       &       &       & 100.0 &  24.4 \\
2.10--2.40 &       &       &       &       &       &       &       &       &       &       & 100.0 \\

\hline
\multicolumn{12}{c}{$\pt > 35\GeV$}\\\hline
0.00--0.20 & 100.0 &   4.6 &   4.8 &   6.4 &   3.4 &   3.6 &   4.7 &   3.4 &   5.4 &   5.8 &   4.3 \\
0.20--0.40 &       & 100.0 &   6.4 &   8.5 &   3.3 &   4.3 &   6.3 &   4.4 &   7.2 &   8.0 &   5.8 \\
0.40--0.60 &       &       & 100.0 &   9.8 &   3.8 &   5.6 &   8.4 &   6.2 &   8.9 &   9.9 &   6.6 \\
0.60--0.80 &       &       &       & 100.0 &   5.1 &   6.9 &  10.7 &   7.8 &  11.9 &  13.5 &   9.7 \\
0.80--1.00 &       &       &       &       & 100.0 &   3.2 &   4.2 &   3.3 &   4.7 &   5.0 &   3.6 \\
1.00--1.20 &       &       &       &       &       & 100.0 &   7.0 &   5.4 &   7.0 &   7.5 &   4.7 \\
1.20--1.40 &       &       &       &       &       &       & 100.0 &   8.1 &  10.8 &  12.0 &   7.8 \\
1.40--1.60 &       &       &       &       &       &       &       & 100.0 &   8.8 &   9.9 &   6.7 \\
1.60--1.85 &       &       &       &       &       &       &       &       & 100.0 &  14.2 &  10.3 \\
1.85--2.10 &       &       &       &       &       &       &       &       &       & 100.0 &  12.6 \\
2.10--2.40 &       &       &       &       &       &       &       &       &       &       & 100.0 \\
\end{scotch}
}
\end{table*}

\section{Results and discussion}
\label{sec:results}

\begin{figure}[hbtp]
\centering
    \includegraphics[width=\cmsFigWidth]{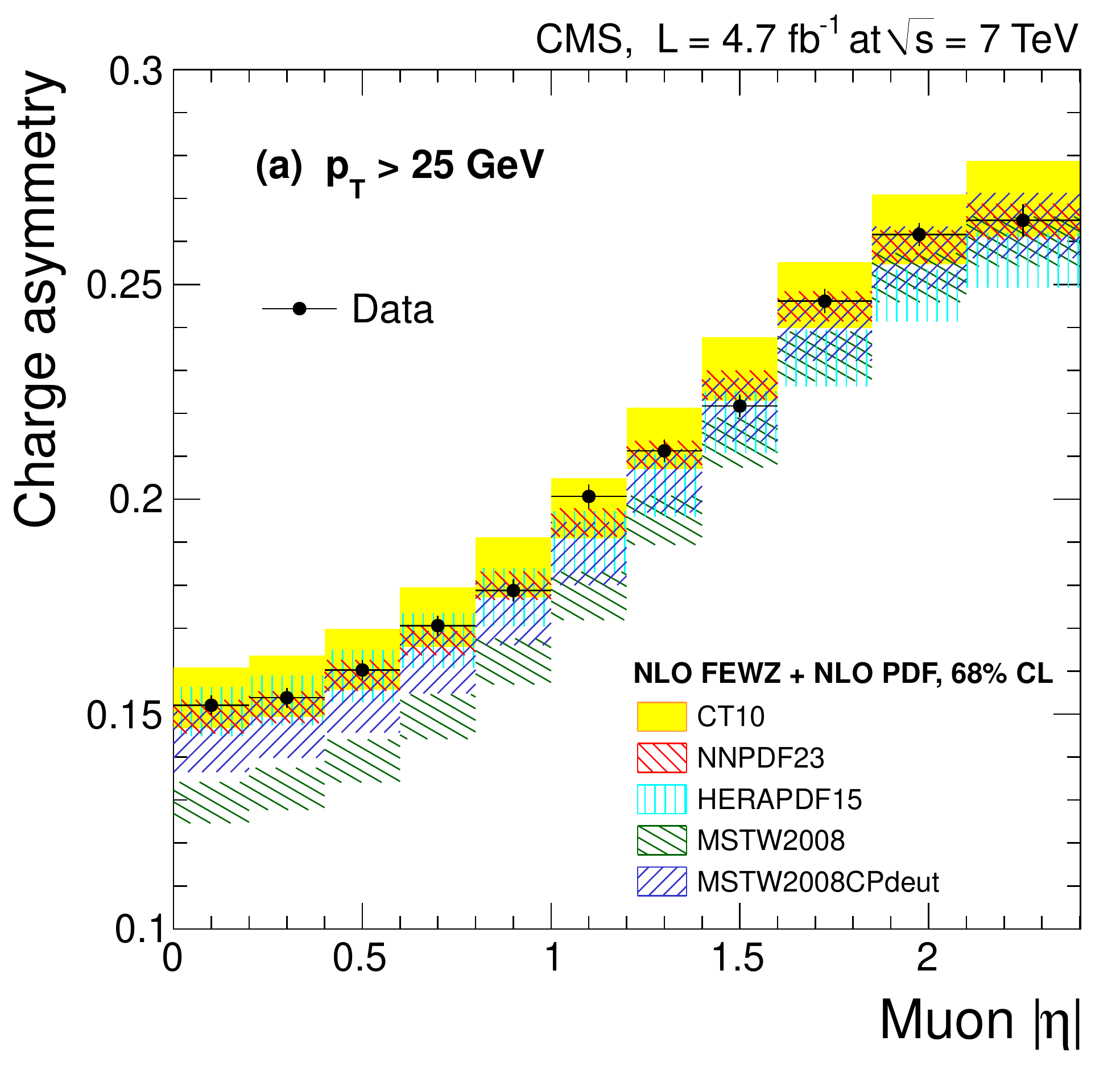}
    \includegraphics[width=\cmsFigWidth]{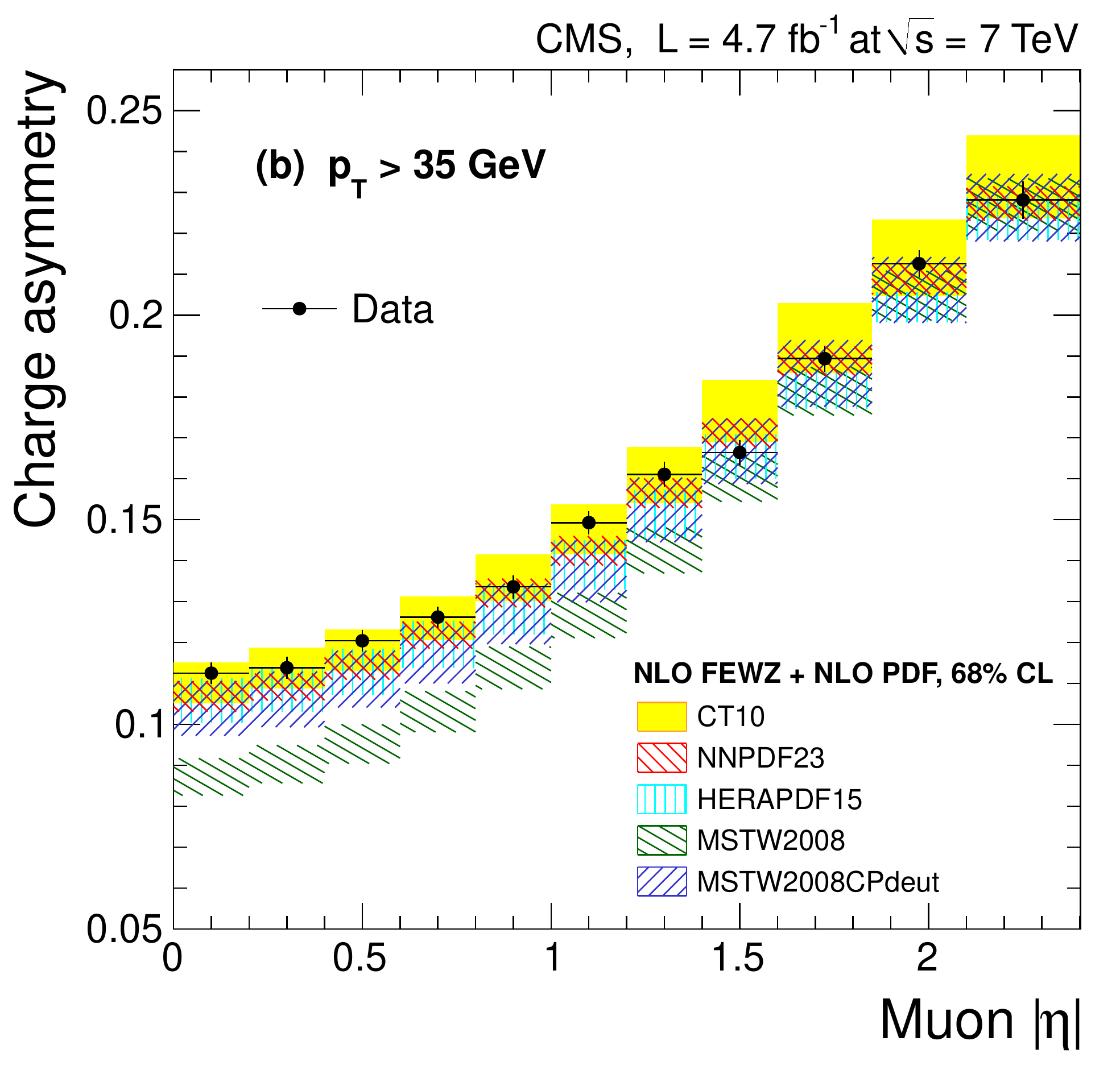}
    \caption{Comparison of the measured muon charge
asymmetries to the NLO predictions calculated using the \textsc{fewz}~3.1~\cite{FEWZ} MC tool interfaced
with the NLO CT10~\cite{CTEQ:1007},  NNPDF2.3~\cite{NNPDF2_3}, HERAPDF1.5~\cite{HERAPDF1_5},
MSTW2008~\cite{Martin:2009ad}, and MSTW2008CPdeut~\cite{MSTW:12}
 PDF sets.  No EW corrections have been considered in these
predictions. Results for
muon $\pt > 25$ and $>$35\GeV are shown in panels~(a)
and (b), respectively.
The vertical error bars on data points
include both statistical and systematic uncertainties.
The data points are shown at  the center of each  $\abs{\eta}$ bin.
The theoretical predictions are calculated  using the    \textsc{fewz}~3.1~\cite{FEWZ} MC tool. The PDF uncertainty for each PDF set is shown by
the shaded~(or hatched) band and corresponds  to 68\% \CLend
\label{fig:results:final}}
\end{figure}

The measured asymmetries~$\mathcal{A}$, after all the  corrections,
are shown in
Fig.~\ref{fig:results:final} as a function of muon $\abs{\eta}$ and
summarized in  Table~\ref{table:results:final}.
 In Fig.~\ref{fig:results:final} both statistical
and systematic uncertainties are included in the error bars.
These asymmetries are compared to predictions based on
several PDF sets. The theoretical predictions
are  obtained using the \textsc{fewz}~3.1~\cite{FEWZ} NLO MC calculation
interfaced with the
CT10~\cite{CTEQ:1007}, NNPDF2.3~\cite{NNPDF2_3}, HERAPDF1.5~\cite{HERAPDF1_5},
   MSTW2008~\cite{Martin:2009ad}, and  MSTW2008CPdeut~\cite{MSTW:12}
PDF sets. No EW corrections are included in these
calculations.
The numerical values of the theoretical predictions are
shown in Table~\ref{table:results:final}. We cross-check the theoretical
predictions using the
\textsc{dynnlo}~1.0~\cite{DYNNLO:a, DYNNLO:b} MC tool and the agreement between the
\textsc{fewz}~3.1  and \textsc{dynnlo}~1.0 is within 1\%.
The predictions using the  CT10 and HERAPDF1.5
PDF sets are in good agreement with the data.
The predictions using the NNPDF2.3 PDF set (which include the previous
CMS electron charge asymmetry result and
other LHC experimental measurements~\cite{NNPDF2_3}) are also  in
good agreement with the data.
The predictions using the MSTW2008 PDF set are not in
agreement with the data, as seen in our previous analyses~\cite{CMS:asym:2010, CMS-PAS-SMP-12-001}.
The more recent MSTW2008CPdeut PDF set is a variant of
the MSTW2008 PDF set with a more flexible input parametrization and
deuteron corrections~\cite{MSTW:12}. This modification has significantly improved the
agreement with the CMS data even though they have not included LHC data,  as shown in Fig.~\ref{fig:results:final}.
\begin{table*}[htm]
\centering
\topcaption{Summary of the  final results for muon charge asymmetry~$\mathcal{A}$.
The first uncertainty is statistical and the second is systematic. The theoretical predictions are obtained using the \textsc{fewz}~3.1~\cite{FEWZ} MC tool interfaced with the NLO CT10~\cite{CTEQ:1007}, NNPDF2.3~\cite{NNPDF2_3},  HERAPDF1.5~\cite{HERAPDF1_5}, and MSTW2008CPdeut~\cite{MSTW:12} PDF sets. The PDF uncertainty is at 68\% \CLend
For each $\abs{\eta}$ bin, the
 theoretical prediction is  calculated using the
averaged differential cross sections for positively
and negatively charged leptons. The numerical precision of the
theoretical predictions is less than 10\% of the statistical uncertainties
of the measurements. The values  are
 expressed as percentages.
}
\label{table:results:final}
\resizebox{\textwidth}{!}{
\renewcommand\arraystretch{1.1}
\begin{scotch}{  l | c  c  c  c c }
$\abs{\eta}$ & $\mathcal{A}$ (~$\pm\stat\pm\syst$~)      &  CT10           & NNPDF2.3           & HERAPDF1.5   &  MSTW2008-CP-deut      \\
\hline

\multicolumn{6}{c}{$\pt > 25\GeV$}\\\hline

0.00--0.20 &  $ 15.21\pm0.10\pm0.20 $  &  $ 15.35^{+0.74}_{-0.68} $  &  $ 14.94\pm0.39 $  &  $ 15.33^{+0.30}_{-0.84} $ & $ 14.34^{+0.75}_{-0.69} $    \\
0.20--0.40 &  $ 15.38\pm0.10\pm0.21 $  &  $ 15.63^{+0.73}_{-0.69} $  &  $ 15.16\pm0.37 $  &  $ 15.58^{+0.32}_{-0.85} $ & $ 14.67^{+0.75}_{-0.69} $    \\
0.40--0.60 &  $ 16.03\pm0.09\pm0.22 $  &  $ 16.27^{+0.71}_{-0.70} $  &  $ 15.90\pm0.36 $  &  $ 16.16^{+0.34}_{-0.88} $ & $ 15.27^{+0.75}_{-0.70} $    \\
0.60--0.80 &  $ 17.06\pm0.09\pm0.22 $  &  $ 17.27^{+0.68}_{-0.71} $  &  $ 16.71\pm0.34 $  &  $ 16.98^{+0.37}_{-0.91} $ & $ 16.19^{+0.74}_{-0.71} $    \\
0.80--1.00 &  $ 17.88\pm0.10\pm0.24 $  &  $ 18.45^{+0.66}_{-0.74} $  &  $ 17.99\pm0.33 $  &  $ 17.98^{+0.42}_{-0.94} $ & $ 17.33^{+0.74}_{-0.73} $    \\
1.00--1.20 &  $ 20.07\pm0.10\pm0.26 $  &  $ 19.85^{+0.64}_{-0.76} $  &  $ 19.46\pm0.33 $  &  $ 19.25^{+0.48}_{-0.95} $ & $ 18.74^{+0.73}_{-0.74} $    \\
1.20--1.40 &  $ 21.13\pm0.10\pm0.25 $  &  $ 21.50^{+0.63}_{-0.80} $  &  $ 21.03\pm0.33 $  &  $ 20.51^{+0.54}_{-0.92} $ & $ 20.45^{+0.72}_{-0.76} $    \\
1.40--1.60 &  $ 22.17\pm0.10\pm0.25 $  &  $ 23.13^{+0.64}_{-0.84} $  &  $ 22.66\pm0.34 $  &  $ 21.92^{+0.59}_{-0.84} $ & $ 22.12^{+0.70}_{-0.78} $    \\
1.60--1.85 &  $ 24.61\pm0.09\pm0.27 $  &  $ 24.87^{+0.65}_{-0.89} $  &  $ 24.49\pm0.35 $  &  $ 23.32^{+0.63}_{-0.70} $ & $ 24.01^{+0.68}_{-0.79} $    \\
1.85--2.10 &  $ 26.16\pm0.09\pm0.26 $  &  $ 26.42^{+0.67}_{-0.95} $  &  $ 25.88\pm0.38 $  &  $ 24.70^{+0.65}_{-0.57} $ & $ 25.70^{+0.65}_{-0.81} $    \\
2.10--2.40 &  $ 26.49\pm0.11\pm0.36 $  &  $ 27.13^{+0.74}_{-1.03} $  &  $ 26.46\pm0.42 $  &  $ 25.40^{+0.81}_{-0.48} $ & $ 26.48^{+0.65}_{-0.87} $    \\
\hline
\multicolumn{6}{c}{$\pt > 35\GeV$}\\\hline
0.00--0.20 &  $ 11.25\pm0.12\pm0.23 $  &  $ 11.00^{+0.52}_{-0.48} $  &  $ 10.68\pm0.37 $  &  $ 10.80^{+0.32}_{-0.76} $ & $ 10.39^{+0.67}_{-0.67} $    \\
0.20--0.40 &  $ 11.38\pm0.12\pm0.24 $  &  $ 11.36^{+0.52}_{-0.49} $  &  $ 10.91\pm0.33 $  &  $ 11.07^{+0.33}_{-0.77} $ & $ 10.61^{+0.68}_{-0.68} $    \\
0.40--0.60 &  $ 12.04\pm0.11\pm0.23 $  &  $ 11.80^{+0.52}_{-0.50} $  &  $ 11.40\pm0.31 $  &  $ 11.51^{+0.34}_{-0.79} $ & $ 11.10^{+0.70}_{-0.69} $    \\
0.60--0.80 &  $ 12.62\pm0.11\pm0.23 $  &  $ 12.59^{+0.53}_{-0.53} $  &  $ 12.18\pm0.33 $  &  $ 12.17^{+0.36}_{-0.80} $ & $ 11.71^{+0.72}_{-0.71} $    \\
0.80--1.00 &  $ 13.36\pm0.12\pm0.26 $  &  $ 13.60^{+0.55}_{-0.58} $  &  $ 13.21\pm0.35 $  &  $ 13.02^{+0.37}_{-0.82} $ & $ 12.70^{+0.74}_{-0.74} $    \\
1.00--1.20 &  $ 14.93\pm0.12\pm0.26 $  &  $ 14.79^{+0.59}_{-0.64} $  &  $ 14.24\pm0.36 $  &  $ 14.10^{+0.40}_{-0.81} $ & $ 13.75^{+0.77}_{-0.77} $    \\
1.20--1.40 &  $ 16.11\pm0.12\pm0.28 $  &  $ 16.14^{+0.64}_{-0.73} $  &  $ 15.65\pm0.36 $  &  $ 15.31^{+0.41}_{-0.77} $ & $ 15.24^{+0.79}_{-0.79} $    \\
1.40--1.60 &  $ 16.64\pm0.12\pm0.28 $  &  $ 17.72^{+0.70}_{-0.83} $  &  $ 17.11\pm0.36 $  &  $ 16.68^{+0.40}_{-0.68} $ & $ 16.69^{+0.79}_{-0.82} $    \\
1.60--1.85 &  $ 18.94\pm0.12\pm0.30 $  &  $ 19.53^{+0.77}_{-0.94} $  &  $ 18.87\pm0.36 $  &  $ 18.22^{+0.40}_{-0.51} $ & $ 18.62^{+0.77}_{-0.86} $    \\
1.85--2.10 &  $ 21.26\pm0.12\pm0.31 $  &  $ 21.52^{+0.82}_{-1.06} $  &  $ 20.89\pm0.38 $  &  $ 20.15^{+0.41}_{-0.32} $ & $ 20.71^{+0.71}_{-0.90} $    \\
2.10--2.40 &  $ 22.81\pm0.14\pm0.44 $  &  $ 23.53^{+0.86}_{-1.17} $  &  $ 22.73\pm0.42 $  &  $ 22.17^{+0.71}_{-0.33} $ & $ 22.79^{+0.66}_{-0.99} $    \\
\end{scotch}
}
\end{table*}

Since the  per-bin total experimental uncertainties are
 significantly smaller than the uncertainty in the
current  PDF parametrizations,
 this measurement can be used to constrain PDFs
 in the next generation of  PDF sets.

\begin{figure}[hbtp]
  \begin{center}
    \includegraphics[width=\cmsFigWidth]{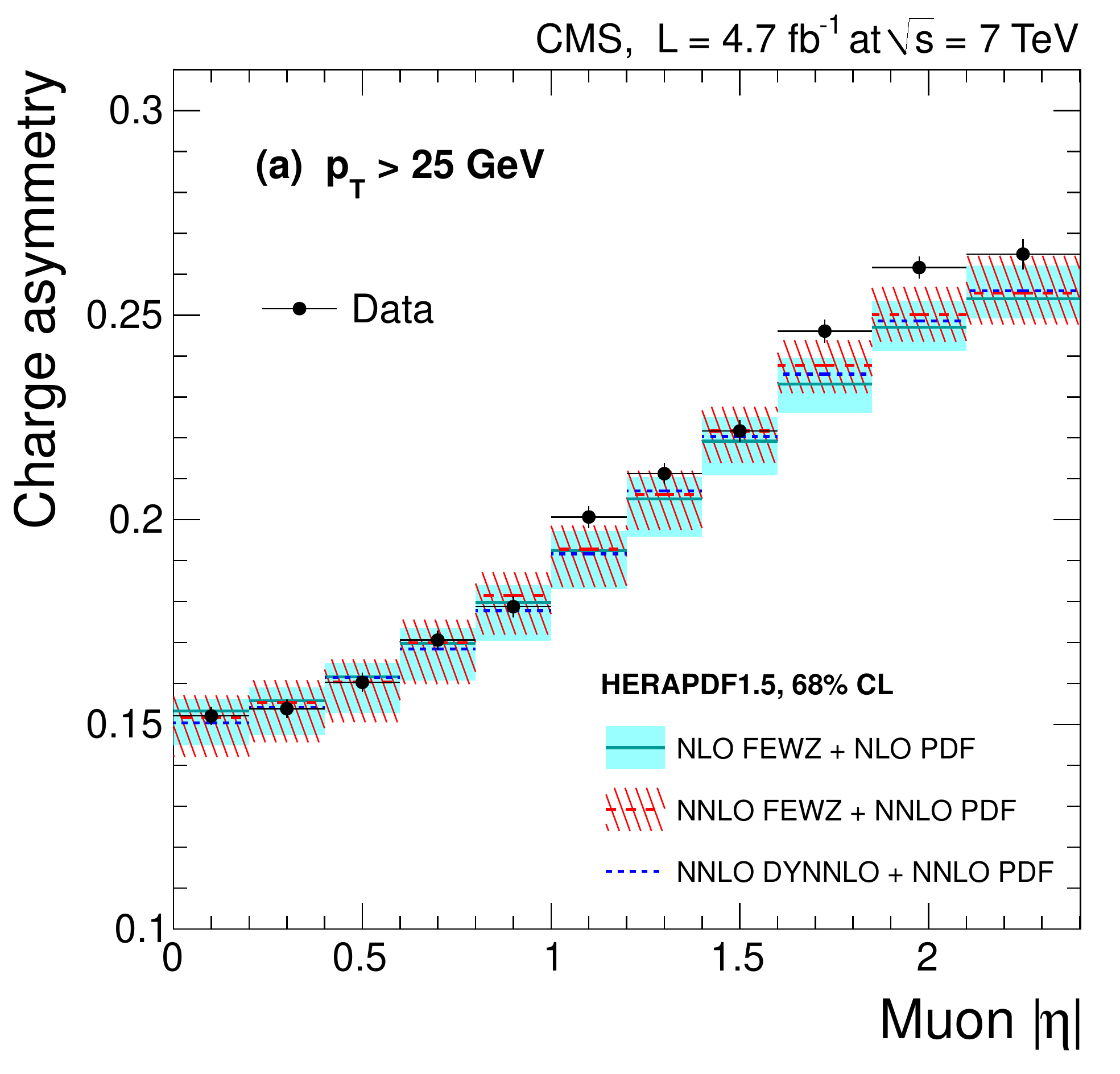}
    \includegraphics[width=\cmsFigWidth]{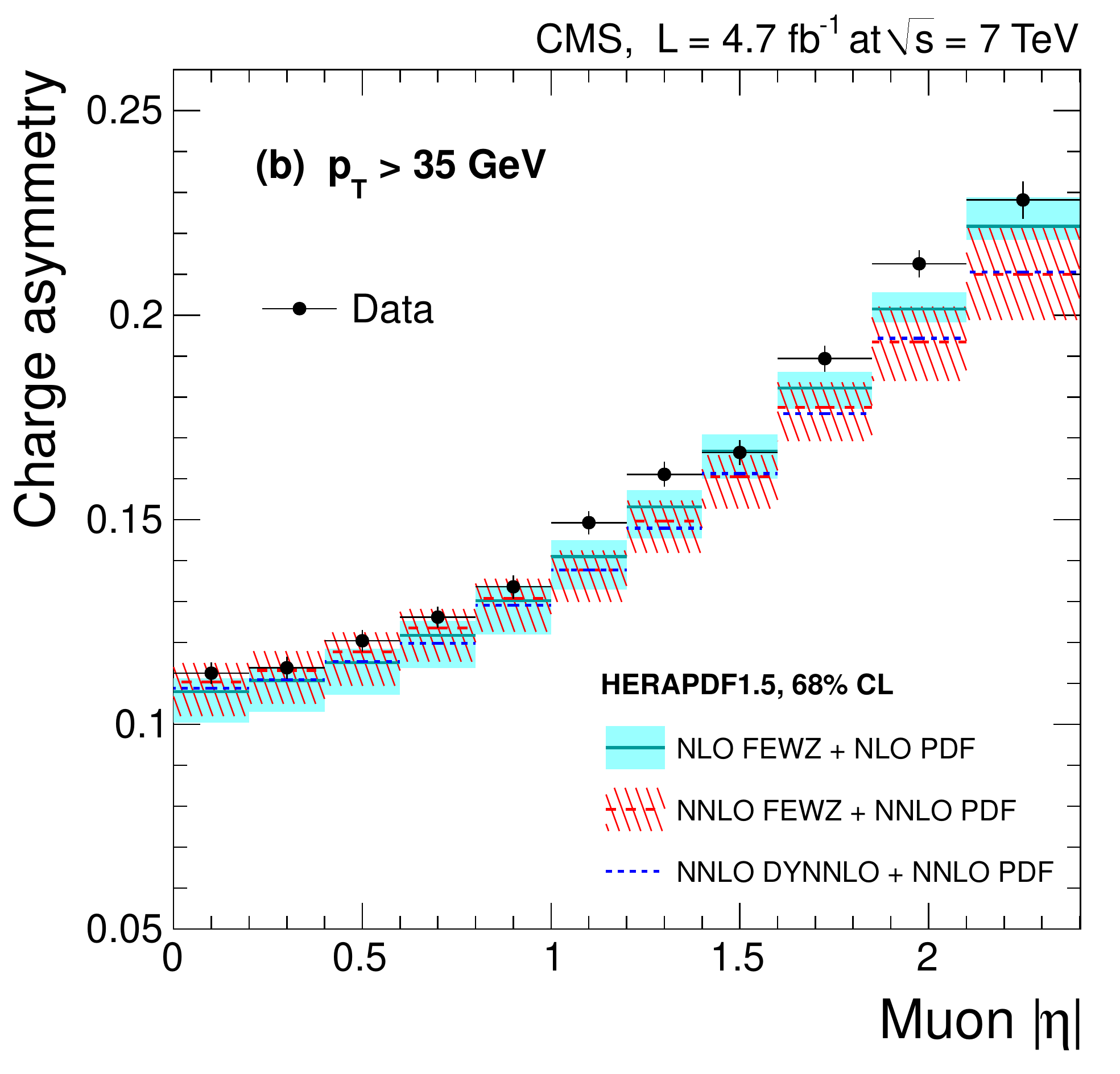}
\caption{Comparison of the measured muon charge
asymmetries to the NNLO predictions  for
muon $\pt > 25$~(a) and  muon $\pt > 35\GeV$~(b). The
 NNLO HERAPDF1.5~\cite{HERAPDF1_5} PDF has been used in the
 NNLO calculations.  The calculations are
performed using both the \textsc{fewz}~3.1~\cite{FEWZ} and \textsc{dynnlo}~1.0~\cite{DYNNLO:a, DYNNLO:b} MC tools. The NLO prediction based on
\textsc{fewz}~3.1 is also shown
here. The HERAPDF1.5 PDF uncertainties are shown
by the shaded~(NLO) and hatched~(NNLO) bands. }
    \label{fig:results:final:nnlo}
  \end{center}
\end{figure}

Figure~\ref{fig:results:final:nnlo} shows a comparison of  the measured muon charge
asymmetries to the NNLO predictions.  The
 NNLO HERAPDF1.5 PDF is used.
The calculations are
performed using the \textsc{fewz}~3.1 and \textsc{dynnlo}~1.0 MC tools.
 Both MC simulations give consistent results with agreement at the
1\% level.
With a \pt threshold of 25\GeV, the NLO and NNLO predictions are very similar.
The NNLO predictions are slightly higher in high-$\abs{\eta}$ regions. In the same high-$\abs{\eta}$ region at a \pt threshold of 35\GeV,
the NNLO
predictions are significantly
lower than the NLO prediction. However, they agree well within
 the quoted
PDF uncertainty in  the HERAPDF1.5 PDFs.

\begin{figure}[hbtp]
\centering
\includegraphics[width=\cmsFigWidth]{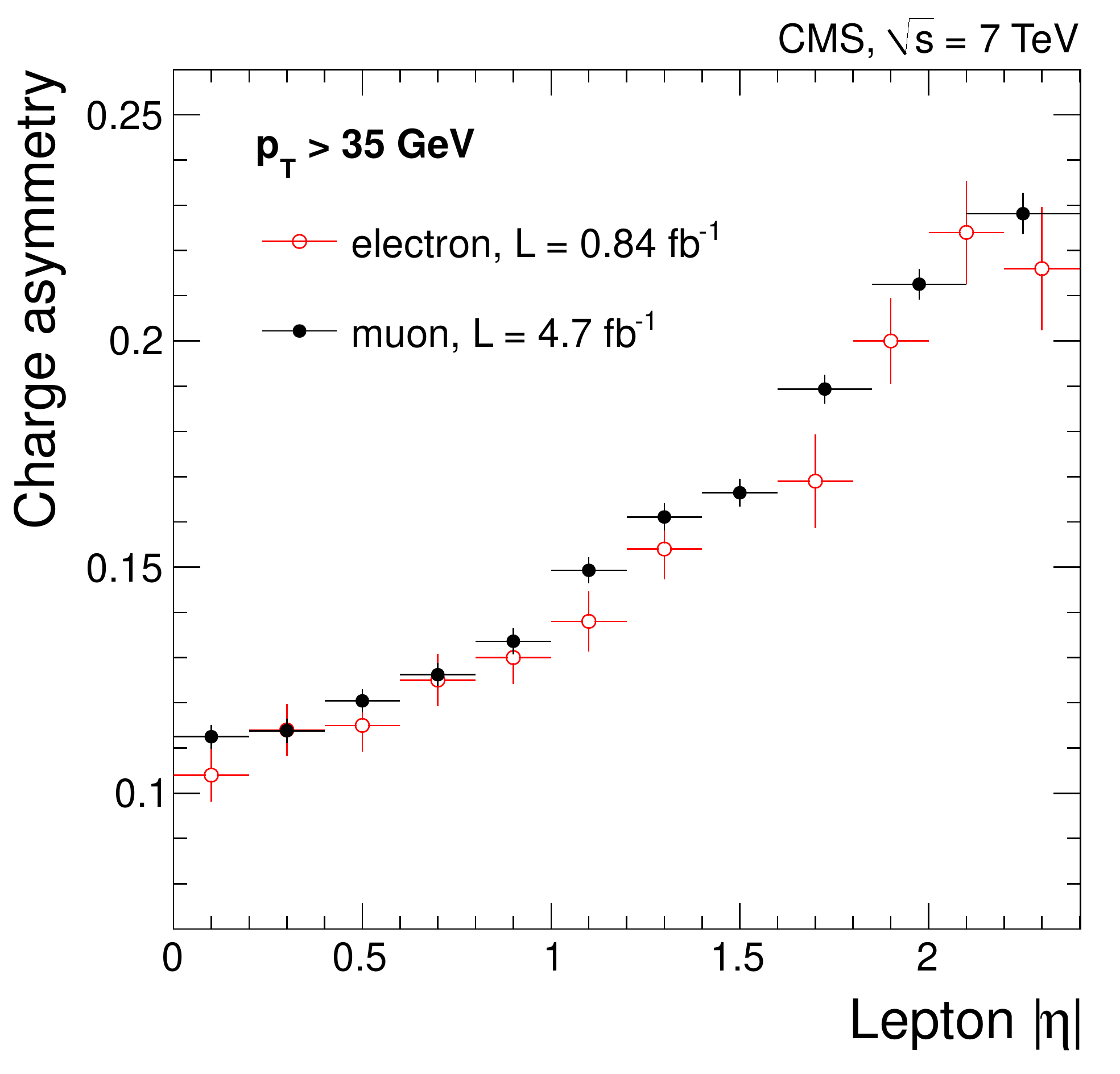}
\caption{Comparison of this measurement to the previous CMS electron charge
asymmetry result~\cite{CMS-PAS-SMP-12-001}. Results are shown for lepton $\pt > 35\GeV$.
\label{fig:results:oldvsnew}
}
\end{figure}

Figure~\ref{fig:results:oldvsnew} shows a comparison of this result to
the previous CMS electron charge
asymmetry measurement extracted from
part of the  2011 CMS data~\cite{CMS-PAS-SMP-12-001}.
The electron charge asymmetry has been measured with a slightly different $\eta$ binning because of the different subdetector geometry in
the calorimeter and  the muon system. We have calculated the bin-by-bin differences between these two measurements
using the first seven $\eta$ bins, where identical bin definitions are used, and the differences are fitted with a constant.
The fitted constant is larger than  zero by about 1.7 sigma,  and the muon
channel exhibits slightly higher asymmetry in these
seven $\eta$ bins than the electron one.
 The electron charge asymmetry  uses a
statistically independent data sample. A combination
of both results can be used to improve the global PDF fits.
The
correlation between the electron charge asymmetry and this result is expected to be small. The
completely correlated systematic sources of uncertainty include the luminosity measurement, $\ttbar$ background,
$\wtaunu$ background, and PDF uncertainty.

\begin{figure}[hbtp]
  \begin{center}
    \includegraphics[width=\cmsFigWidth]{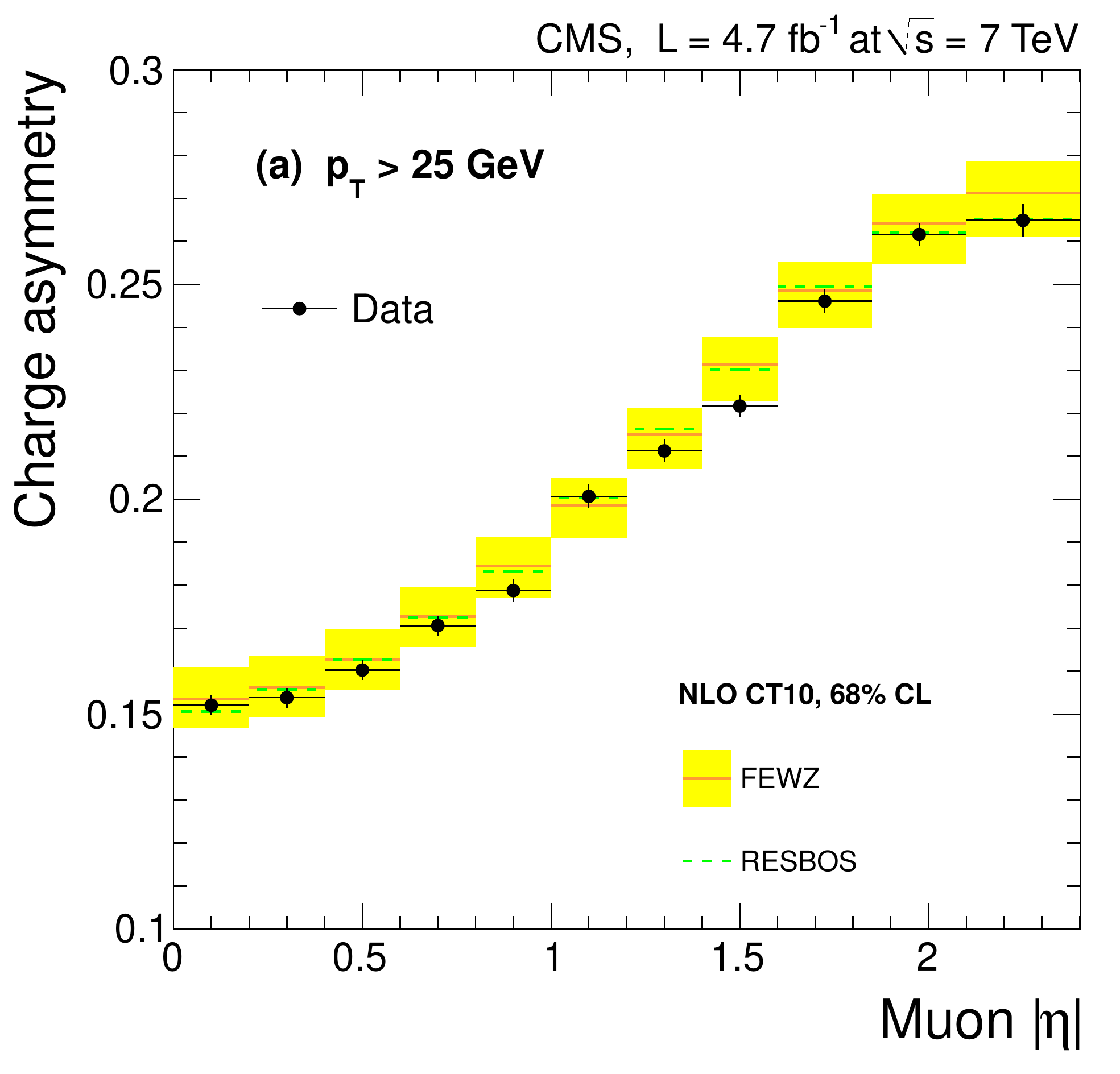}
    \includegraphics[width=\cmsFigWidth]{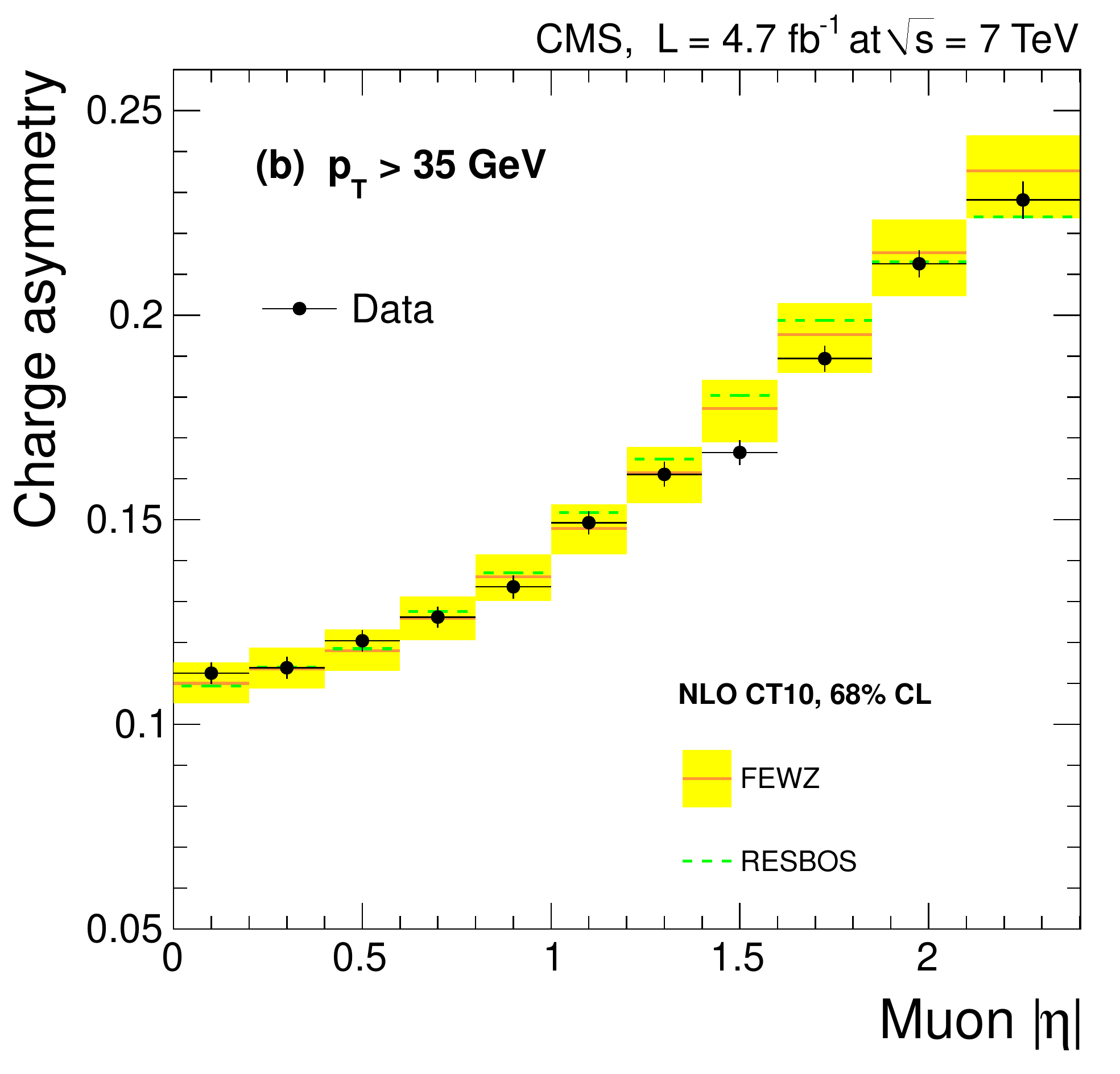}
    \caption{Comparison of the measured  muon charge asymmetry  to theoretical predictions based on the \textsc{fewz}~3.1~\cite{FEWZ} and \textsc{resbos}~\cite{RES1, RES2, RES3, Guzzi:2013aja} tools.  The NLO CT10 PDF is used in both predictions. Results are shown for
 muon $\pt > 25$~(a) and muon $\pt > 35\GeV$~(b). The CT10 PDF
uncertainty is shown by the shaded bands.
}
    \label{fig:results:fixedvsres}
  \end{center}
\end{figure}

The theoretical predictions for  the lepton charge asymmetry are given for the
 kinematic region specified by
 the lepton \pt threshold.
The \pt distribution of the $\PW$ boson affects the acceptance, and hence,
the predicted charge asymmetry.
However,
the effect on $\PWp$ and $\PWm$ is largely
correlated. Therefore, the impact on the  lepton charge asymmetry
measurement mostly cancels.
Figure~\ref{fig:results:fixedvsres} shows the comparison of these
 results to the NLO CT10 PDF
predictions based on the \textsc{fewz}~3.1  and
\textsc{resbos}~\cite{RES1, RES2, RES3, Guzzi:2013aja}. \textsc{resbos} does a
resummation in boson $\qT$ at NLO (and approximate NNLO) plus
next-to-next-to-leading logarithm which yields a  more realistic
description of boson $\qT$ than a
fixed-order calculation such as the \textsc{fewz}~3.1.
The difference between the \textsc{fewz}~3.1 and \textsc{resbos} predictions
 is  negligible and our measurement, however precise, is not  sensitive to
the difference.

\section{The QCD analysis of HERA and CMS results of \texorpdfstring{\PW}{W}-boson production}
\label{sec:QCD}

The main objective of the QCD analysis presented in this section
is to exploit the constraining power and the interplay
of the muon charge asymmetry measurements, presented in this paper, and
the recent measurements of $\PW + \text{charm}$ production at CMS ~\cite{wpluscharm_cms} to determine the PDFs of the proton. These two data sets, together with the combined HERA
inclusive cross section measurements~\cite{Aaron:2009aa}, are used in an NLO perturbative QCD (pQCD) analysis.

Renormalization group equations, formulated in terms of DGLAP evolution equations
~\cite{Gribov:1972ri,Altarelli:1977zs,Curci:1980uw,Furmanski:1980cm,Moch:2004pa,Vogt:2004mw},
predict the dependence of the PDFs on the energy scale $Q$ of the process in pQCD. The dependence
on the partonic fraction $x$ of the proton momentum cannot be derived from first principles and must be constrained by experimental
measurements. Deep inelastic lepton-proton scattering (DIS) experiments cover a broad range of the $(x,Q^2)$ kinematic plane.
The region of small and intermediate $x$ is probed primarily by the precise data of HERA,
which impose the tightest constraints on the existing PDFs.
However, some details of flavor composition, in particular the light-sea-quark content and
the strange-quark distribution of the proton, are still poorly known. Measurements of
the $\PW$- and $\cPZ$-boson production cross sections in proton-(anti)proton collisions are sensitive
to the light-quark distributions, and the constraining power of the $\PW$-boson measurements is applied in this analysis.

The muon charge asymmetry measurements probe the valence-quark distribution in the kinematic
range $10^{-3} \leq x \leq 10^{-1}$ and have indirect sensitivity to the strange-quark distribution.
The measurements of the total and differential cross sections of $\PW+\text{charm}$ production have the potential to
 access the strange-quark distribution directly through the LO process $\cPg + \cPqs \to \PW + \cPqc$.
This reaction was proposed as a way to determine the strange-quark and antiquark distributions~\cite{PhysRevD.40.83,Baur,Lai:2007dq}.

Before the LHC era, constraints on the strange-quark distribution were obtained from semi-inclusive charged-current scattering
at the NuTeV~\cite{Goncharov:2001qe,Mason:2007zz} and CCFR~\cite{Bazarko:1994tt} experiments.
Dimuon production in neutrino-nucleus reactions is sensitive to strangeness at LO in QCD
in reactions such as $\PWp + \cPqs \to \cPqc$. These measurements probe the (anti)strange-quark
density at $x\approx 10^{-1}$ and $Q^2$ of approximately $10\GeV^2$, but their interpretation is
complicated by nuclear corrections and uncertainties in the charm-quark fragmentation function.
The NOMAD Collaboration reported a recent determination of the strange-quark suppression factor
\begin{equation}
\kappa_s (Q^2)= \frac{\int^1_0 x\left[\cPaqs (x, Q^2) + \cPqs(x,Q^2)\right]\rd{}x }{\int^1_0 x \left[\cPaqu(x,Q^2) + \cPaqd(x,Q^2)\right]\rd{}x},
\end{equation}
where the value $\kappa_s\, (Q^2 = 20\GeV{}^2) = 0.591 \pm 0.019$ is determined at NNLO
by using dimuon production~\cite{Samoylov:2013xoa}.
The measurements of semi-inclusive hadron production on a deuteron target at HERMES~\cite{Airapetian:2012ki}
have been recently reevaluated~\cite{jackson_dis13} to obtain the $x$ dependence of the strange-quark distribution at LO
at an average $\langle Q^2\rangle =2.5\GeV{}^2$.
In that analysis the strange-quark distribution is found to vanish above $x=0.1$, but this result depends
strongly on the assumptions of the kaon fragmentation function.

In a recent analysis by the ATLAS Collaboration~\cite{Aad:2012sb}, the inclusive cross section measurements
of $\PW$- and $\cPZ$-boson production were used in conjunction with DIS inclusive data from HERA. The result
supports the hypothesis of a symmetric composition of the light-quark sea in the kinematic region probed, \ie, $\cPaqs=\cPaqd$.

The LHC measurements of associated production of $\PW$ bosons and charm quarks probe
the strange-quark distribution in the kinematic region of $x\approx 0.012$ at the scale $Q^2=m^2_{\PW}$.
The cross sections for this process were recently measured by the CMS Collaboration~\cite{wpluscharm_cms}
at a center-of mass energy $\sqrt{s} = 7\TeV$ with a total integrated luminosity of 5\fbinv.
The results of the QCD analysis presented here use the absolute differential cross sections of $\PW+\text{charm}$ production,
measured in bins of the pseudorapidity of the lepton from the $\PW$ decay, for transverse momenta larger than 35\GeV.

\subsection{Details of the QCD analysis}

The NLO QCD analysis is based on the inclusive DIS data~\cite{Aaron:2009aa} from HERA, measurements of the muon charge
asymmetry in $\PW$ production for $\pt > 25\GeV$, and measurements of associated $\PW + \text{charm}$ production~\cite{wpluscharm_cms}.
The treatment of experimental uncertainties for the HERA data follows the prescription of HERAPDF1.0~\cite{Aaron:2009aa}.
The correlations of the experimental uncertainties for the muon charge asymmetry and $\PW + \text{charm}$ data are taken into account.

The theory predictions for the muon charge asymmetry and $\PW+\text{charm}$ production are calculated at NLO by using the
\textsc{mcfm} program~\cite{Campbell:1999ah,Campbell:2010ff}, which is interfaced to \textsc{applgrid}~\cite{Carli:2010rw}.

The open source QCD fit framework for PDF determination \textsc{HERAFitter}~\cite{Aaron:2009kv, Aaron:2009aa, herafitter} is
used and the partons are evolved by using the \textsc{qcdnum} program~\cite{Botje:2010ay}.
The TR' \cite{Thorne:2006qt,Martin:2009ad} general mass variable flavor number
scheme is used for the treatment of heavy-quark contributions with the
following conditions: (i) heavy-quark masses are chosen as $m_{\cPqc} = 1.4\GeV$ and $m_{\cPqb} = 4.75\GeV$,
(ii) renormalization and factorization scales are set to $\mu_r = \mu_f = Q$,
and (iii) the strong coupling constant is set to $\alpha_s (m_\cPZ) = 0.1176$.

The $Q^2$ range of HERA data is restricted to $Q^2 \geq Q^2_{\min} = 3.5\GeV{}^2$ to assure the
applicability of pQCD over the kinematic range of the fit. The procedure for the determination of
the PDFs follows the approach used in the HERAPDF1.0 QCD fit~\cite{Aaron:2009aa}.

The following independent combinations of parton distributions are chosen in the fit procedure at the initial scale of the QCD
evolution $Q_0^2 = 1.9\GeV{}^2$: $x\cPqu_{\mathrm{v}}(x)$, $x\cPqd_{\mathrm{v}}(x)$, $x\cPg(x)$ and $x\overline{\mathrm{U}}(x)$, $x\overline{\mathrm{D}}(x)$ where
$x\overline{\mathrm{U}}(x) = x\cPaqu(x)$, $x\overline{\mathrm{D}}(x) = x\cPaqd(x) + x\cPaqs(x)$. At $Q_0$, the parton distributions are represented by
\begin{align}
x \cPqu_\mathrm{v}(x) &= A_{\cPqu_{\mathrm{v}}} ~  x^{B_{\cPqu_{\mathrm{v}}}} ~ (1-x)^{C_{\cPqu_{\mathrm{v}}}} ~(1+E_{\cPqu_{\mathrm{v}}} x^2) ,
\label{eq:uv}\\
x \cPqd_\mathrm{v}(x) &= A_{\cPqd_{\mathrm{v}}} ~ x^{B_{\cPqd_{\mathrm{v}}}} ~ (1-x)^{C_{\cPqd_{\mathrm{v}}}},
\label{eq:dv}\\
x \overline {\mathrm{U}}(x) &= A_{\overline {\mathrm{U}}} ~ x^{B_{\overline {\mathrm{U}}}} ~ (1-x)^{C_{\overline {\mathrm{U}}}},
\label{eq:Ubar}\\
x \overline {\mathrm{D}}(x) &= A_{\overline{\mathrm{D}}} ~ x^{B_{\overline{\mathrm{D}}}} ~ (1-x)^{C_{\overline{\mathrm{D}}}},
\label{eq:Dbar}\\
x \cPg(x) &= A_{\cPg} ~ x^{B_{\cPg}} ~ (1-x)^{C_{\cPg}}
+ A'_{\cPg} ~ x^{B'_{\cPg}} ~ (1-x)^{C'_{\cPg}}.
\label{eq:g}
\end{align}
The normalization parameters $A_{\cPqu_{\mathrm{v}}}$ , $A_{\cPqd_\mathrm{v}}$ , $A_\cPg$ are determined by the QCD sum rules, the $B$ parameter is responsible for
small-$x$ behavior of the PDFs, and the parameter $C$ describes the shape of the distribution as $x \to 1$.
A flexible form for the gluon distribution is adopted here, where the choice of $C'_\cPg=25$ is motivated by the
approach of the MSTW group~\cite{Thorne:2006qt,Martin:2009ad}.

Two types of analyses are made. The first is denoted as ``fixed-$\cPqs$ fit'' and is performed by fitting 13 parameters
in Eqs.~(\ref{eq:uv}-\ref{eq:g}) to analyze the impact of the muon charge asymmetry measurements
on the valence-quark distributions. Additional constraints $B_{\overline{\mathrm{U}}} = B_{\overline{\mathrm{D}}}$ and $A_{\overline{\mathrm{U}}} = A_{\overline{\mathrm{D}}}(1 - f_\cPqs)$
are imposed with $f_\cPqs$ being the strangeness fraction, $f_\cPqs = \cPaqs/( \cPaqd + \cPaqs)$, which is fixed
to $f_\cPqs=0.31\pm0.08$ as in Ref.~\cite{Martin:2009ad}.

The second analysis is denoted as ``free-$\cPqs$ fit'', in which the interplay between the muon charge asymmetry
measurements and $\PW+\text{charm}$ production data is analyzed. The strange-quark distribution is determined by fitting
15 parameters
in Eqs.~(\ref{eq:uv}-\ref{eq:g}). Here, instead of Eq.~(\ref{eq:Dbar}) $\cPaqd$ and $\cPaqs$ are fitted separately by using
the functional forms
\begin{align}
 x \cPaqd(x) &=  A_{\cPaqd} ~ x^{B_{\cPaqd}} ~ (1-x)^{C_{\cPaqd}} \,,
\\
 x \cPaqs(x) &=  A_{\cPaqs} ~ x^{B_{\cPaqs}} ~ (1-x)^{C_{\cPaqs}}.
\end{align}
Additional constraints $A_{\cPaqu} = A_{\cPaqd}$ and $B_{\cPaqu} = B_{\cPaqd}$ are applied to ensure the same
normalization for $\cPaqu$ and $\cPaqd$ densities at $x \to 0$. The strange-antiquark parameter $B_{\cPaqs}$ is set equal
to $B_{\cPaqd}$, while $A_{\cPaqs}$ and $C_{\cPaqs}$ are treated as free parameters of the fit, assuming  $x\cPqs = x\cPaqs$.
This parametrization cannot be applied to HERA DIS data alone, because those data do not have sufficient sensitivity to
the strange-quark distribution.

\subsection{The PDF uncertainties}

The PDF uncertainties are estimated according to the general approach of
HERAPDF1.0~\cite{Aaron:2009aa} in which experimental, model, and
parametrization uncertainties are taken into account. A tolerance criterion of $\Delta\chi^2 =1$ is adopted for
defining the experimental uncertainties that originate from the measurements included in the analysis.
 Model uncertainties arise from the variations in the values assumed for the heavy-quark masses
$m_\cPqb$, $m_\cPqc$ with $4.3\leq m_\cPqb\leq 5$\GeV, $1.35\leq m_\cPqc\leq 1.65$\GeV, and the value of $Q^2_{\min}$ imposed on
the HERA data, which is varied in the interval $2.5 \leq Q^2_{\min}\leq 5.0\GeV{}^2$.
The parametrization uncertainty is estimated similarly to the HERAPDF1.0 procedure:
for all parton densities, additional parameters are added one by one in the functional form of
the parametrizations such that Eqs.~(\ref{eq:uv}-\ref{eq:Dbar}) are generalized
to $A ~ x^B ~ (1-x)^C (1+Dx)$ or $A~ x^B ~ (1-x)^C ~ (1+Dx+Ex^2)$.
In the free-$\cPqs$ fit, in addition, the parameters $B_{\cPaqs}$ and $B_{\cPaqd}$ are decoupled.
Furthermore, the starting scale is varied within $1.5\leq Q^2_0 \leq 2.5\GeV{}^2$.
The parametrization uncertainty is constructed as an envelope built from the maximal
differences between the PDFs resulting from all the parametrization variations and the
central fit at each $x$ value.
The total PDF uncertainty is obtained by adding experimental,
model, and parametrization uncertainties in quadrature.
In the following, the quoted uncertainties correspond to 68\% \CLend

\subsection{Results of the QCD analysis}

\begin{figure}[htbp]
\center
   \includegraphics[width=\cmsFigWidth]{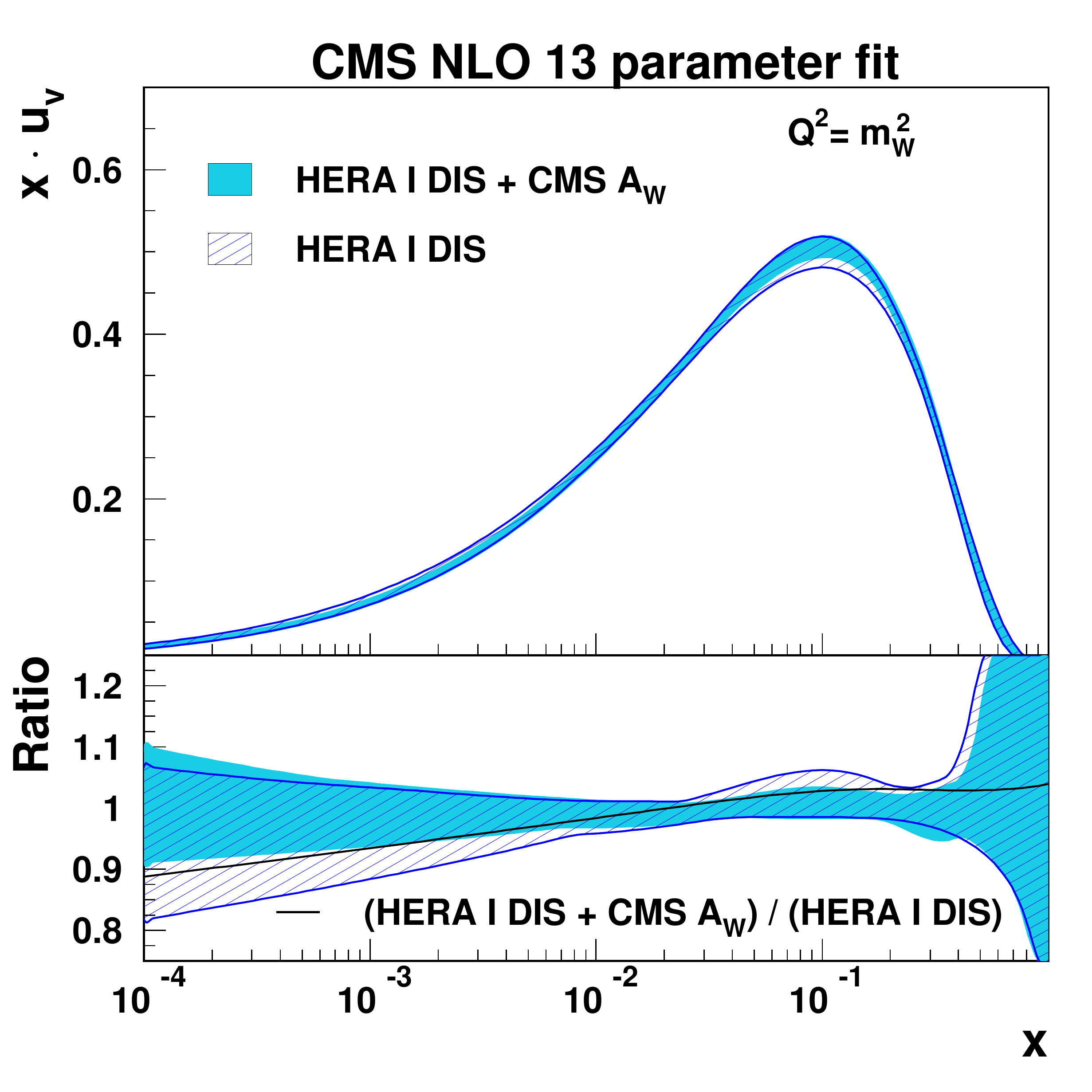}
   \includegraphics[width=\cmsFigWidth]{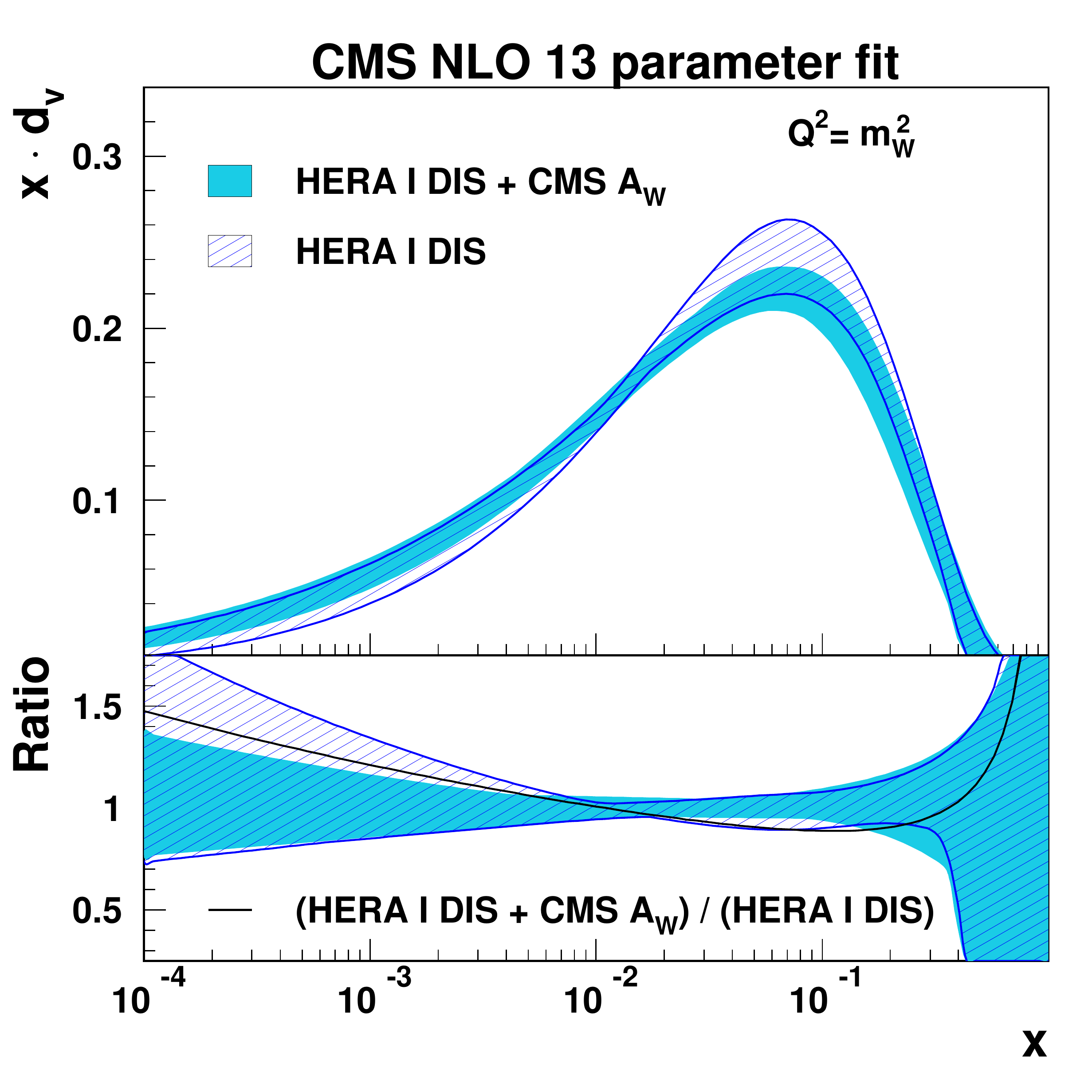}
\setlength{\unitlength}{1cm}
\caption{Distributions of $\cPqu$ valence (\cmsLeft) and $\cPqd$ valence (\cmsRight) quarks as functions of $x$ at
the scale $Q^2=m^2_\PW$. The results of the 13-parameter fixed-$\cPqs$ fit to the HERA data and muon asymmetry
measurements (light shaded band), and to HERA only (dark hatched band) are compared. The total PDF uncertainties
are shown. In the bottom panels the distributions are normalized to one for a direct comparison of the
 uncertainties. The change of the PDFs with respect to the HERA-only fit is represented by a solid line.}
\label{herapluswasym}
\end{figure}

The muon charge asymmetry measurements, together with HERA DIS cross section data,
improve the precision of the valence quarks over the entire $x$ range in the fixed-$\cPqs$ fit. This is illustrated in
Fig.~\ref{herapluswasym}, where the $\cPqu$ and $\cPqd$ valence-quark distributions are shown
at the scale relevant for the W-boson production, $Q^2= m^2_\PW$. The results at
$Q^2 = 1.9\GeV{}^2$ can be found in supplemental material.
A change in the shapes of the light-quark distributions within the total uncertainties is observed.
The details of the effect on the experimental PDF uncertainty of $\cPqu$ valence, $\cPqd$ valence, and
$\cPqd/\cPqu$ distributions are also given in supplemental material.

In the next step of the analysis, the CMS $\PW+\text{charm}$ measurements are used together with the HERA DIS data and
the CMS muon charge asymmetry. Since both CMS $\PW$-boson production measurements are sensitive to the
strange-quark distribution, a free-$\cPqs$ fit can be performed.
The advantage of including these two CMS data sets in the 15-parameter fit occurs because the $\cPqd$-quark distribution is
significantly constrained by the muon charge asymmetry data, while the strange-quark distribution is directly probed by the
associated $\PW+\text{charm}$ production measurements. In the free-$\cPqs$ fit, the strange-quark distribution
$\cPqs(x,Q^2)$,
and the strange-quark fraction $R_{\cPqs}(x,Q^2)=(\cPqs+\cPaqs)/(\cPaqu+\cPaqd)$ are determined.
The global and partial $\chi^2$ values for each data set are listed in Table~\ref{chi2_paper_table},
where the $\chi^2$ values illustrate a general agreement among all the data sets.

\begin{figure}[htbp]
\centering
   \includegraphics[width=\cmsFigWidth]{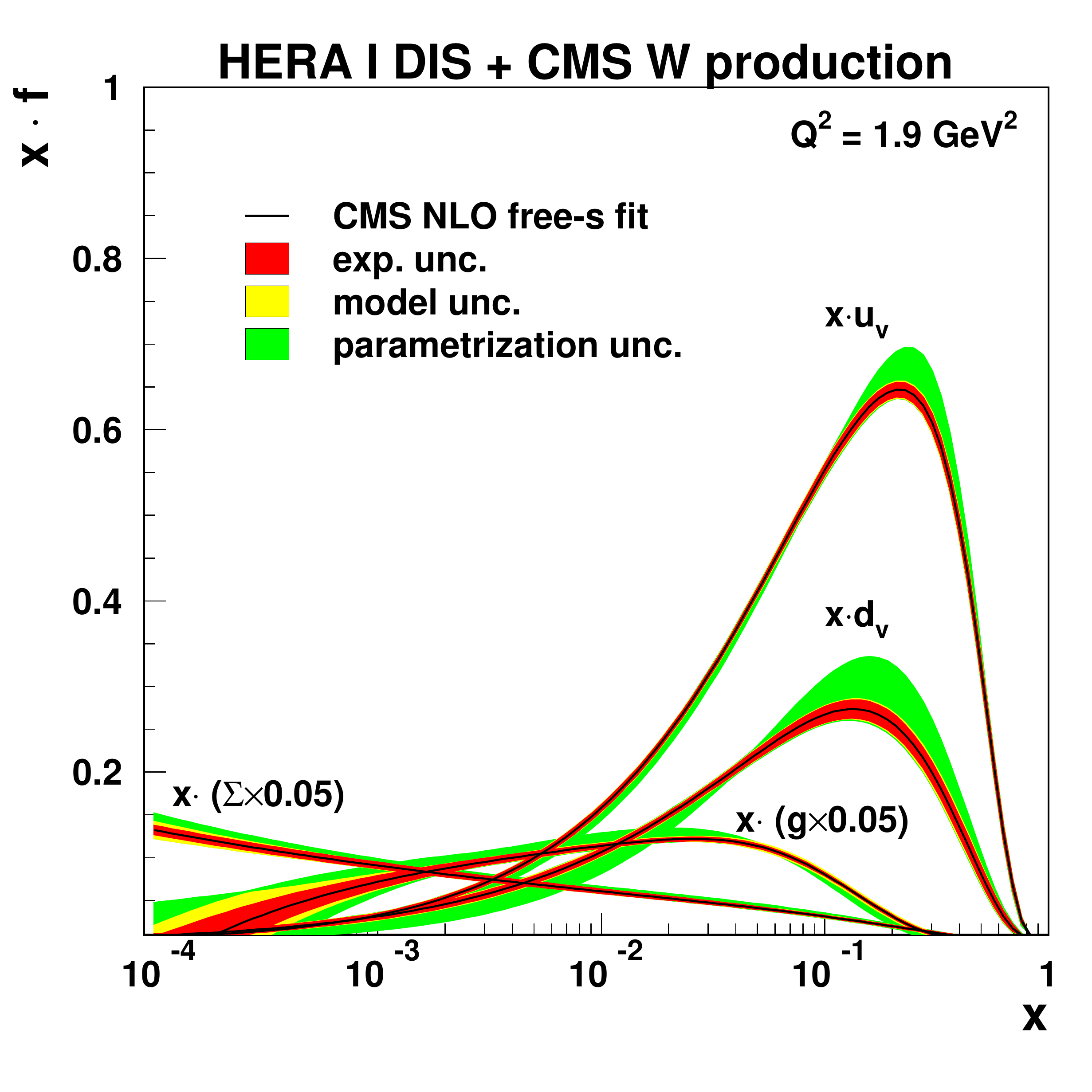}
   \includegraphics[width=\cmsFigWidth]{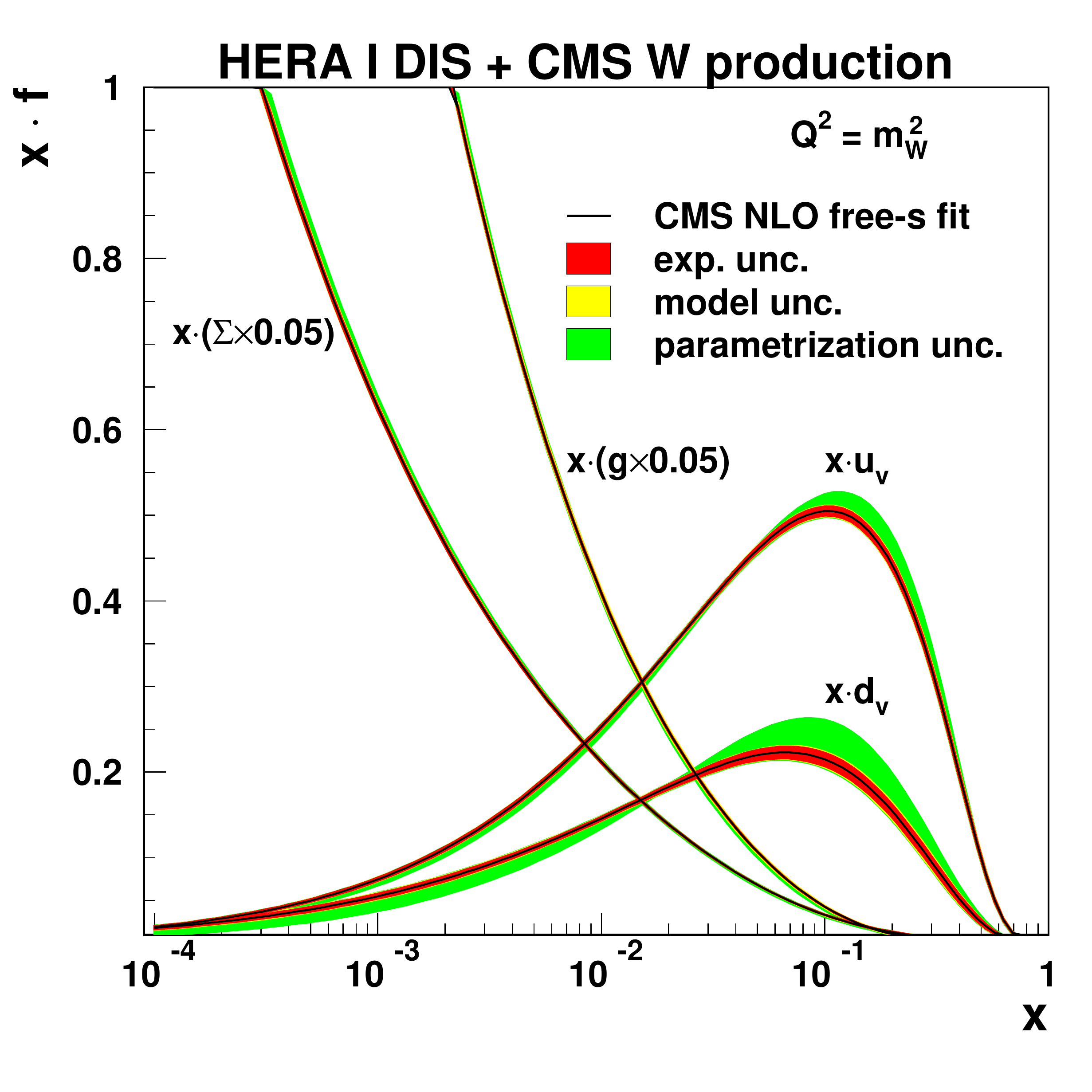}
\caption{Parton distribution functions, shown as functions of $x$, obtained by using HERA DIS data and CMS measurements of $\PW$-boson
production in the free-$\cPqs$ NLO QCD analysis. Gluon, valence, and sea
distributions are presented at the starting scale $Q^2_0=1.9\GeV^2$ of the PDF evolution (\cmsLeft) and the
mass squared of the $\PW$ boson (\cmsRight).
The sea distribution is defined as
$\Sigma = 2 \cdot (\cPaqu + \cPaqd  + \cPaqs)$. The full band represents the total uncertainty.
The individual contributions from the experimental, model, and parametrization uncertainties are represented
by the bands of different shades. The gluon and sea distributions are scaled down by a factor of 20.}
\label{pdf_fig1}
\end{figure}

\begin{figure}[htbp]
\centering
   \includegraphics[width=\cmsFigWidth]{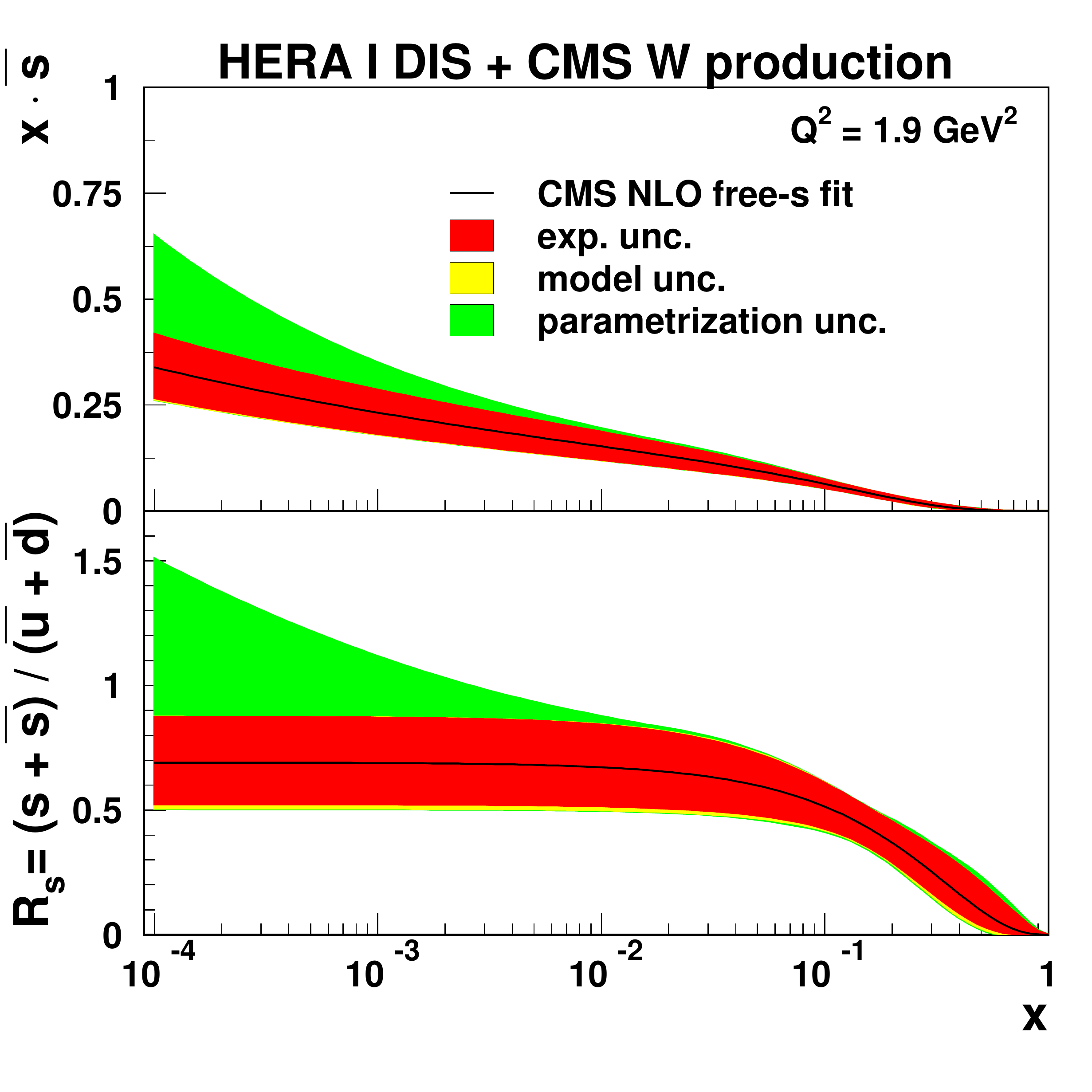}
\includegraphics[width=\cmsFigWidth]{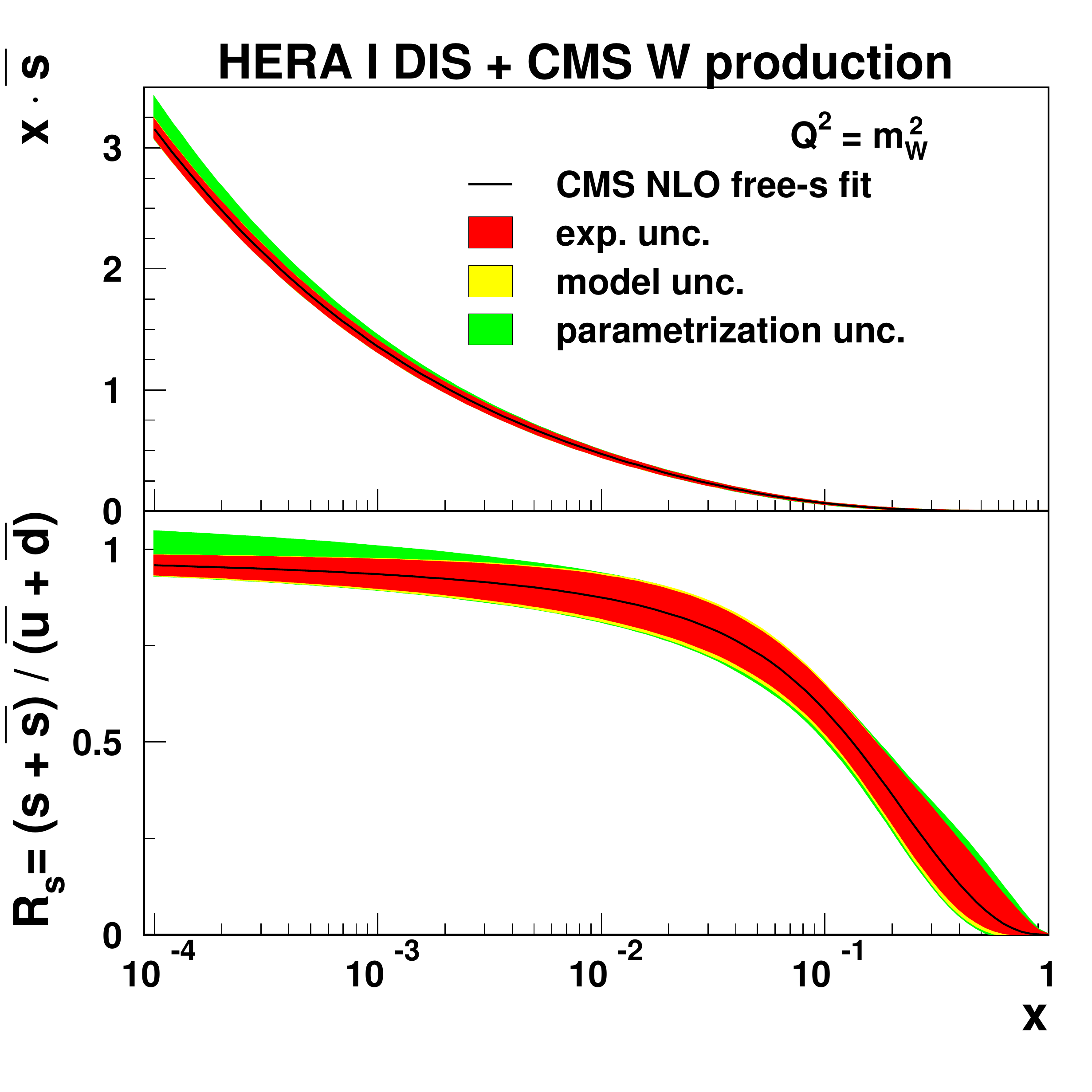}
\caption{Antistrange-quark distribution $\cPaqs(x,Q)$ and the ratio $R_{\cPqs}(x,Q)$, obtained in the QCD analysis of
HERA and CMS data, shown as functions of $x$ at the scale $Q^2=1.9\GeV{}^2$ (\cmsLeft) and $Q^2=m^2_\PW$ (\cmsRight).
The full band represents the total uncertainty. The individual contributions from the experimental, model, and
parametrization uncertainties are represented by the bands of different shades.
}
\label{Rs_fs_fig2}
\end{figure}

\begin{table*}[!htbp]
\centering
\renewcommand{\arraystretch}{1.25}
\caption{Global $\chi^2/n_{\text{dof}}$ and partial $\chi^2$ per number
of data points $n_{\textrm{dp}}$ for the data sets used in the 15-parameter QCD analysis.}
\begin{scotch}{l |c c}
 Data sets & Global $\chi^2/n_{\text{dof}}$ & Partial $\chi^2/n_{\mathrm{dp}}$ \\
\hline
DIS, $\frac{\rd\sigma_{\PW+\cPqc}}{\rd\eta^{l}}$, $\mathcal{A}(\eta_\Pgm)$ &  $598/593$ & \\
\hline
NC cross section HERA I H1+ZEUS $\Pem \Pp$ &  &$107/145$  \\
NC cross section HERA I H1+ZEUS $\Pep \Pp$&   &$417/379$ \\
CC cross section HERA I H1+ZEUS $\Pem \Pp$&   &$20/34$ \\
CC cross section HERA I H1+ZEUS $\Pep \Pp$&   &$36/34$ \\
CMS $\PW^\pm$ muon charge asymmetry   ${\cal A}(\eta_\Pgm)$&  & $14/11$ \\
CMS $\PW+\cPqc$ cross section  $\frac{\rd\sigma_{\PW+\cPqc}}{\rd\eta^{l}}$&   &$5/5$ \\
\end{scotch}
\label{chi2_paper_table}
\end{table*}

In Fig.~\ref{pdf_fig1}, the resulting NLO parton distributions are presented at $Q_0^2=1.9\GeV{}^2$ and $Q^2=m^2_\PW$.
The strange quark distribution $\cPqs(x,Q^2)$ and the ratio $R_{\cPqs}(x,Q^2)$ are illustrated in Fig.~\ref{Rs_fs_fig2} at the
same values of $Q$ as in Fig.~\ref{pdf_fig1}.
The total uncertainty in Fig.~\ref{pdf_fig1} is dominated by the parametrization uncertainty in
which most of the expansion in the envelope is caused by the decoupling parameter choice $B_{\cPaqs} \neq B_{\cPaqd}$.
The strange-quark fraction rises with energy and reaches a value comparable to that of $\cPqu$ and $\cPqd$ antiquarks at
intermediate to low $x$. Also, a suppression of $R_\cPqs$ at large $x$ is observed, which scales differently with the
energy. This result is consistent with the prediction provided by the ATLAS Collaboration~\cite{Aad:2012sb},
where inclusive $\PW$- and $\cPZ$-boson production measurements were used to determine
$r_{\cPqs}=0.5(\cPqs+\cPaqs)/\cPaqd$. In Ref.~\cite{Aad:2012sb}, the NLO value of
 $r_{\cPqs}= 1.03$ with the experimental uncertainty ${\pm}0.19_{\text{exp}}$ is quoted at $x= 0.023$ and $Q^2 = 1.9\GeV{}^2$. In the framework used, the two definitions of the strange-quark fraction are very
similar at the starting scale $Q^2_0$ and the values $R_\cPqs$ and $r_\cPqs$ can be directly compared.

In the free-$\cPqs$ fit, the strangeness suppression factor is determined at Q$^2 = 20\GeV{}^2$ to be
$\kappa_\cPqs = 0.52^{+0.12}_{-0.10}\,\text{(exp.)}^{+0.05}_{-0.06}\,\text{(model)} ^{+0.13}_{-0.10}\,\text{(parametrization)}$, which
is in agreement with the value~\cite{Samoylov:2013xoa} obtained by the NOMAD experiment at NNLO.

The impact of the measurement of differential cross sections of $\PW+\text{charm}$ production on the strange-quark distribution
and strangeness fraction $R_\cPqs$ is also examined by using the Bayesian reweighting~\cite{Ball:2010gb, Ball:2011gg} technique.
The results qualitatively support the main conclusions of the current NLO QCD analysis.
Details can be found in supplemental material.

\section{Summary}
\label{sec:summary}

The  $\wmunu$ lepton charge asymmetry  is measured in $\Pp \Pp$ collisions
at $\sqrt{s} = 7\TeV$
using a data sample corresponding to
 an integrated luminosity
of  4.7\fbinv collected with the CMS detector at the LHC
(a sample of more than 20 million  $\wmunu$  events).
The asymmetry is measured in 11 bins in absolute muon pseudorapidity, $\abs{\eta}$,
  for two different muon \pt thresholds,
 25 and 35\GeV.
 Compared to the previous
CMS measurement, this measurement
significantly reduces both the statistical and the systematic uncertainties.
The total uncertainty per bin is  0.2--0.4\%. The data are in good agreement with
 the theoretical predictions using CT10, NNPDF2.3, and HERAPDF1.5 PDF sets. The data are in poor agreement with the prediction based on the
MSTW2008 PDF set, although the agreement is significantly improved when
using the MSTW2008CPdeut PDF set.
 The experimental uncertainties are smaller than the current PDF uncertainties
 in the present QCD calculations. Therefore,
this measurement can be used  to significantly improve
the determination of PDFs in future fits.

This precise measurement of the  $\wmunu$ lepton charge asymmetry and the recent CMS measurement of associated  $\PW + \text{charm}$
production are used together with the cross sections for inclusive deep inelastic scattering at HERA
in an NLO QCD analysis of the proton structure. The muon charge asymmetry in $\PW$-boson production imposes strong
constraints on the valence-quark distributions, while the $\PW + \text{charm}$  process is directly sensitive to the strange-quark
distribution.

\section*{Acknowledgments}

\hyphenation{Bundes-ministerium Forschungs-gemeinschaft Forschungs-zentren} We congratulate our colleagues in the CERN accelerator departments for the excellent performance of the LHC and thank the technical and administrative staffs at CERN and at other CMS institutes for their contributions to the success of the CMS effort. In addition, we gratefully acknowledge the computing centers and personnel of the Worldwide LHC Computing Grid for delivering so effectively the computing infrastructure essential to our analyses. Finally, we acknowledge the enduring support for the construction and operation of the LHC and the CMS detector provided by the following funding agencies: the Austrian Federal Ministry of Science and Research and the Austrian Science Fund; the Belgian Fonds de la Recherche Scientifique, and Fonds voor Wetenschappelijk Onderzoek; the Brazilian Funding Agencies (CNPq, CAPES, FAPERJ, and FAPESP); the Bulgarian Ministry of Education and Science; CERN; the Chinese Academy of Sciences, Ministry of Science and Technology, and National Natural Science Foundation of China; the Colombian Funding Agency (COLCIENCIAS); the Croatian Ministry of Science, Education and Sport, and the Croatian Science Foundation; the Research Promotion Foundation, Cyprus; the Ministry of Education and Research, Recurrent financing contract SF0690030s09 and European Regional Development Fund, Estonia; the Academy of Finland, Finnish Ministry of Education and Culture, and Helsinki Institute of Physics; the Institut National de Physique Nucl\'eaire et de Physique des Particules~/~CNRS, and Commissariat \`a l'\'Energie Atomique et aux \'Energies Alternatives~/~CEA, France; the Bundesministerium f\"ur Bildung und Forschung, Deutsche Forschungsgemeinschaft, and Helmholtz-Gemeinschaft Deutscher Forschungszentren, Germany; the General Secretariat for Research and Technology, Greece; the National Scientific Research Foundation, and National Innovation Office, Hungary; the Department of Atomic Energy and the Department of Science and Technology, India; the Institute for Studies in Theoretical Physics and Mathematics, Iran; the Science Foundation, Ireland; the Istituto Nazionale di Fisica Nucleare, Italy; the Korean Ministry of Education, Science and Technology and the World Class University program of NRF, Republic of Korea; the Lithuanian Academy of Sciences; the Mexican Funding Agencies (CINVESTAV, CONACYT, SEP, and UASLP-FAI); the Ministry of Business, Innovation and Employment, New Zealand; the Pakistan Atomic Energy Commission; the Ministry of Science and Higher Education and the National Science Centre, Poland; the Funda\c{c}\~ao para a Ci\^encia e a Tecnologia, Portugal; JINR, Dubna; the Ministry of Education and Science of the Russian Federation, the Federal Agency of Atomic Energy of the Russian Federation, Russian Academy of Sciences, and the Russian Foundation for Basic Research; the Ministry of Education, Science and Technological Development of Serbia; the Secretar\'{\i}a de Estado de Investigaci\'on, Desarrollo e Innovaci\'on and Programa Consolider-Ingenio 2010, Spain; the Swiss Funding Agencies (ETH Board, ETH Zurich, PSI, SNF, UniZH, Canton Zurich, and SER); the National Science Council, Taipei; the Thailand Center of Excellence in Physics, the Institute for the Promotion of Teaching Science and Technology of Thailand, Special Task Force for Activating Research and the National Science and Technology Development Agency of Thailand; the Scientific and Technical Research Council of Turkey, and Turkish Atomic Energy Authority; the Science and Technology Facilities Council, UK; the US Department of Energy, and the US National Science Foundation.

Individuals have received support from the Marie-Curie programme and the European Research Council and EPLANET (European Union); the Leventis Foundation; the A. P. Sloan Foundation; the Alexander von Humboldt Foundation; the Belgian Federal Science Policy Office; the Fonds pour la Formation \`a la Recherche dans l'Industrie et dans l'Agriculture (FRIA-Belgium); the Agentschap voor Innovatie door Wetenschap en Technologie (IWT-Belgium); the Ministry of Education, Youth and Sports (MEYS) of Czech Republic; the Council of Science and Industrial Research, India; the Compagnia di San Paolo (Torino); the HOMING PLUS programme of Foundation for Polish Science, cofinanced by EU, Regional Development Fund; and the Thalis and Aristeia programmes cofinanced by EU-ESF and the Greek NSRF.

\bibliography{auto_generated}   
\clearpage
\ifthenelse{\boolean{cms@external}}{}{
\clearpage
\appendix
\section{Additional PDF distributions}
\label{sec:Supplemental material}
\begin{figure}[htb]
\centering
   \includegraphics[width=\cmsFigWidth]{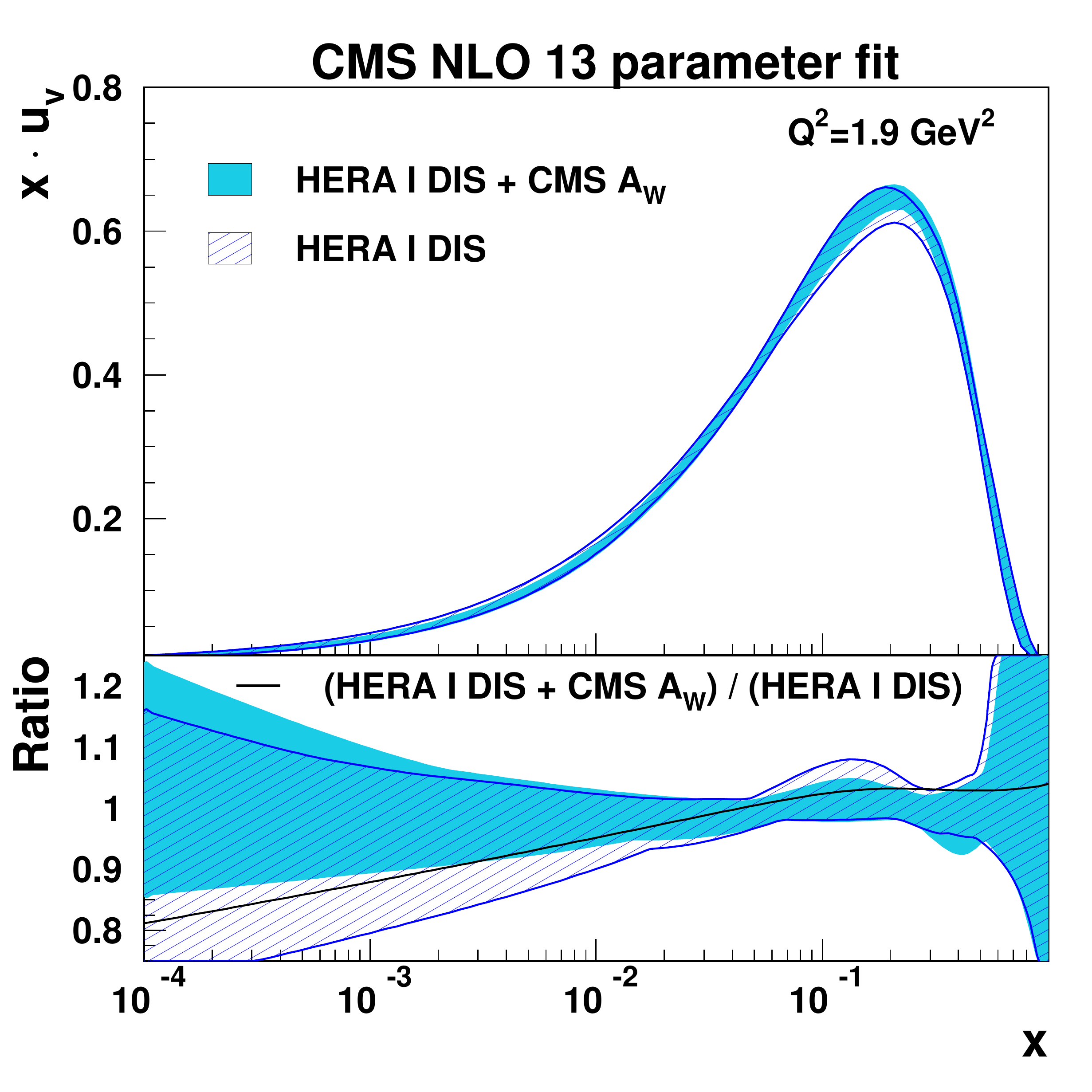}
   \includegraphics[width=\cmsFigWidth]{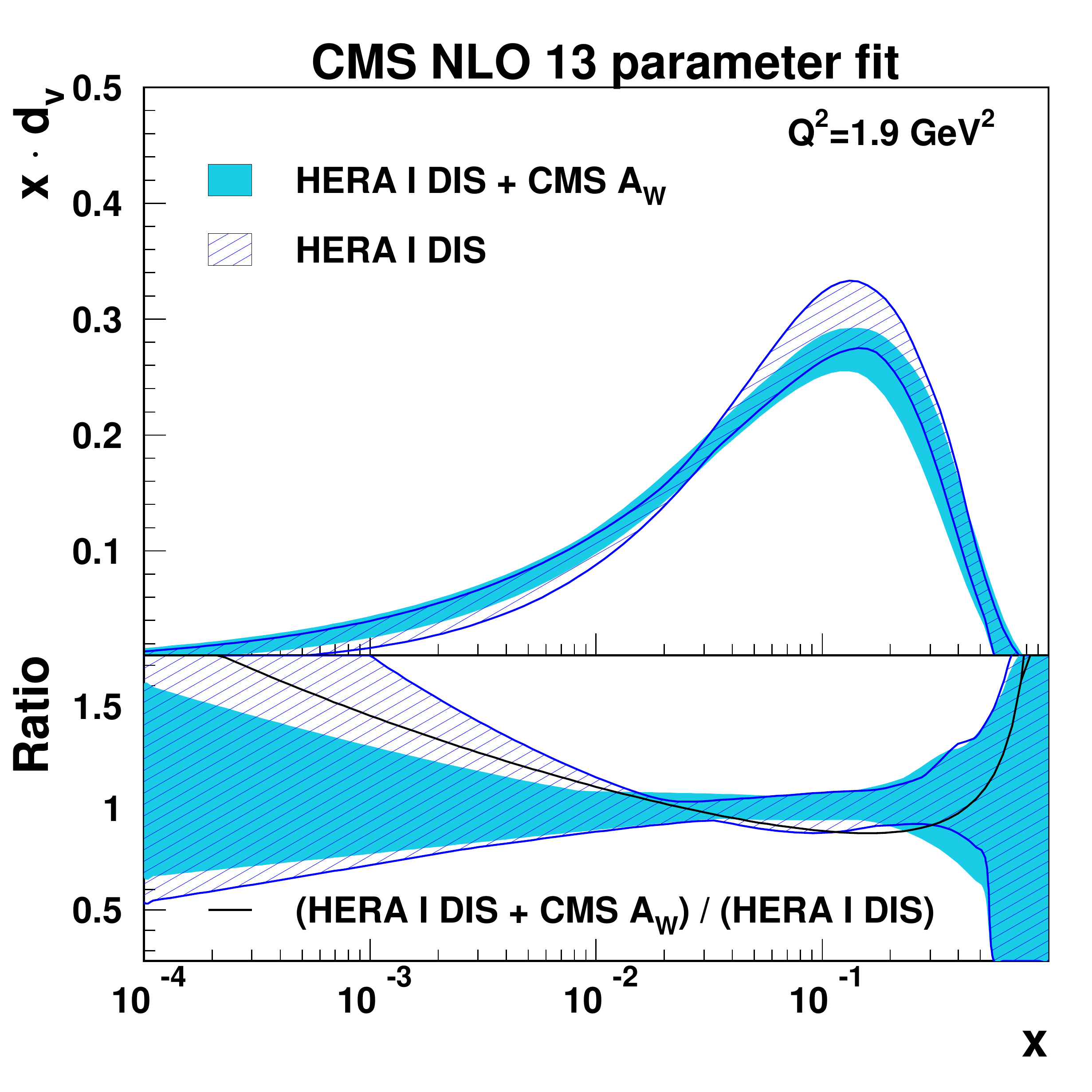}
\caption{Distributions of $\cPqu$ valence (\cmsLeft) and $\cPqd$ valence (\cmsRight) quarks as functions of $x$ at
the scale $Q^2=1.9\GeV{}^2$. The results of the 13-parameter fixed-$\cPqs$ fit to the HERA data and muon asymmetry measurements
(light shaded band), and to HERA only (dark hatched band) are compared. The total PDF uncertainties are shown. In the bottom
panels the distributions are normalized to one for a direct comparison of the uncertainties. The change of the PDFs
with respect to the HERA-only fit is represented by a solid line.}
\label{herapluswasym-tot-1.9Gev2}
\end{figure}

\begin{figure}[htbp]
\centering
   \includegraphics[width=\cmsFigWidth]{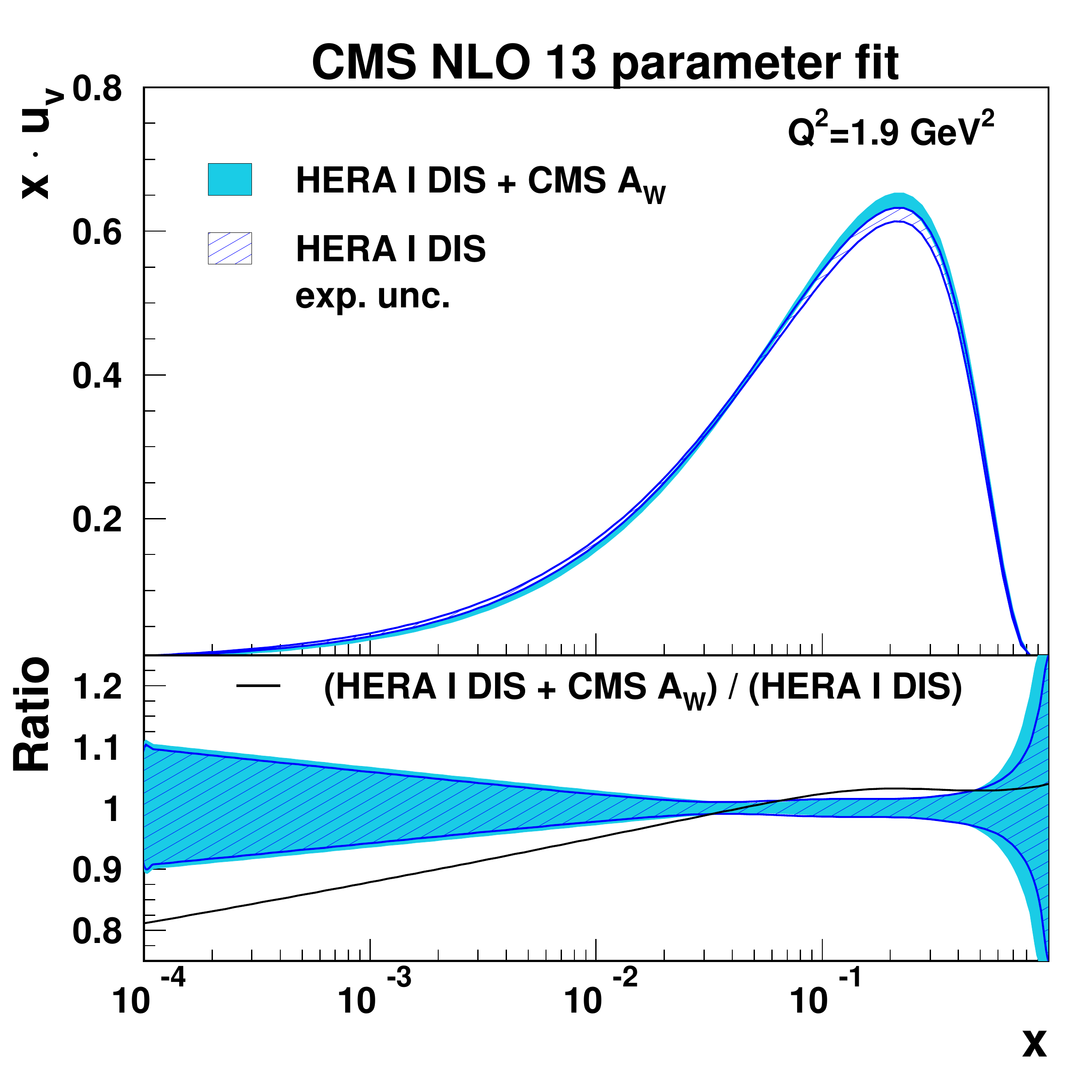}
   \includegraphics[width=\cmsFigWidth]{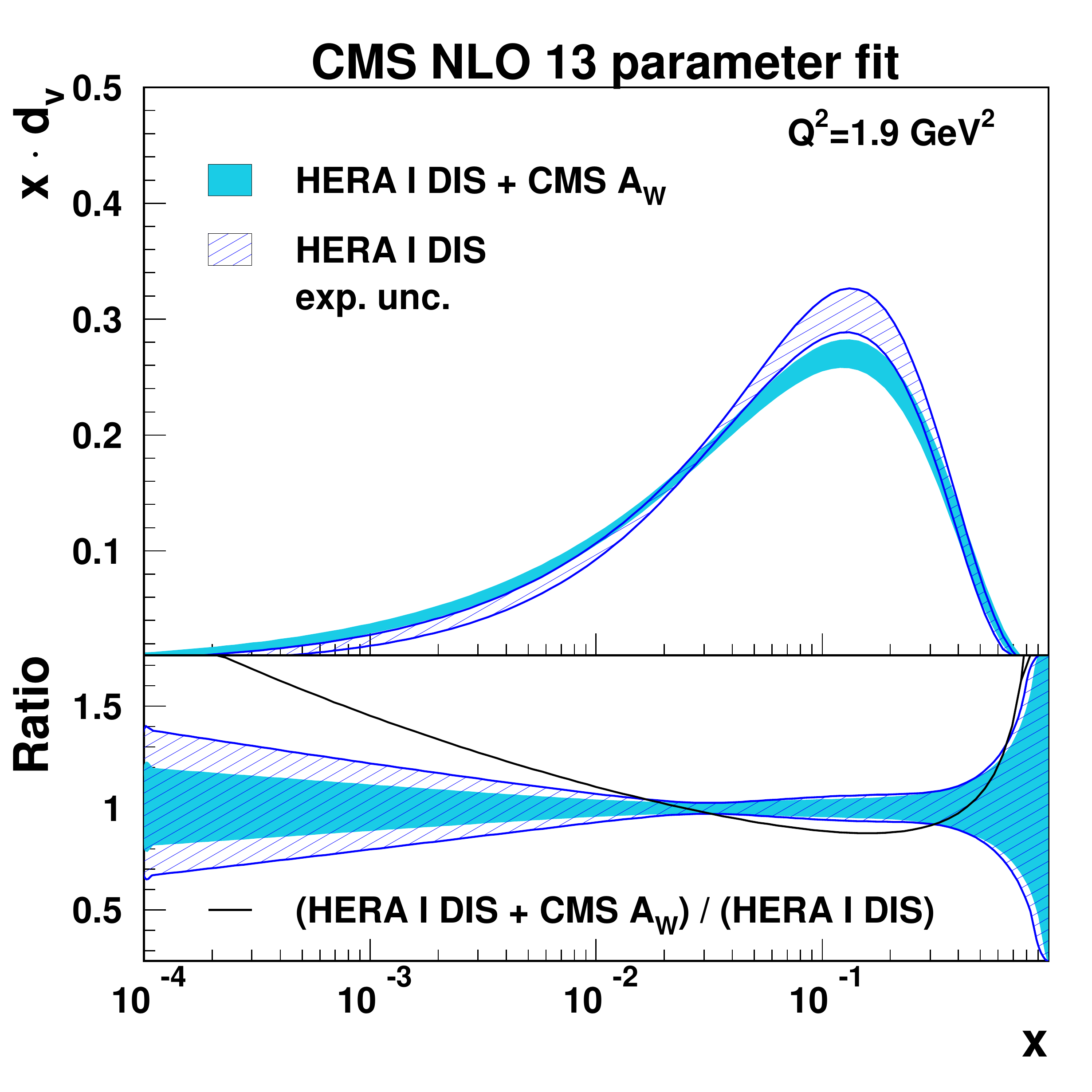}
\caption{Distributions of $\cPqu$ valence (\cmsLeft) and $\cPqd$ valence (\cmsRight) quarks as functions of $x$ at
the scale $Q^2=1.9\GeV{}^2$. The results of the 13-parameter fixed-$\cPqs$ fit to the HERA data and muon asymmetry
measurements (light shaded band), and to HERA only (dark hatched band) are compared. The experimental PDF
uncertainties are shown. In the bottom panels the distributions are normalized to one for a direct comparison
of the uncertainties. The change of the PDFs with respect to the HERA-only fit is represented by a solid line.}
\label{herapluswasym-exp-1.9Gev2}
\end{figure}

\begin{figure}[htb]
\centering
   \includegraphics[width=\cmsFigWidth]{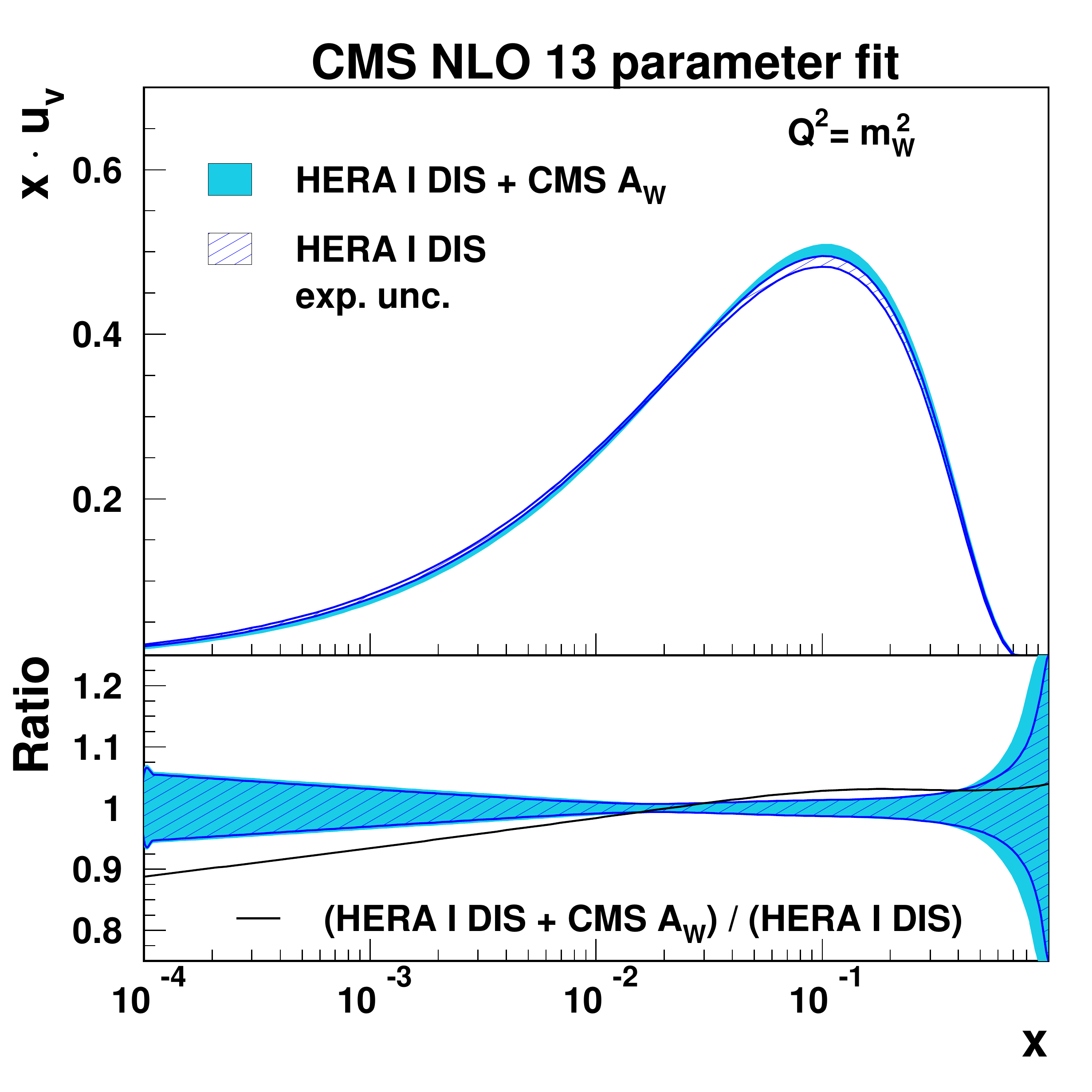}
   \includegraphics[width=\cmsFigWidth]{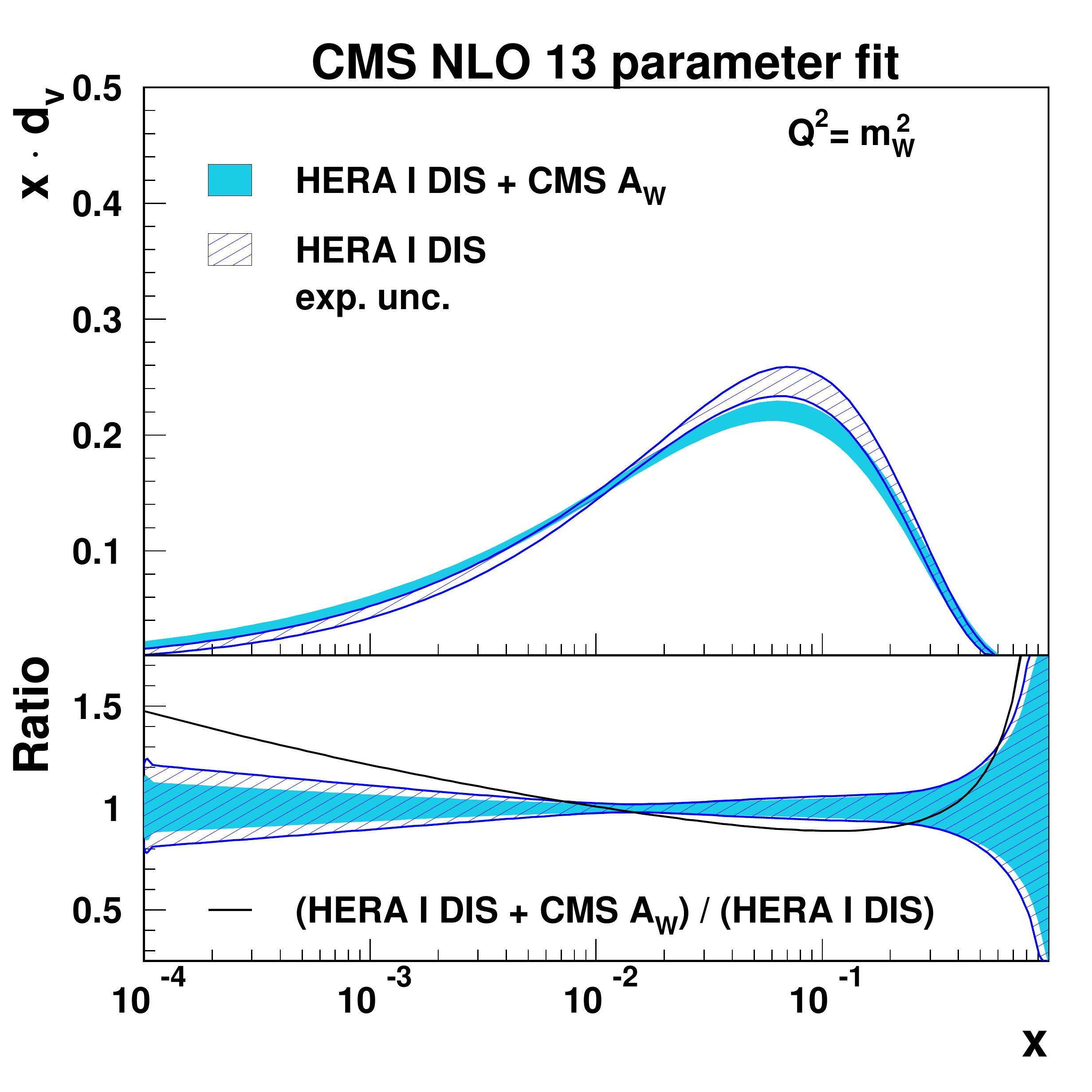}
\caption{Distributions of $\cPqu$ valence (\cmsLeft) and $\cPqd$ valence (\cmsRight) quarks as functions of $x$ at
the scale $Q^2=m_{\PW}^2$. The results of the 13-parameter fixed-$\cPqs$ fit to the HERA data and muon asymmetry
measurements (light shaded band), and to HERA only (dark hatched band) are compared. The experimental PDF
uncertainties are shown. In the bottom panels the distributions are normalized to one for a direct comparison
of the uncertainties. The change of the PDFs with respect to the HERA-only fit is represented by a solid line.}
\label{herapluswasym-exp-mw2}
\end{figure}

\begin{figure}[htbp]
\centering
   \includegraphics[width=\cmsFigWidth]{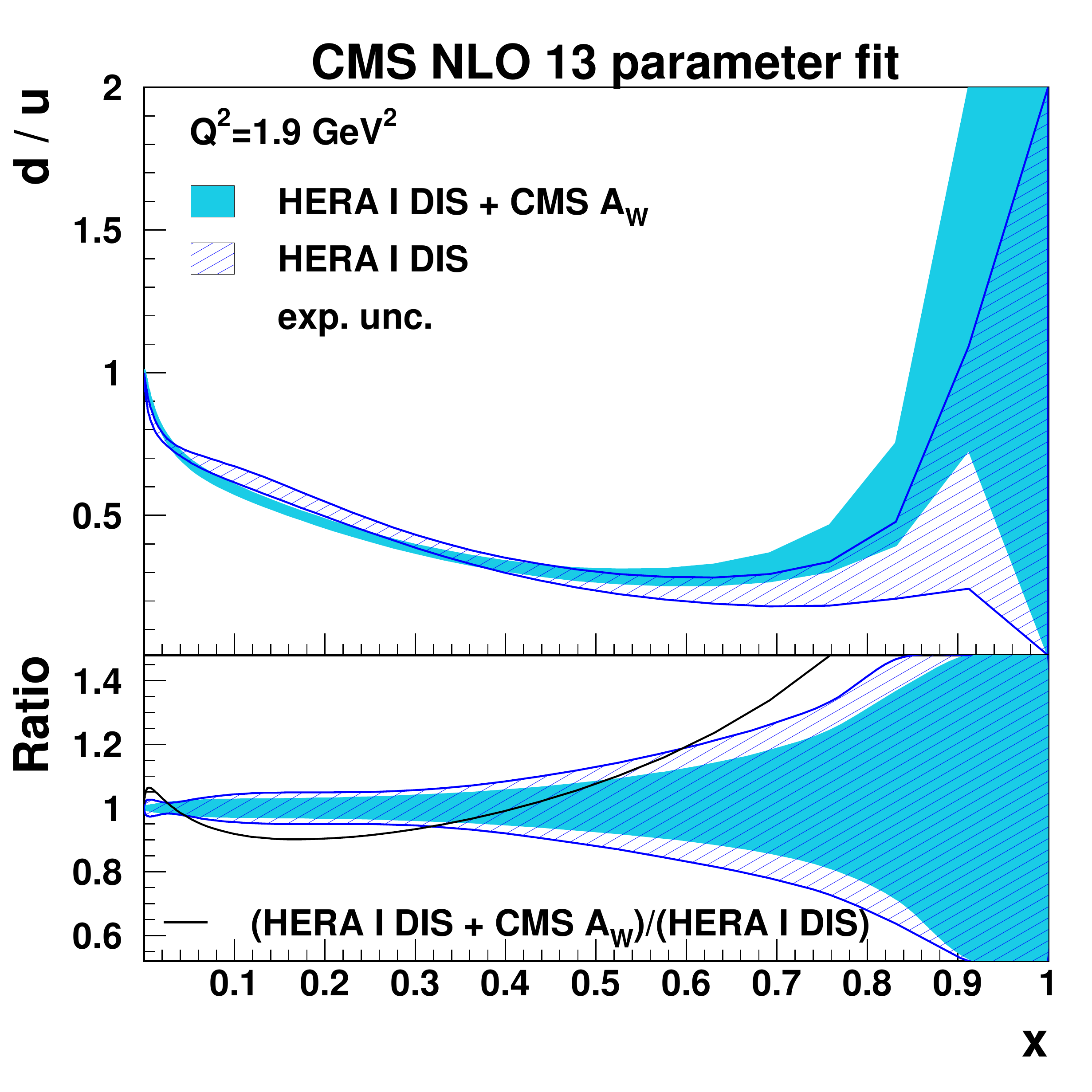}
   \includegraphics[width=\cmsFigWidth]{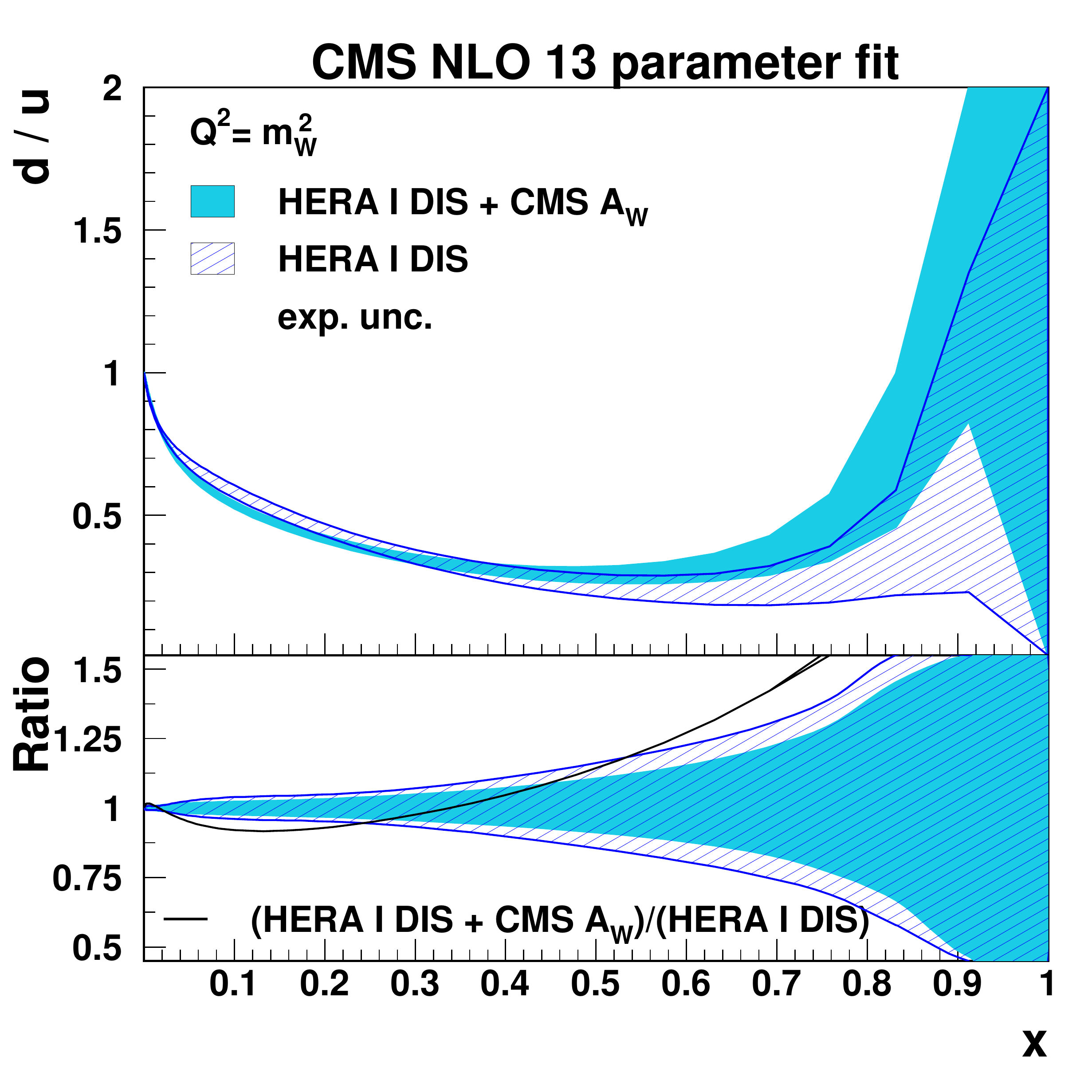}
\caption{Ratio of $\cPqd$- and $\cPqu$-quark distributions, $\cPqd/\cPqu$, presented as functions of $x$ at the
scales $Q^2=1.9\GeV{}^2$ (\cmsLeft) and $Q^2=m_{\PW}^2$ (\cmsRight). The results of the 13-parameter fixed-$\cPqs$ fit to the
HERA data and muon asymmetry measurements (light shaded band), and to HERA only (dark hatched band) are compared.
The experimental PDF uncertainties are shown. In the bottom panels the distributions are normalized to one for a
direct comparison of the uncertainties. The change of the $\cPqd/\cPqu$ ratio with respect to the result of the
HERA-only fit is represented by a solid line.}
\label{herapluswasym-exp-dtou}
\end{figure}

\begin{figure}[htb]
\centering
   \includegraphics[width=\cmsFigWidth]{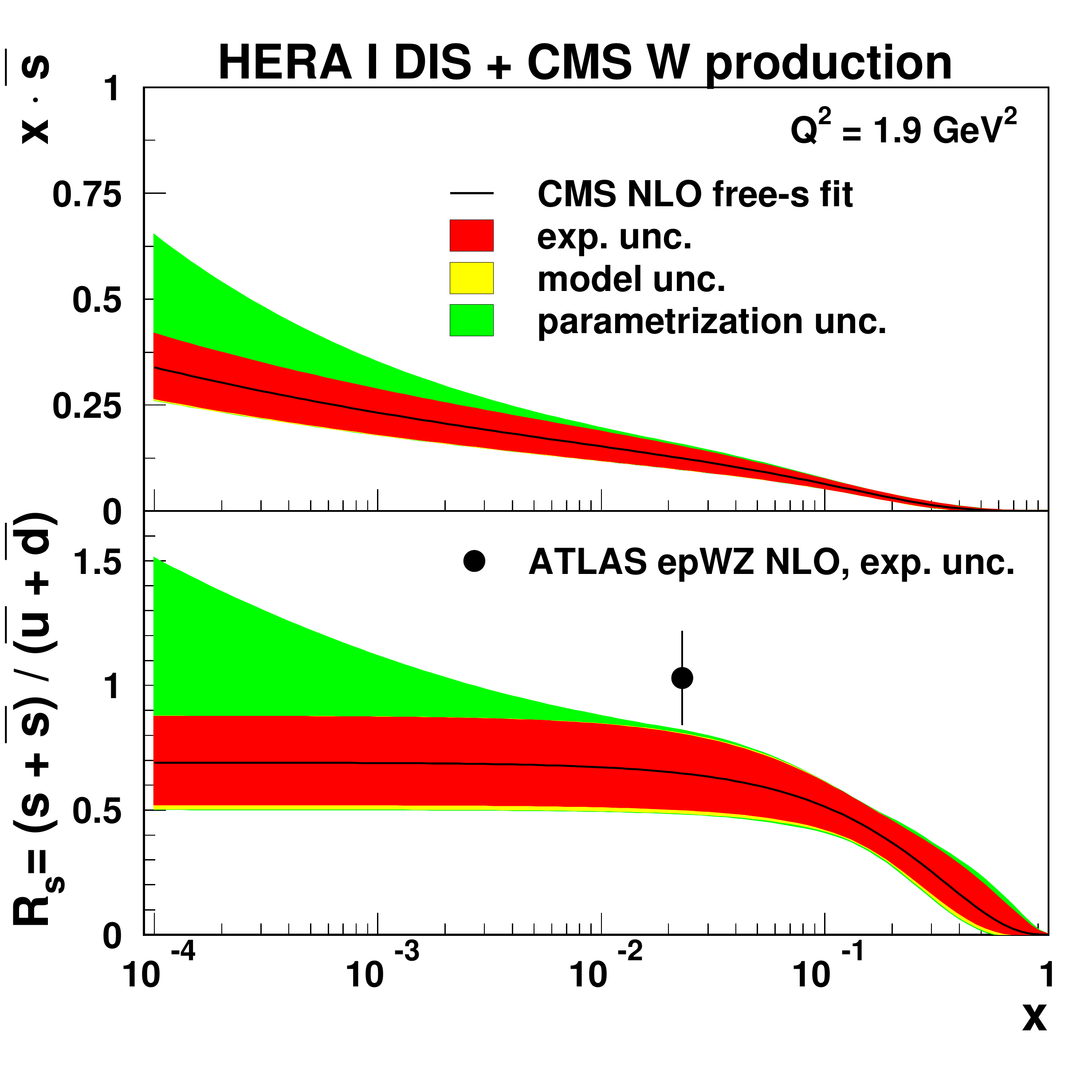}
\caption{Antistrange-quark distribution $\cPaqs(x,Q)$ (top) and the ratio $R_{\cPqs}(x,Q)$ (bottom), obtained
in the NLO QCD analysis of HERA and CMS data, shown as functions of $x$ at the scale $Q^2=1.9\GeV{}^2$.
The full band represents the total uncertainty. The individual contributions from the experimental, model,
and parametrization uncertainties are represented by the bands of different shades. For comparison, the NLO
result of the ATLAS analysis~\cite{Aad:2012sb} of $r_{\cPqs}=0.5(\cPqs+\cPaqs)/\cPaqd$ using inclusive
$\PW$-and $\cPZ$-boson production, is presented by a closed symbol. Only the experimental uncertainty from
ATLAS is available and is shown by the vertical error bar.}
\label{Rs_fs_fig2-atlas}
\end{figure}

\begin{figure}[htbp]
\centering
   \includegraphics[width=\cmsFigWidth]{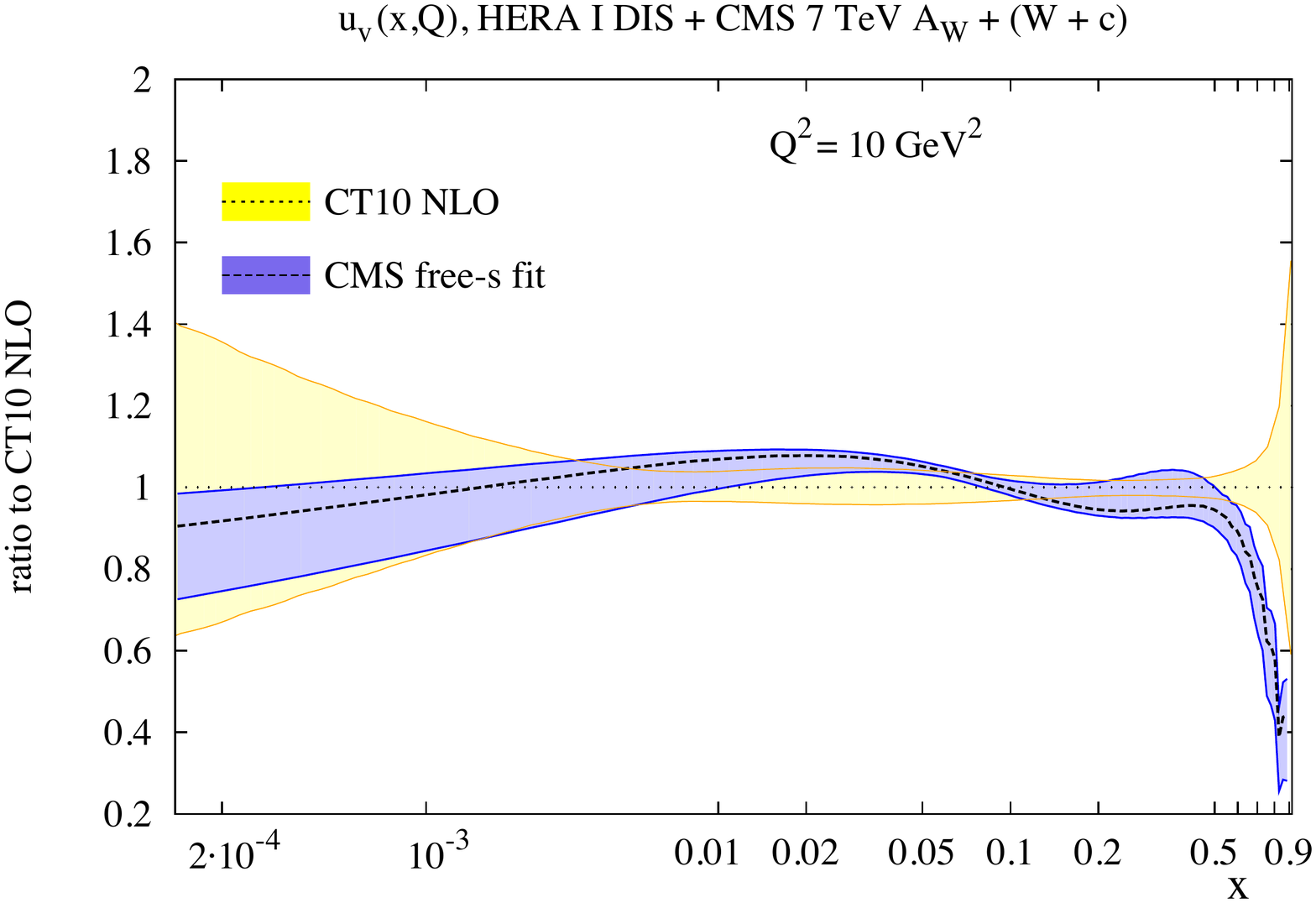}
   \includegraphics[width=\cmsFigWidth]{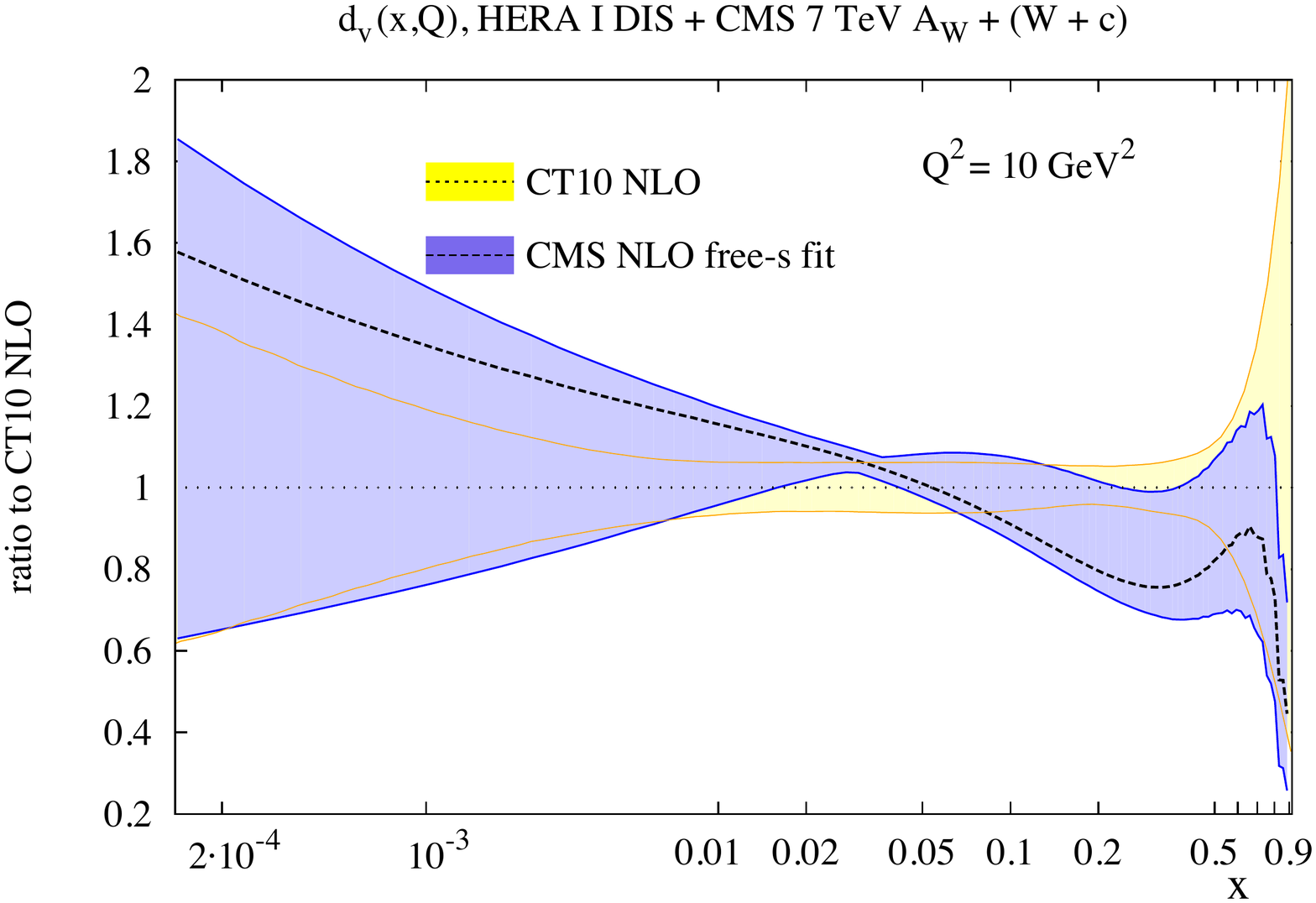}
   \includegraphics[width=\cmsFigWidth]{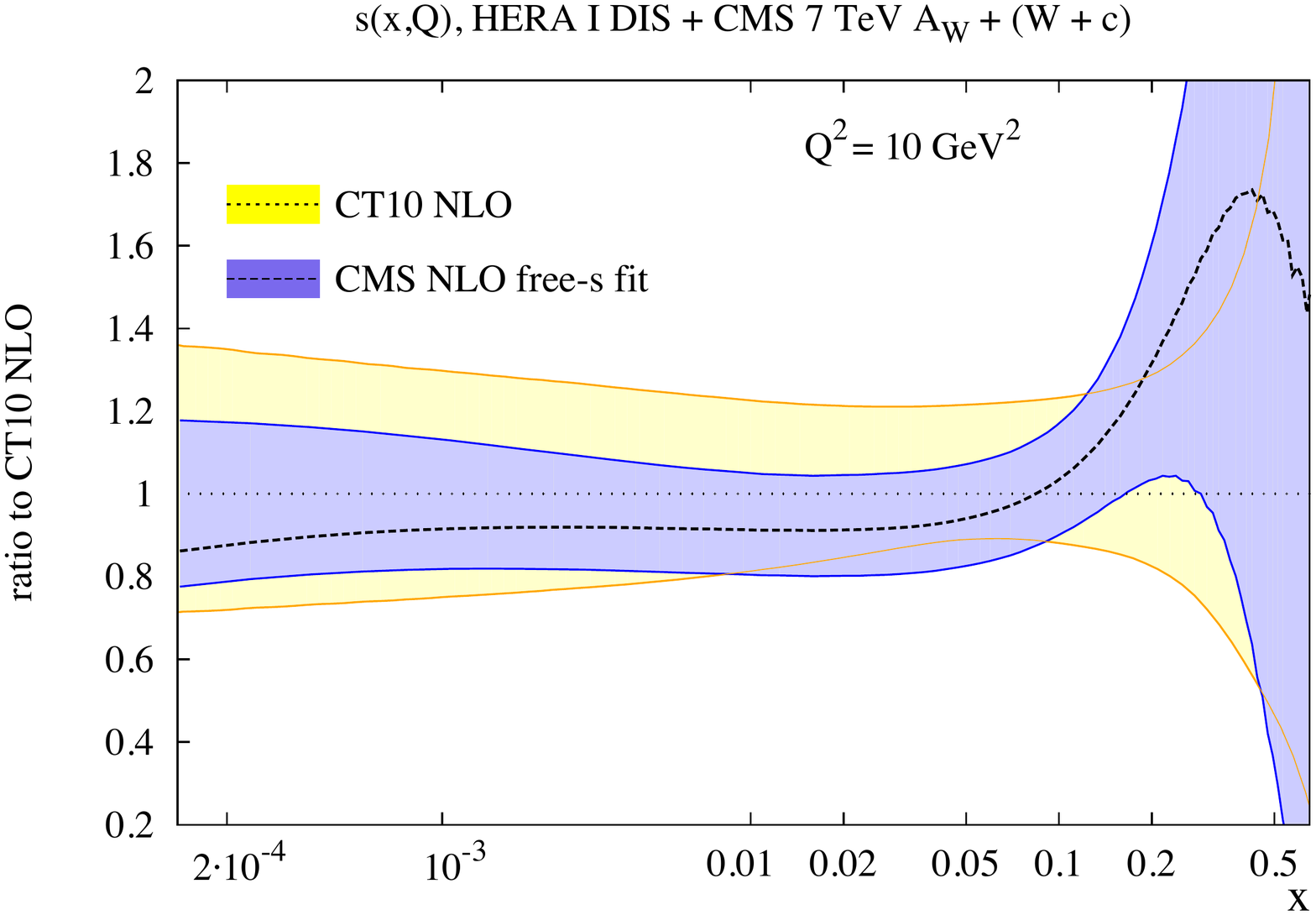}
\caption{The distributions of $\cPqu$ (top left), $\cPqd$ (top right), and $\cPqs$ (bottom) quarks, resulting from
the NLO QCD analysis of HERA and CMS data, shown as functions of $x$ at the scale $Q^2=10\GeV{}^2$ in comparison
to CT10NLO. The dark shaded band represents the total PDF uncertainty of the current fit, which is normalized to the
CT10NLO central value. The light hatched band represents the CT10NLO uncertainty normalized to one. All
uncertainties are given at 68\% CL.}
\label{uv-dv-s-vs-CT10}
\end{figure}

\begin{figure}[htbp]
\centering
   \includegraphics[width=\cmsFigWidth]{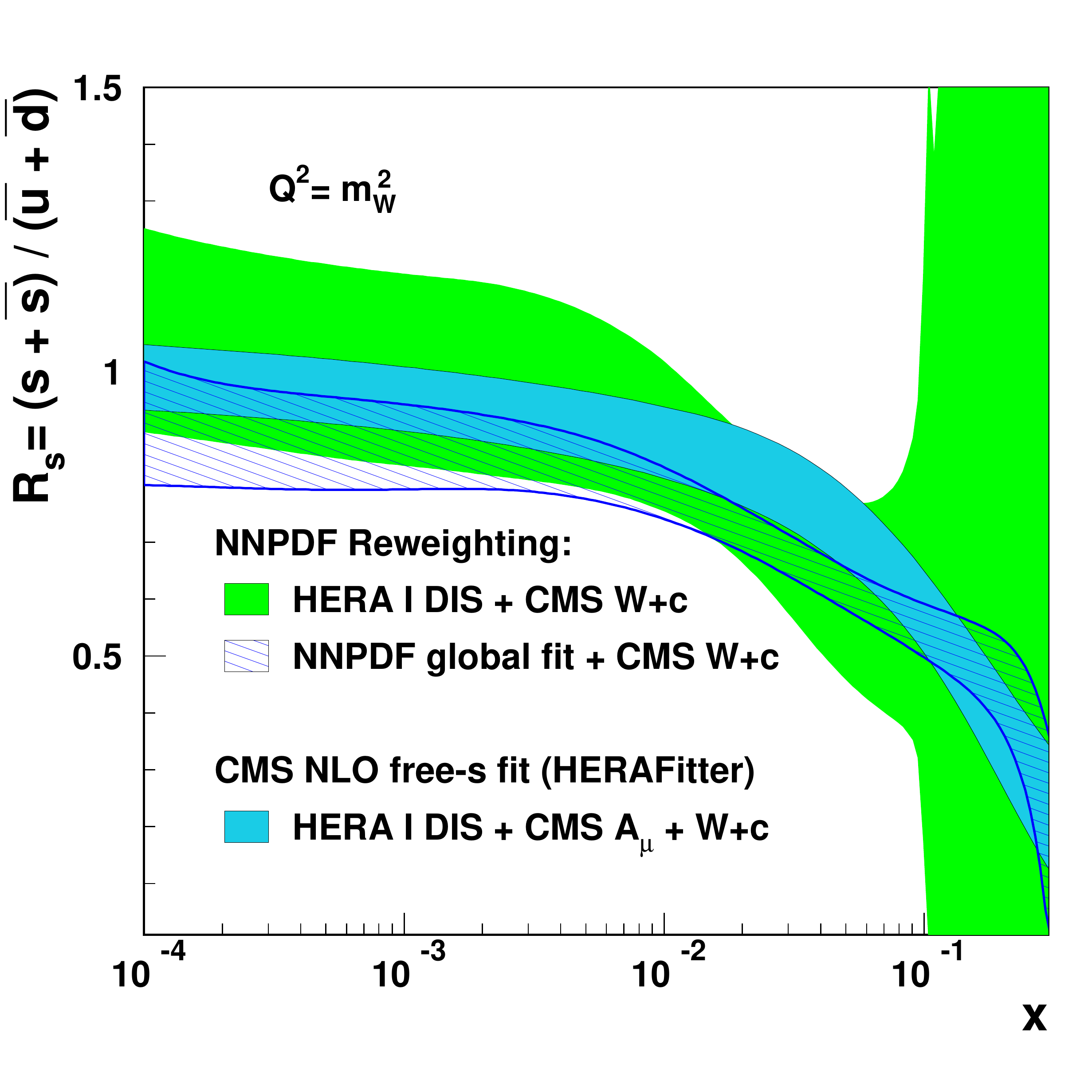}
\caption{Ratio $R_{\cPqs}(x,Q^2)$ resulting from the NLO QCD analysis of HERA and CMS data, presented as a function
of $x$ at the scale $Q^2=m^2_{\PW}$. The light shaded band represents the total PDF uncertainty of the CMS result. For comparison,
results of Bayesian reweighting using HERA I inclusive DIS data and the CMS measurement of $\PW + \text{charm}$ production
(dark shaded band). The reweighting results based on the data used in the global NNPDF2.3 fit and the CMS $\PW$ + charm production are represented by a hatched band.}
\label{nnpdf-comp-01}
\end{figure}

\begin{figure}[thbp]
\centering
   \includegraphics[width=\cmsFigWidth]{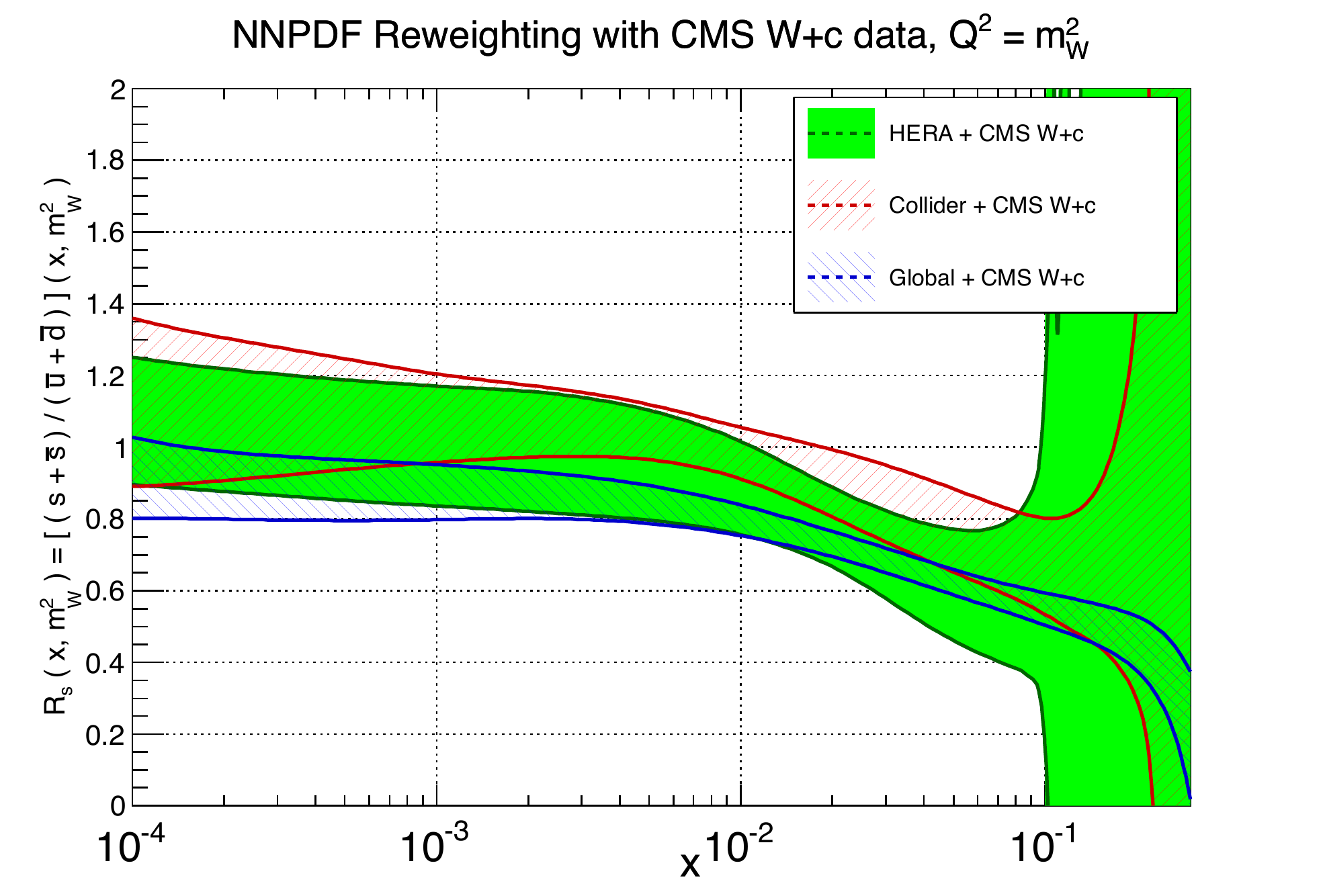}
\caption{Ratio $R_{\cPqs}(x,Q^2)$, obtained by using Bayesian reweighting, shown as a function of $x$ at the
scale $Q^2=m^2_{\PW}$. The dark shaded band represents the result based on the HERA I DIS and CMS $\PW$ + charm
data. The results of the reweighing obtained by using the CMS $\PW$ + charm measurements in addition to
collider-only data, and in addition to the data used in the global NNPDF2.3 analysis, are illustrated by
 bands of different hatches.}
\label{nnpdf-comp-02}
\end{figure}

}

\cleardoublepage \section{The CMS Collaboration \label{app:collab}}\begin{sloppypar}\hyphenpenalty=5000\widowpenalty=500\clubpenalty=5000\textbf{Yerevan Physics Institute,  Yerevan,  Armenia}\\*[0pt]
S.~Chatrchyan, V.~Khachatryan, A.M.~Sirunyan, A.~Tumasyan
\vskip\cmsinstskip
\textbf{Institut f\"{u}r Hochenergiephysik der OeAW,  Wien,  Austria}\\*[0pt]
W.~Adam, T.~Bergauer, M.~Dragicevic, J.~Er\"{o}, C.~Fabjan\cmsAuthorMark{1}, M.~Friedl, R.~Fr\"{u}hwirth\cmsAuthorMark{1}, V.M.~Ghete, C.~Hartl, N.~H\"{o}rmann, J.~Hrubec, M.~Jeitler\cmsAuthorMark{1}, W.~Kiesenhofer, V.~Kn\"{u}nz, M.~Krammer\cmsAuthorMark{1}, I.~Kr\"{a}tschmer, D.~Liko, I.~Mikulec, D.~Rabady\cmsAuthorMark{2}, B.~Rahbaran, H.~Rohringer, R.~Sch\"{o}fbeck, J.~Strauss, A.~Taurok, W.~Treberer-Treberspurg, W.~Waltenberger, C.-E.~Wulz\cmsAuthorMark{1}
% ...
\vskip\cmsinstskip
% ...
1:~~Also at Vienna University of Technology, Vienna, Austria\\
2:~~Also at CERN, European Organization for Nuclear Research, Geneva, Switzerland\\
% ...

\end{sloppypar}
\end{document}